\documentclass{aa}
\usepackage{graphicx}
\usepackage{txfonts}
\usepackage{natbib}
\usepackage{color}
\usepackage{url}

\begin{document}

   \title{The lithium isotopic ratio in very metal-poor stars}

   \author{K. Lind\inst{1,2}\and
           J. Melendez\inst{3}\and
           M. Asplund\inst{4}\and
           R. Collet\inst{4}\and
           Z. Magic\inst{1}
           }

   \institute{Max Plack Institute for Astrophysics, Karl-Schwarzschild-Strasse 1,
              857 41 Garching bei M\"unchen, Germany\and
              Institute of Astronomy, University of Cambridge, Madingley Road, Cambridge, CB3 0HA, United Kingdom\\
              \email{klind@ast.cam.ac.uk}\and
	    Departamento de Astronomia do IAG/USP, Universidade de S\~{a}o Paulo, Rua do Mat\~{a}o 1226, Cidade Universit\`{a}ria, 05508-900 S\~{a}o Paulo, SP, Brazil\and
	    Research School of Astronomy \& Astrophysics, Australian National University, Cotter Road, Weston Creek, ACT 2611, Australia
             }

   \date{Received 5 March 2013; accepted 25 May 2013}

\authorrunning{K.Lind et al.}  \titlerunning{The lithium isotopic ratio in very metal-poor stars}

  \abstract
   {Un-evolved, very metal-poor stars are the most important tracers of the cosmic abundance of lithium in the early universe. Combining the standard Big Bang nucleosynthesis model with Galactic production through cosmic ray spallation, these stars at $\rm[Fe/H]<-2$ are expected to show an undetectably small $\rm^6Li/^7Li$ isotopic signature. Evidence to the contrary may necessitate an additional pre-galactic production source or a revision of the standard model of Big Bang nucleosynthesis. It would also cast doubts on Li depletion from stellar atmospheres as an explanation for the factor 3--5 discrepancy between the predicted primordial $\rm^7Li$ from the Big Bang and the observed value in metal-poor dwarf/turn-off stars.}
   {We revisit the isotopic analysis of four halo stars, two with claimed $\rm^6Li$-detections in the literature, to investigate the influence of improved model atmospheres and line formation treatment.}
   {For the first time, a combined 3D, NLTE (non-local thermodynamic equilibrium) modelling technique for Li, Na, and Ca lines is utilised to constrain the intrinsic line-broadening and to determine the Li isotopic ratio. We discuss the influence of 3D NLTE effects on line profile shapes and assess the realism of our modelling using the Ca excitation and ionisation balance.}
   {By accounting for NLTE line formation in realistic 3D hydrodynamical model atmospheres, we can model the Li resonance line and other neutral lines with a consistency that is superior to LTE, with no need for additional line asymmetry caused by the presence of $\rm^6Li$. Contrary to the results from 1D and 3D LTE modelling, no star in our sample has a significant ($2\sigma$) detection of the lighter isotope in NLTE. Over a large parameter space, NLTE modelling systematically reduces the best-fit Li isotopic ratios by up to five percentage points. As a bi-product, we also present the first ever 3D NLTE Ca and Na abundances of halo stars, which reveal significant departures from LTE.}
   {The observational support for a significant and non-standard $\rm^6Li$ production source in the early universe is substantially weakened by our findings.}


   \maketitle

\section{Introduction}
To explain the complex abundance patterns of Li isotopes in the Galaxy, chemical evolution modelling must consider several poorly constrained production and destruction mechanisms. Stars, in particular, can act as both Li sources and sinks depending sensitively on the timescales and efficiencies of mixing processes in the stellar interior \citep[for reviews see][]{Pinsonneault97,Dantona10}. The most well-constrained production source is that of standard Big Bang nucleosynthesis based on the high-precision calculations of the baryon density in the early universe, as inferred from WMAP observations \citep{Komatsu11}. Using extended and improved nuclear networks, the most recent predictions by \citet{Coc12} imply $A(\rm^7Li)=\log(N(\rm^7Li)/N(\rm H))+12=2.72$ and $A\rm(^6Li)=-1.91$. These primordial abundance ratios have proven highly nontrivial to reconcile with the those measured in Population II stars  born shortly after the Big Bang. The well-documented, universal shortage of $^7\rm Li$ in these stars and the claimed detection of $^6\rm Li$ in at least some of them,  form the basis for two separate -- and opposite -- cosmological Li problems \citep[see e.g.][]{Asplund06,Fields11}. 

A key aspect is that many proposed solutions to one of the problems aggravate, rather than alleviate, the other. The best example, which has slowly been gaining traction over the last decade, is the suggestion that low-mass stars deplete $^7$Li from their atmospheres by gravitational settling over the course of their main-sequence life times \citep[e.g][]{Richard05b}. To constrain the physical processes that would regulate the depletion, many authors have endeavored to carefully map $^7$Li abundances in low-mass stars as function of age and/or mass and/or metallicity \citep{Korn07,Lind09b,Melendez10}. While the final word has not yet been said on the subject, it is considered a likely explanation that stellar depletion accounts for part or all of the missing $^7\rm Li$. 

For the lighter and more fragile $^6\rm Li$ isotope, the surface drainage would have a similar or even greater impact and hence it is of fundamental importance to constrain its abundance levels in metal-poor stellar atmospheres. The standard Big Bang scenario produces insignificant amounts of $^6\rm Li$, not detectable through spectral analysis of stars. Specifically, the primordial isotopic number density ratio, here denoted by $\rm^6Li/^7Li$, amounts to no more than $2.35\times10^{-5}$ \citep{Coc12}. Cosmic ray spallation is a more efficient production mechanism that may be constrained by the evolution of Be as traced by un-evolved metal-poor stars. Still, at $\rm[Fe/H]<-2$, the predicted level of $\rm^6Li/^7Li\la1\%$ \citep{Prantzos12} fall below the detection limit in stars even with the highest quality observational data. Hence, $^6\rm Li$-detections in any stars at this low metallicity call for either a revision of one of the known production channels, e.g. non-standard physics in the Big Bang \citep{Jedamzik09}, or for a third production channel, such as cosmic-ray production by Pop\,III stars \citep{Rollinde06} or during the shocks of large-scale structure formation \citep{Suzuki02}. \\

While upper limits to the $^6\rm Li$ abundances of metal-poor halo stars were determined already by \citet{Maurice84}, the first significant detection was claimed by \citet{Smith93} for the turn-off star HD84937 (see also \citealt{Hobbs94} and \citealt{Cayrel99}). When the isotopic analyses were extended to greater samples \citep{Smith98,Asplund06}, additional stars were found to have significant ($>2\sigma$) detections of $^6\rm Li$. Indeed, evidence for an upper envelope emerged, with several very metal-poor stars, two of them having $\rm[Fe/H]<-2.0$, clustered around $A\rm(^6Li)\approx1$. Correcting for the depletion factors of $^6\rm Li$ expected from standard stellar evolution theory before and during the main sequence, the highest initial abundances measured fall in the range $A\rm(^6Li)=1.1-1.3$. For very metal-poor stars, this is approximately one order of magnitude higher than expected from Galactic cosmic-ray production \citep{Prantzos12}. Such standard Li depletion is however not sufficient to explain the cosmological $^7$Li problem, which requires non-canonical processes to act below the convective envelope (e.g. gravitational settling of Li, rotation-induced shears, gravity waves). The predicted initial abundances of $^6\rm Li$ may therefore be even higher, further aggravating the problem. 

A revised analysis of the Li isotopic abundances of metal-poor stars is prompted by the considerable improvement in line formation modelling that has recently become possible. Specifically, realistic 3D radiation-hydrodynamical models of stellar surface convection can be conveniently tailored to the stellar parameters of individual stars and the expected strong departures from local thermodynamic equilibrium (LTE) in the metal-poor atmospheres can be properly resolved. That the simplifying assumptions of 1D, hydrostatic equilibrium and LTE influence the line profiles and may bias the determination of the isotopic ratio has been shown e.g. by \citet{Cayrel07}. The novel approach of our study is to account for non-LTE (NLTE) effects in 3D atmospheric models for both the Li line and Na and Ca lines used for calibration of rotational line broadening, a parameter that is highly degenerate with the isotopic ratio. This is a substantial improvement with respect to previous analyses that either subjected the Li line and the calibration lines to LTE modelling in 1D or 3D \citep[e.g.][]{Smith98,Asplund06}, or applied a 3D, NLTE method only to the Li line without any additional constraint on external line broadening \citep{Cayrel07,Steffen10a,Steffen10b}. Note that the results presented in this paper supersede the preliminary findings presented by \citet{Lind12c}.

\section{Observations}

Because of the high computational demand of 3D, NLTE analysis, we limit our first study on this subject to four metal-poor stars, spanning a range in well-constrained stellar parameters. HD19445 is a main-sequence star, HD84937 and G64-12 are turn-off stars,  and HD140283 a subgiant. 

For the above stars we have obtained spectra of superb quality using the HIRES spectrograph \citep{Vogt94} on the 10\,m Keck\,I telescope. We used the decker E4 that has a length of $7\arcsec$ (allowing a good sky subtraction) and a slit width of $0.4\arcsec$ (achieving a resolving power of about 10$^5$). The wavelength coverage is about $400-800$\,nm using a mosaic of three CCDs optimized for the blue, green and red spectral regions. The exposure time ranges from 10 min for the brightest star (HD140283) to six hours for G64-12, as detailed in Table \ref{tab:hires}. 

We extracted the spectral orders by hand using the IRAF package\footnote{\url{http://iraf.noao.edu/}} and also using the data reduction package MAKEE\footnote{\url{http://www.astro.caltech.edu/~tb/ipac\_staff/tab/makee/index.html}}, which was developed by T.~A.~Barlow specifically for reduction of Keck HIRES data and is optimized for spectral extraction of single point sources. The manual reduction with IRAF followed the standard procedures, correcting for bias, flat-fielding, cosmic rays and scattered light, then extraction of the orders (and sky subtraction) and finally wavelength calibration using the ThAr frames. We verified that both IRAF and MAKEE gave similar results and adopted the orders extracted with the MAKEE package, which was specifically designed for HIRES. For each individual frame the barycentric correction was applied. Further data reduction (continuum normalization and combining the different frames) was performed with IRAF.

Before combining the spectra we corrected each individual exposure to the rest frame, thus cancelling shifts due to guiding errors at the slit, changes to the optics of the spectrograph (e.g., due to temperature variations), or even small intrinsic radial velocity shifts. We used robust, outlier-resistant statistics to find the trimean radial velocity of each frame, achieving an internal line-to-line scatter of about 0.4 km/s, which is about 7 times better than the spectral resolution of the spectrograph  ($\sim$ 3 km/s), and a standard error of the mean of $\sigma_{m}$ = 0.06 km/s. Without discarding any potential outlier the achieved internal precision is $\sigma_{m}$ = 0.09 km/s, still significantly better than the resolution. Thus, precise zero-point corrections were applied to different exposures. Due to variations within the individual exposures, the resolving power of the combined spectrum is slightly lower than that measured using the ThAr frames (see below). We take this effect into account, so that our adopted resolving power is slightly lower than the measured value.

As the stars are very metal-poor and relatively warm (see Table\,\ref{tab:param}) there is a large number of continuum points to normalise the spectra. We tried different approaches to normalise them and the best results were obtained when selecting small spectral regions (about $\pm$3 \AA\ around the relevant lines) for continuum normalization, so that the lowest possible spline was used. The absolute continuum placement was further fine-tuned in the line-profile analysis (see Sect. 3.3).

We measured the resolving power in the different spectral orders using the ThAr exposures, so that we can precisely estimate the instrumental broadening for a given line.  The achieved resolving power in the wavelength region around the Li\,6707\AA\ line ranges from $R=90\,000-100\,000$, already taking into account the small corrections mentioned above. We verified that the ThAr lines can be reproduced very well with a Gaussian, thus our corrected resolving power was used to convolve the synthetic spectra using a Gaussian profile. The combined spectra have a signal-to-noise (S/N) per pixel of approximately $\rm S/N=800-1100$ around the Li\,6707\AA\ line. Thus, our data have both the required spectral resolution and S/N for the analysis of the Li isotopic ratio.

\begin{table*}
      \caption{Keck/HIRES observing log. The S/N value has been estimated in the region around the Li 6707\AA\ line.}
         \label{tab:hires}
         \centering
         \begin{tabular}{lccccc}
                \hline\hline
Star     &   V   & Observing date (UT) & Exp. time [s]  & Total Exp. time [s] & S/N \\
   \hline
   \noalign{\smallskip} 
HD19445  &  8.06 & 2005/10/22       & 3 x 500              & 1500 &  740 \\ 
HD84937  &  8.32 & 2006/01/19       & 2 x 400 + 3 x 500 + 4 x 600 & 4700 & 1030 \\ 
HD140283 &  7.21 & 2005/06/16       & 2 x 300              & 600  & 990 \\ 
G64-12 & 11.45 & 2005/06/16-17    & 18 x 1200  & 21600 & 820 \\ 
  \noalign{\smallskip}  
  \hline
         \end{tabular}
\end{table*} 

To investigate the Ca ionisation balance we complement the optical spectra with near infra-red FOCES spectra, covering the wavelength regions of the Ca\,II triplet lines at $\sim$8600\AA. The data were acquired in several observing runs between 1995 and 1999\footnote{The FOCES spectra have been collected by the members of the group lead by Prof. Thomas Gehren (LMU, Munich) and were kindly provided to us by M. Bergemann.}. FOCES is a fiber-fed spectrograph mounted on the 2.2\,m telescope on Calar Alto observatory \citep{Pfeiffer98}. The data have a $R\approx30\,000$ and $S/N\approx400$, with excellent continuum definition necessary to retain the shapes of the broad Ca lines.\\

\section{Analysis}

\subsection{Model atmospheres and stellar parameters}

For the spectral line formation calculations presented here, we adopted a set of time-dependent, 3D, hydrodynamical model stellar atmospheres of the halo stars in our sample \citep{Magic13}.
The models are based on radiation-hydrodynamical stellar-surface convection simulations generated with a custom version of the \textsc{Stagger}-code originally developed by Nordlund and Galsgaard\footnote{\url{http://www.astro.ku.dk/~kg/Papers/MHD_code.ps.gz}}. The simulations are part of a set of simulations of standard stars computed by \citet{Collet11a} and two of them were used by \citet{Bergemann12} for a NLTE study of Fe lines with average 3D model atmospheres. For a detailed account of the procedure used to generate such simulations, we refer to \citet{Magic13}.  We used the \textsc{Stagger}-code to solve the discretized, time-dependent, 3D, hydrodynamical equations for the conservation of mass, momentum, and energy in a representative volume located at the stellar surface. The simulation domains comprise the photosphere and the upper portion of the convection zone; more specifically, they cover typically about twelve pressure scale heights vertically and about ten granules at the surface at any one time. The corresponding spatial scale on the surface ranges from $6^2$\,Mm for the dwarf to $35^2$\,Mm for the subgiant. The simulation domains are discretized using a Cartesian mesh with a numerical resolution of $240^{3}$. The total stellar time covered by the simulations is 3--4\,h for the turn-off and subgiant and 0.5\,h for the dwarf. Open boundary conditions are assumed in the vertical direction and periodic ones horizontally.

During the 3D atmosphere modelling, energy exchange between gas and radiation is accounted for by solving the radiative transfer equation along the vertical and eight ($2\,\mu~\times~4\,\phi$) inclined rays cast through the simulation domain using a Feautrier-like method \citep{Feautrier64}. The non-grey character of radiative transfer in stellar atmospheres is approximated using an opacity-binning method \citep{Nordlund82,Skartlien00} with twelve opacity bins in the implementation described by \citet{Collet11a}. We use an updated version of the realistic equation of state by \citet{Mihalas88} and state-of-the-art continuous and line opacities for preparing the opacity bins \citep[see][for the complete references]{Magic13}.

We tailored the 3D models of our stars to reflect their expected parameters. Three stars have accurate parallax measurements that can be used to constrain the surface gravity to within $\pm0.05\rm\,dex$. For G64-12 we have instead considered the spectroscopic gravity estimate from Fe ionisation balance by \citet{Bergemann12} and from Str\"{o}mgren photometry by \citet{Nissen07}. We have targetted effective temperatures based on a comparison between Balmer line analysis and direct application of the infra-red flux method (IRFM), as summarised in Table \ref{tab:param}.  Only our model for HD84937 is somewhat cooler than the trusted indicators imply, which we account for in the error determination. 

\begin{table}
      \caption{Stellar parameters of the sample assumed in our 3D models and compared to literature values.}
         \label{tab:param}
         \centering
         \begin{tabular}{lrrrr}
                \hline\hline
               &    G64-12  & HD140283    &   HD84937    &   HD19445 \\
          \hline
	\noalign{\smallskip}	
	$T_{\rm eff}$  H$\alpha^a$     & ... & 5753 & ... &  5980  \\ 
	$T_{\rm eff}$  H$\beta^b$      & 6435 & 5849 & 6357 &  ...  \\ 
	$T_{\rm eff}$  IRFM$^c$         & 6464 & 5777 & 6408 &  6135 \\          
	\noalign{\smallskip}	
	$T_{\rm eff}$   model  & 6428 & 5780 & 6238 &  6061  \\  	
	\noalign{\smallskip}	
	\hline
	\noalign{\smallskip}	
	$\log(g)$  ast.$^{a,b}$           & (4.26) &  3.72 &  4.07 &  4.42 \\
	$\log(g)$  spec.$^d$              & 4.34 &  3.63 &  4.28 &  ...  \\
	\noalign{\smallskip}	
	$\log(g)$   model         & 4.20  &  3.70    &  4.00 &  4.50  \\  
	\noalign{\smallskip}	
                   \hline
	\noalign{\smallskip}	
	$\rm[Fe/H]$  spec. $^{a,b}$                  & -3.24 &  -2.38 & -2.11  &  -2.02  \\
	$\rm[Fe/H]$ spec.   $^{d}$                    & -3.16 &  -2.41 &  -2.04 &  ...  \\	
	\noalign{\smallskip}	
	$\rm[Fe/H]$  model               & -3.00  &  -2.50    &  -2.00 &  -2.00  \\  
	\noalign{\smallskip}	
                   \hline
         \noalign{\smallskip}	
          \multicolumn{5}{l}{$^a$ \citet{Asplund06}}\\
          \multicolumn{5}{l}{$^b$ \citet{Nissen07}. Note that $\log(g)$ of G64-12  was not derived}\\
          \multicolumn{5}{l}{ from parallax measurements, but from Str\"{o}mgren photometry.}\\
          \multicolumn{5}{l}{$^c$ \citet{Casagrande10}}\\
          \multicolumn{5}{l}{$^d$ \citet{Bergemann12}}\\          
         \noalign{\smallskip}	
         \end{tabular}
\end{table}

\subsection{LTE and NLTE line formation}

\begin{figure}
\begin{minipage}[b]{\linewidth}
\centering
\includegraphics[width=\textwidth]{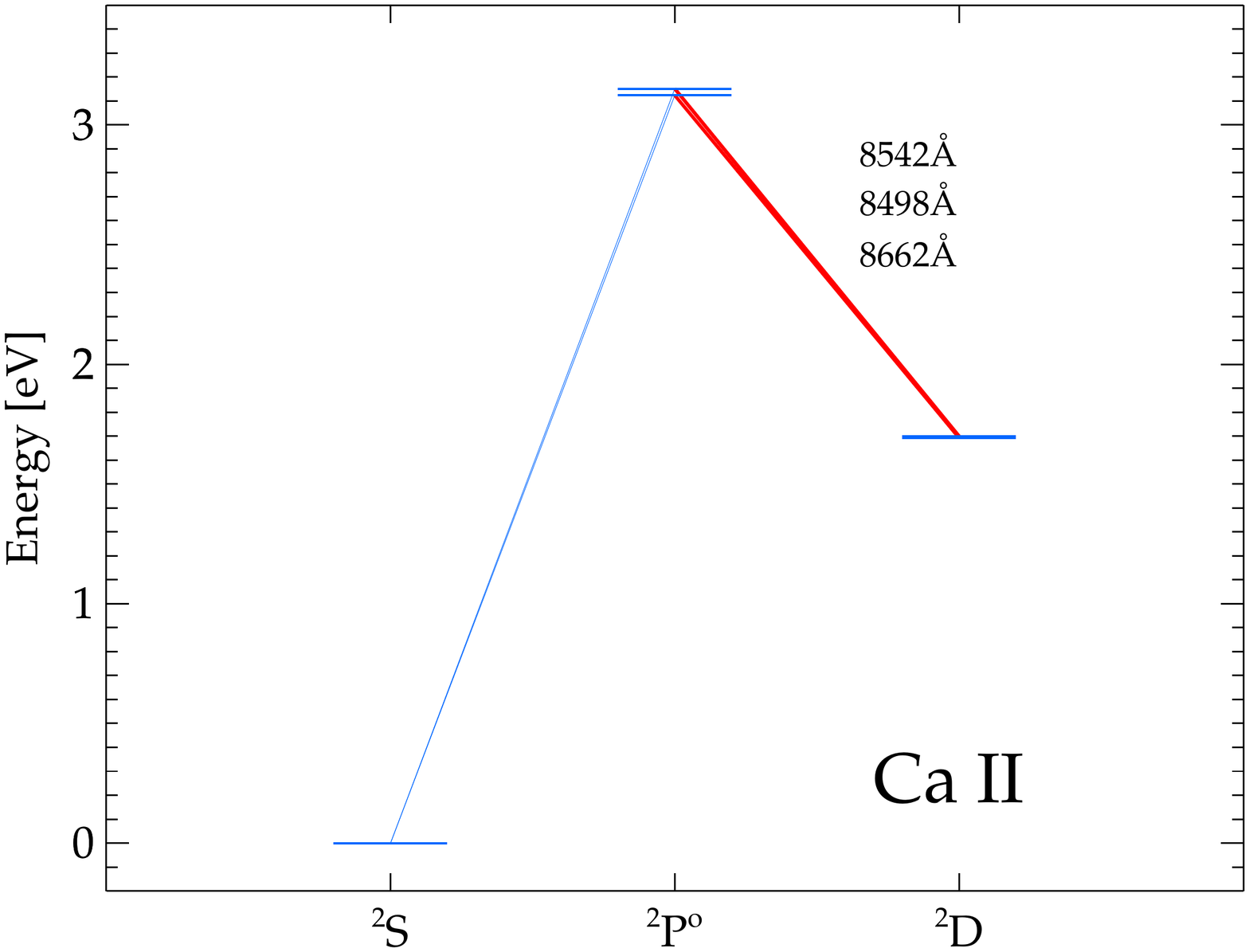}
\end{minipage}
\hspace{0.5cm}
\begin{minipage}[b]{\linewidth}
\centering
\includegraphics[width=\textwidth]{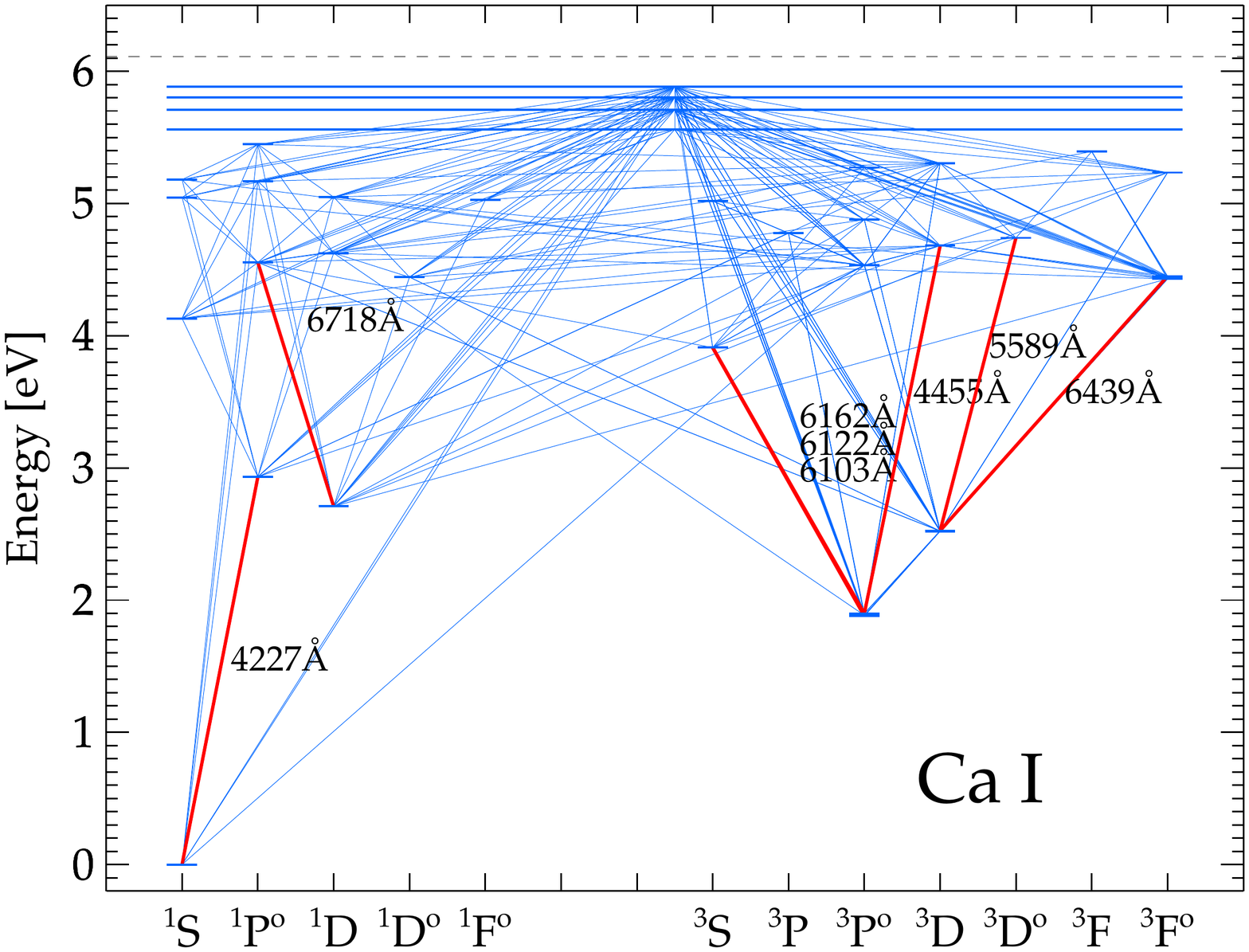}
\caption{Term diagrams of the Ca model atom used in the NLTE analysis. The dashed horisontal line marks the ground state of Ca\,II. The four states of Ca\,I with highest excitation potential are super-levels, corresponding to all or some of the individual fine-structure levels for $n=5-8$. The transitions used for detailed spectral analysis are marked with thick lines, and their approximate wavelengths given in \AA .}
\label{fig:termdiag}
\end{minipage}
\end{figure}

To compute synthetic spectra of Li and Ca lines, we utilized two different 3D spectrum synthesis codes; \textsc{Scate} \citep{Hayek11} and \textsc{Multi3d} \citep{Botnen97,Leenaarts10}. The technique was described in \citet{Lind12c}, but we reiterate the main points here. 

The spatial and temporal average flux profile in LTE was computed with \textsc{Scate} solving the radiative transfer equation for $6\mu\times4\phi$-angles for 20 snapshots in time. For each line, the profile was resolved by 100 wavelength points equidistant by 0.5\,km/s in velocity space and the logarithmic oscillator strength was varied by $\pm0.3\,$dex to allow for accurate interpolation and $\chi^2$-analysis with respect to observed data. Four different isotopic ratios were assumed, spanning the range: $\rm^6Li/^7Li=0.0-0.06$. Negative ratios and intermediate ratios were obtained by inter/extrapolation of the line profiles. Negative values are unphysical, but required for accurate $\chi^2$-minimisation. The adopted line data for the main transitions are summarized in Table \ref{tab:abund} and the fine- and hyper-fine subcomponents of the Li lines are detailed in \citet{Lind09a}.  

NLTE calculations are significantly more time consuming than LTE calculations and a few simplifications are necessary to make these tractable. We chose to calculate the mean NLTE/LTE profile ratio from four snapshots using \textsc{Multi3d} and thereafter multiply the ratio with LTE profiles computed with \textsc{Scate} for the same abundances. The calculations were performed for a range of abundances for each individual line, except that $\rm^6Li/^7Li=0.0$ for the Li line, i.e. we assume that the NLTE/LTE profile ratio is independent of the isotopic ratio. In analogy with LTE calculations,  $6\mu\times4\phi$-angles were used. We have confirmed that the agreement between the two codes is satisfactory in LTE. 

The three model atoms include, respectively, 40 levels of Ca\,I, five levels of Ca\,II, and the Ca\,III ground state, 20 levels of Li\,I and the Li\,II ground state, and 20 levels of Na\,I and the Na\,II ground state. A more detailed description of the atomic data can be found in \citet{Lind09a,Lind11b,Lind12c}. We note that the statistical equilibrium calculations for Li and Na are well constrained thanks to rigorous quantum mechanical calculations of radiative and collisional transition probabilities \citep[see e.g.][]{Barklem10}. However, as is often the case in NLTE modelling \citep{Asplund05}, the main uncertainty in the calculations for Ca arises from the unknown efficiency of inelastic collisions between Ca and H\,I atoms. We have here assumed a scaling factor $S_{\rm H}=0.1$ to the rates computed with the traditional \citet{Drawin68} recipe, as suggested by \citet{Mashonkina07}, and investigate the consistency between NLTE abundances derived from different lines (see Sect. 4.2). Schematic term diagrams are illustrated for Ca in Fig.\,\ref{fig:termdiag}, with the important transitions marked with thick lines.    

\begin{figure*}
\centering
\includegraphics[width=19cm,viewport=0cm 0cm 25cm 20.7cm]{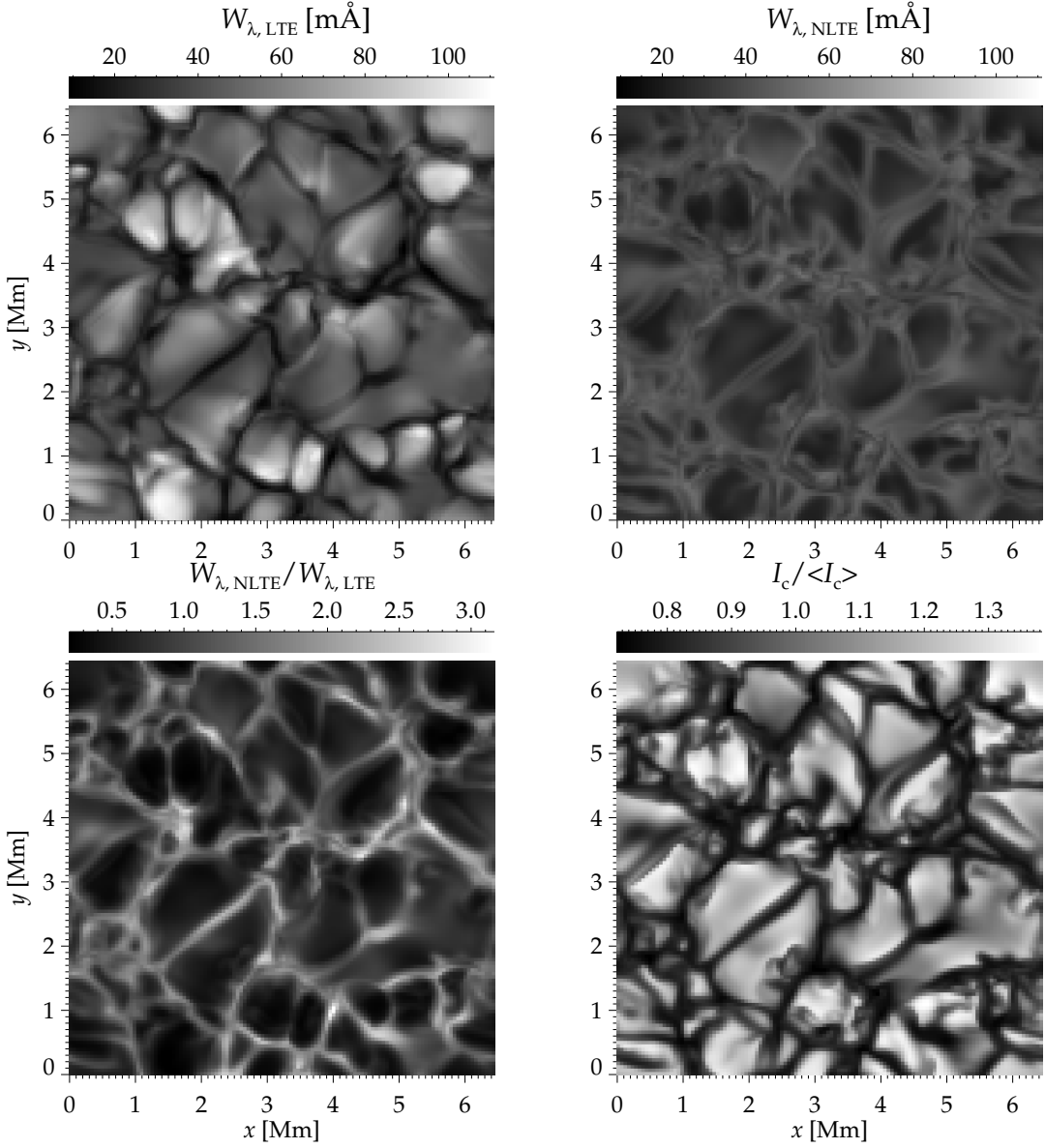}
\caption{\textit{Top left panel:} Spatially resolved LTE equivalent widths at central intensity ($\mu=1$), of the Li\,I 6707\AA\ line in one snapshot of the 3D model of HD19445. \textit{Top right panel:}  The same for NLTE. \textit{Bottom left panel:} Ratio between NLTE and LTE equivalent widths. \textit{Bottom right panel:}  The simulated relative continuum intensity at 6707\AA .}
\label{fig:xy}
\end{figure*}

The NLTE effects on Li, Na, and Ca lines are qualitatively very similar in these atmospheres. The steep temperature stratification of the bright granules of up-flowing gas efficiently boosts the over-ionisation, by giving rise to a strongly super-thermal radiation field in the ultra-violet wavelength regions. This is particularly true at low metallicity where 3D hydrodynamical models have much cooler outer atmospheric layers than classical 1D hydrostatic models \citep[see e.g.][]{Asplund99,Asplund03}. Vice versa, the intergranular lanes of downflowing gas suffer from the opposite effect, which is over-recombination of the neutral atoms compared to LTE. As shown in Fig.\,\ref{fig:xy}, the Li line can be both strengthened and weakened by up to as much as a factor of three in NLTE with respect to LTE. Because the bright regions have the dominant influence over the spatially averaged line profile, the net effect is a considerable weakening of the lines, which must be compensated for by a significant abundance increase. Resonance lines are particularly sensitive to the effects of over-ionisation, because of their higher temperature sensitivity. 

\begin{figure}[htbp]
\begin{center}
\includegraphics[scale=0.35,viewport=5cm 0cm 25cm 21cm]{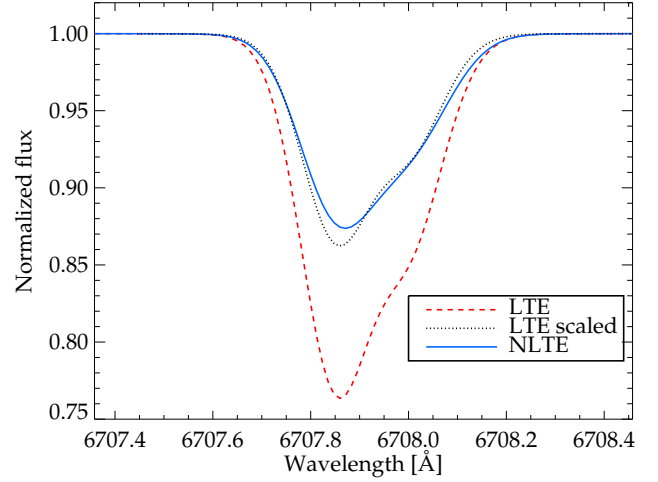}
\caption{Example synthetic profiles in LTE (\textit{dashed}) and NLTE (\textit{solid}) of the Li resonance line, computed with the same abundance ($A\rm(^7Li)=2.0$, $\rm^6Li/^7Li=0.0$, $v_{\rm rot}\sin{i}=0.0$) for the model of HD140283. Also shown is an LTE line profile interpolated to meet the same equivalent width as the NLTE line (\textit{dotted}). More absorption appears in the red wing relative to the blue in NLTE due to strongly differential effects in granules and inter-granular lanes. }
\label{fig:linediff}
\end{center}
\end{figure}

While the total line strength of the Li resonance line determines the $^7$Li-abundance, it is the shape of the line profile that determines the isotopic ratio due to the shift between $^6\rm Li$ and $^7\rm Li$ isotopic components. It is therefore critical to resolve the strongly differential NLTE effects on the granules and inter-granular lanes, because they have a preferential influence over the blue- and red-shifted part of the line profile, respectively. As seen in Fig.\,\ref{fig:xy}, the line is stronger in the granules in LTE, compared to the lanes, while the opposite is true in NLTE \citep{Asplund03}. At a given line strength, the net result is a stronger depression in the red wing of the line, as shown in Fig.\,\ref{fig:linediff}. The contrast in line strength over the surface has also decreased significantly in NLTE, because the line formation is decoupled from the strong temperature and density inhomogeneities that depict the LTE line formation. 

Because of the added absorption in the red wing, the spatially averaged line profiles are broadened in NLTE. Consequently, when constraining the unknown external line broadening that is caused by the rotation of the star, the resulting $v_{\rm rot}\sin{i}$-values decrease by $0.7-1.9\rm\,km/s$ (see Table\,\ref{tab:isotope}). Less additional broadening is needed to reproduce the same observed line profile. Except for the difference in width that can be compensated for by external broadening, there is a small, but not completely negligible difference in the shape of the line profile when lifting the LTE assumption. This influences the isotopic ratios, as demonstrated in the following section.

\begin{figure}[htbp]
\begin{center}
\includegraphics[scale=0.35,viewport=5cm 0cm 25cm 21cm]{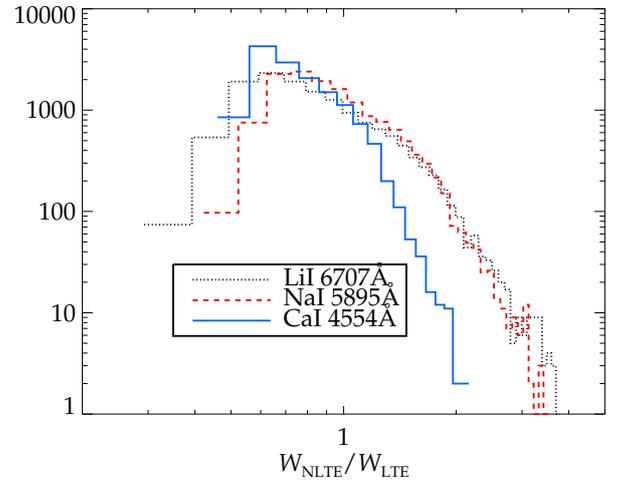}
\caption{Histograms of the ratio between the spatially resolved equivalent widths at disk center intensity found in NLTE and LTE for a snapshot of G64-12. }
\label{fig:hist}
\end{center}
\end{figure}

Similar behavior is seen for all elements, but compared to Li the balance is shifted to a greater degree of over-recombination for Na and over-ionisation for Ca. The net result is smaller positive abundance corrections for Na and larger for Ca than for Li. In Fig.\,\ref{fig:hist} we demonstrate with histograms how the line strength changes in NLTE for the different elements in a selected snapshot of the G64-12 model. Evidently, the important line-strengthening that occurs in the lanes due to over-recombination has an almost identical influence on Li and Na. For Ca, this effect is not as strong.

\subsection{$\chi^2$-minimisation}

The Li isotopic abundances were determined by $\chi^2$-minimisation between observed and synthetic spectra in a region extending $\pm1.7$ \AA\ from the line centre. A $\chi^2$-matrix was formed, with the number of dimensions equal to the number of free parameters, either four or five, as described below. The analysis was performed using the standard definition of $\chi^2$:

   \begin{equation}
      \chi^2 = \sum\limits_{N_{\rm data}}\frac{(F_{\rm obs}-F_{\rm mod})^2}{\sigma^2} \\
   \end{equation}

Here, $F$ denotes the normalised fluxes of model and observation, $\sigma=(S/N)^{-1}$ is the estimated error of the observed flux per pixel and $N_{\rm data}$ the number of pixels used to fit the full line profile. In order to make a fair comparison between solutions obtained using different number of free parameters, we define also the reduced $\chi^2$-statistic:

        \begin{equation}
      \chi_{\rm red}^2 = \frac{1}{N_{\rm d.o.f.}}\chi^2 \\
   \end{equation}
   
$N_{\rm d.o.f.} = N_{\rm data}-N_{\rm free}-1$ is the number of degrees of freedom. The number of free parameters is either $N_{\rm free}=5$ or $N_{\rm free}=4$, depending on the method. In the first case, the Li line itself was used to determine all five free parameters, i.e. continuum normalisation ($C_{\rm norm}$), relative radial velocity shift at central wavelength ($\Delta v_{\rm rad}$), $v_{\rm rot}\sin i$, $A\rm(^7Li)$, and $\rm^6Li/^7Li$, and in the latter case only four of these parameters, trusting $v_{\rm rot}\sin i$ determined using Na and Ca calibration lines. A $\chi_{\rm red}^2$-value close to unity is indicative of a good match between model and observations and a realistic estimate of the observational uncertainties.
              
\begin{figure}
\begin{minipage}[b]{0.495\linewidth}
\centering
\includegraphics[scale=0.19,viewport=5cm 0cm 25cm 20cm]{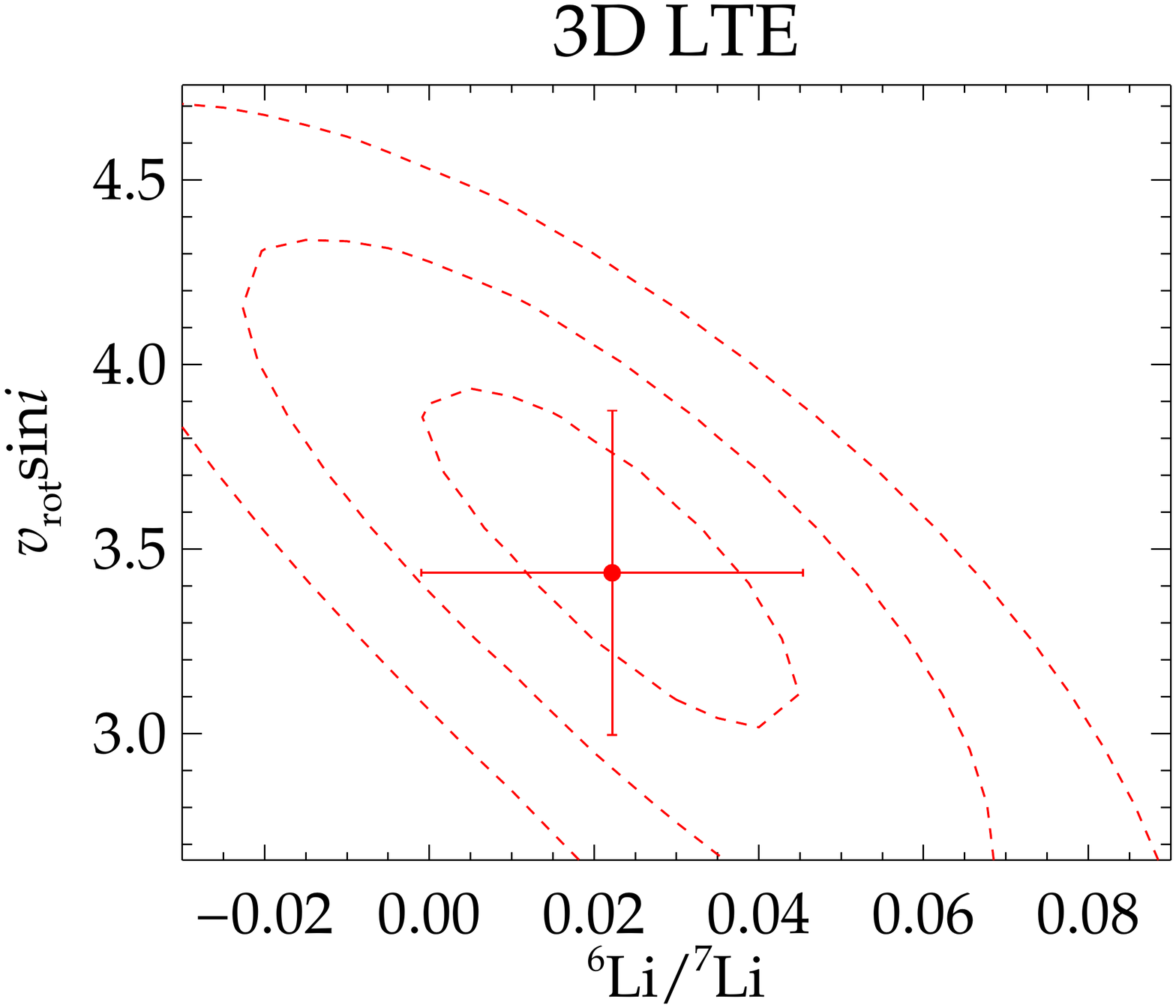}
\end{minipage}
\begin{minipage}[b]{0.495\linewidth}
\centering
\includegraphics[scale=0.19,viewport=5cm 0cm 25cm 20cm]{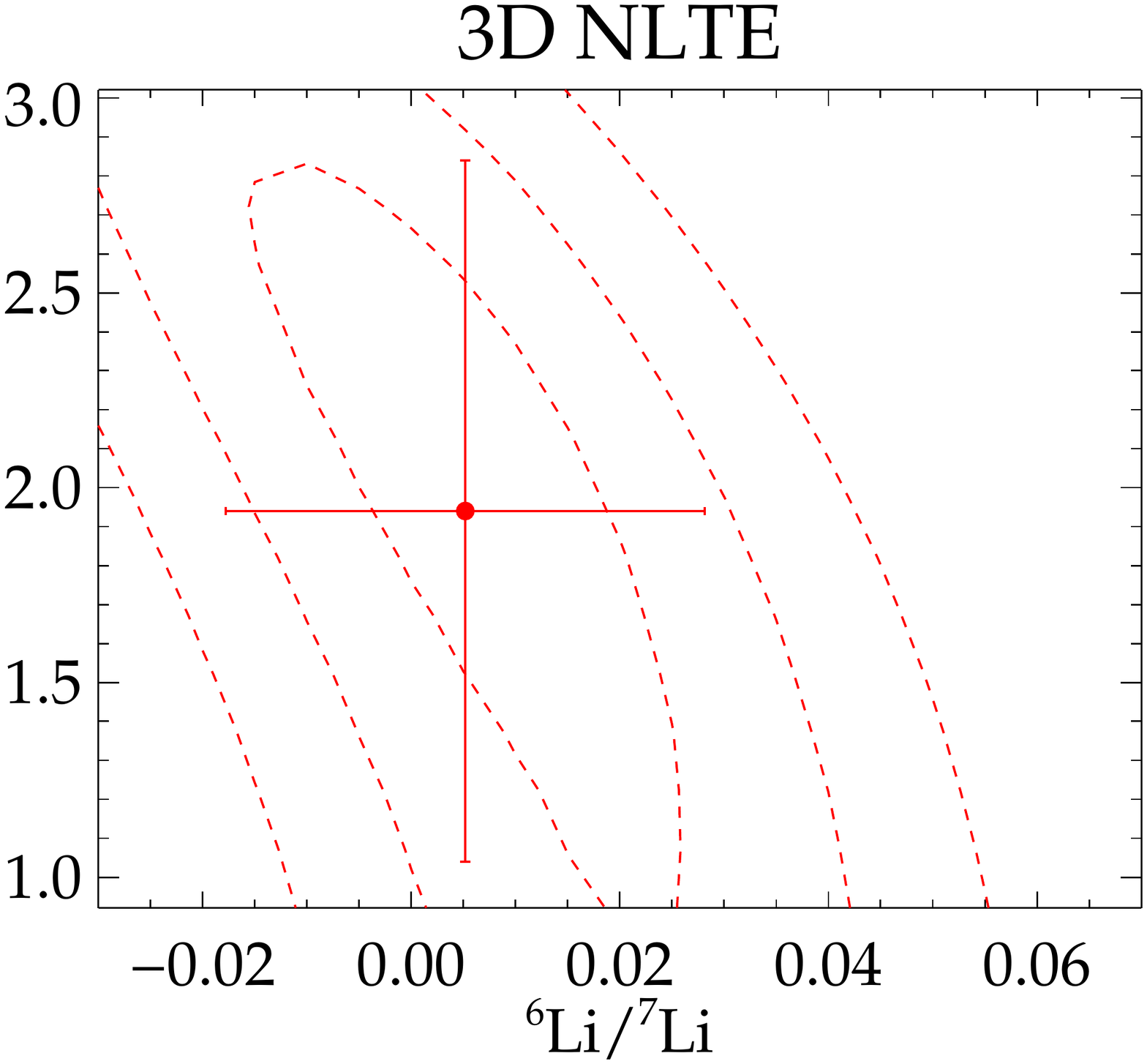}
\end{minipage}
\begin{minipage}[b]{0.495\linewidth}
\centering
\includegraphics[scale=0.19,viewport=5cm 0cm 25cm 20cm]{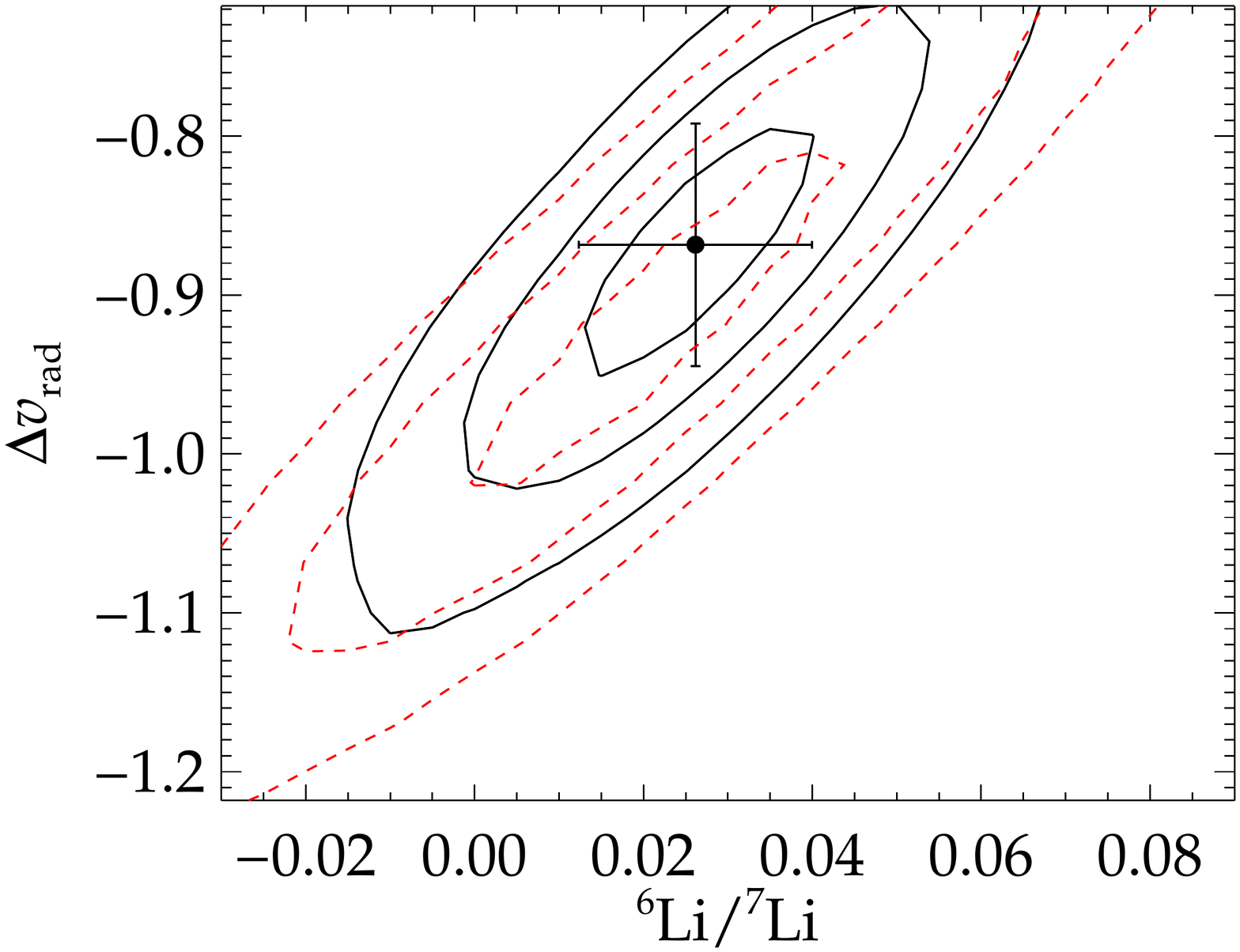}
\end{minipage}
\begin{minipage}[b]{0.495\linewidth}
\centering
\includegraphics[scale=0.19,viewport=5cm 0cm 25cm 20cm]{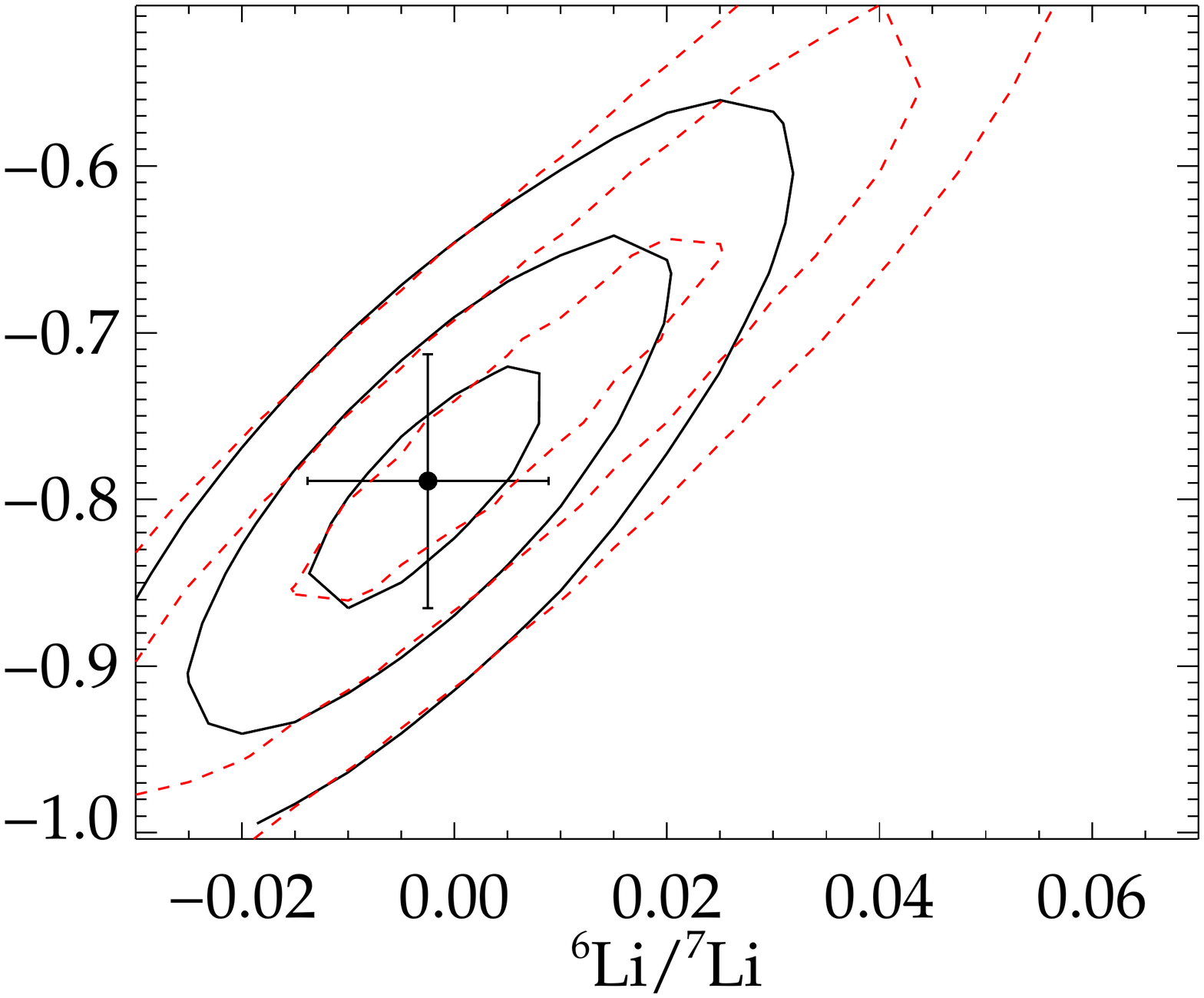}
\end{minipage}
\begin{minipage}[b]{0.495\linewidth}
\centering
\includegraphics[scale=0.19,viewport=5cm 0cm 25cm 20cm]{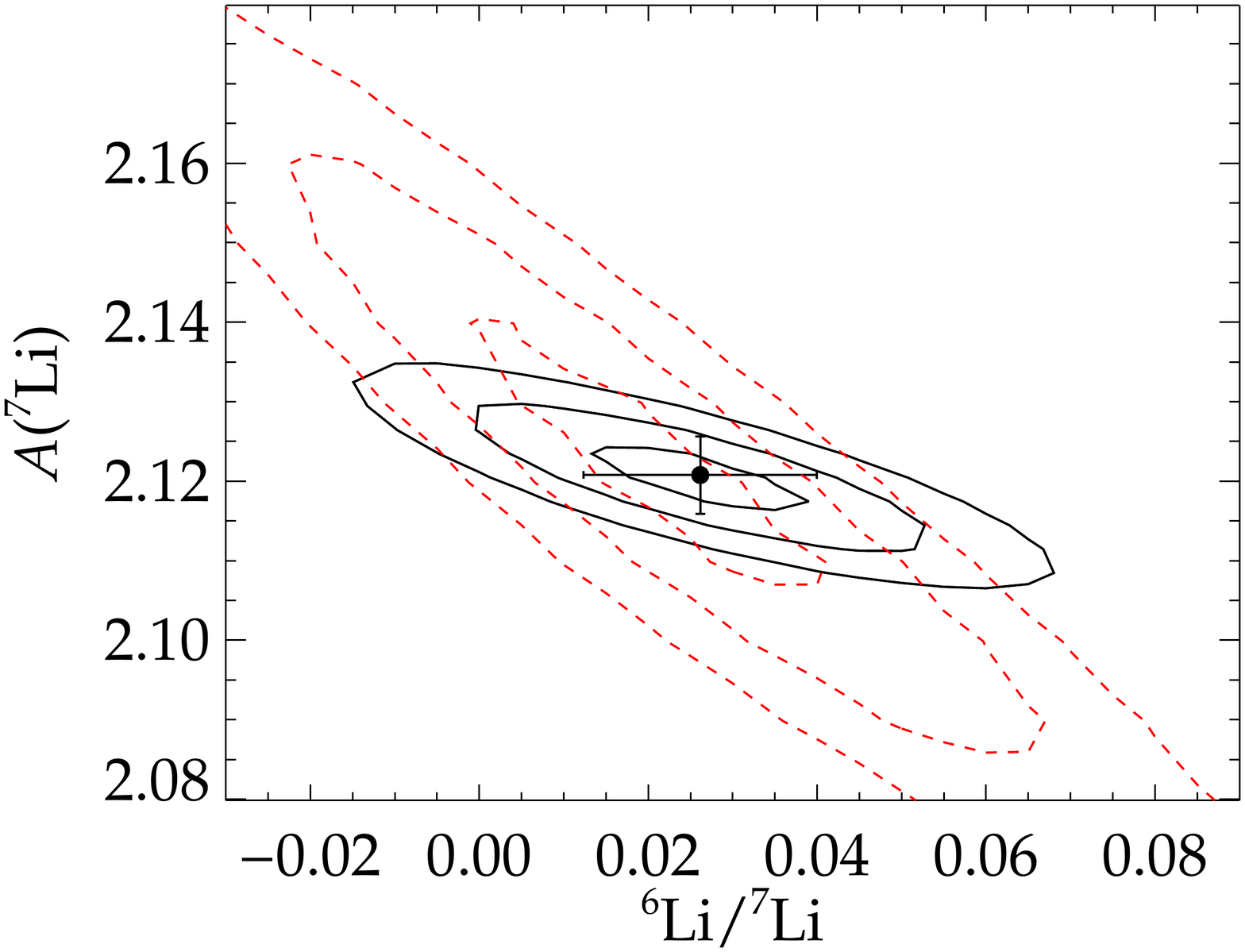}
\end{minipage}
\begin{minipage}[b]{0.495\linewidth}
\centering
\includegraphics[scale=0.19,viewport=5cm 0cm 25cm 20cm]{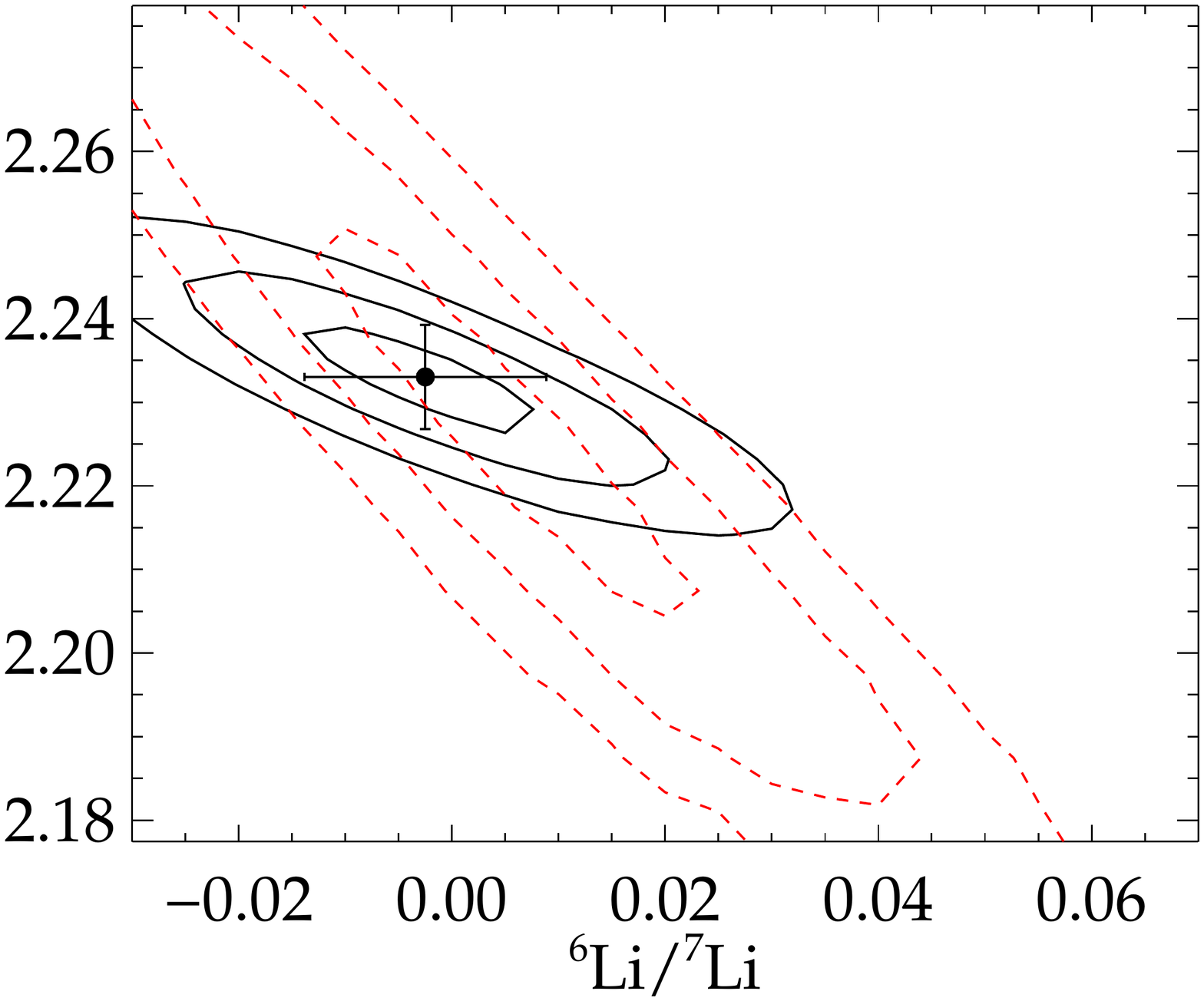}
\end{minipage}
\caption{$\chi^2$-surfaces (1$\sigma$, 2$\sigma$, and 3$\sigma$) obtained for G64-12 by varying two line parameters at the time, as indicated on the respective axes; \textit{red dashed lines:} $N_{\rm free}=5$ and \textit{black solid lines:} $N_{\rm free}=4$. The other free parameters have been optimised at each grid point. The best-fit value and associated error bars are indicated for $N_{\rm free}=5$ in the top panel and $N_{\rm free}=4$ in the two lower panels \textit{(bullets)}.}
\label{fig:chisquare1}
\end{figure}

\begin{figure}
\begin{minipage}[b]{0.495\linewidth}
\centering
\includegraphics[scale=0.19,viewport=5cm 0cm 25cm 20cm]{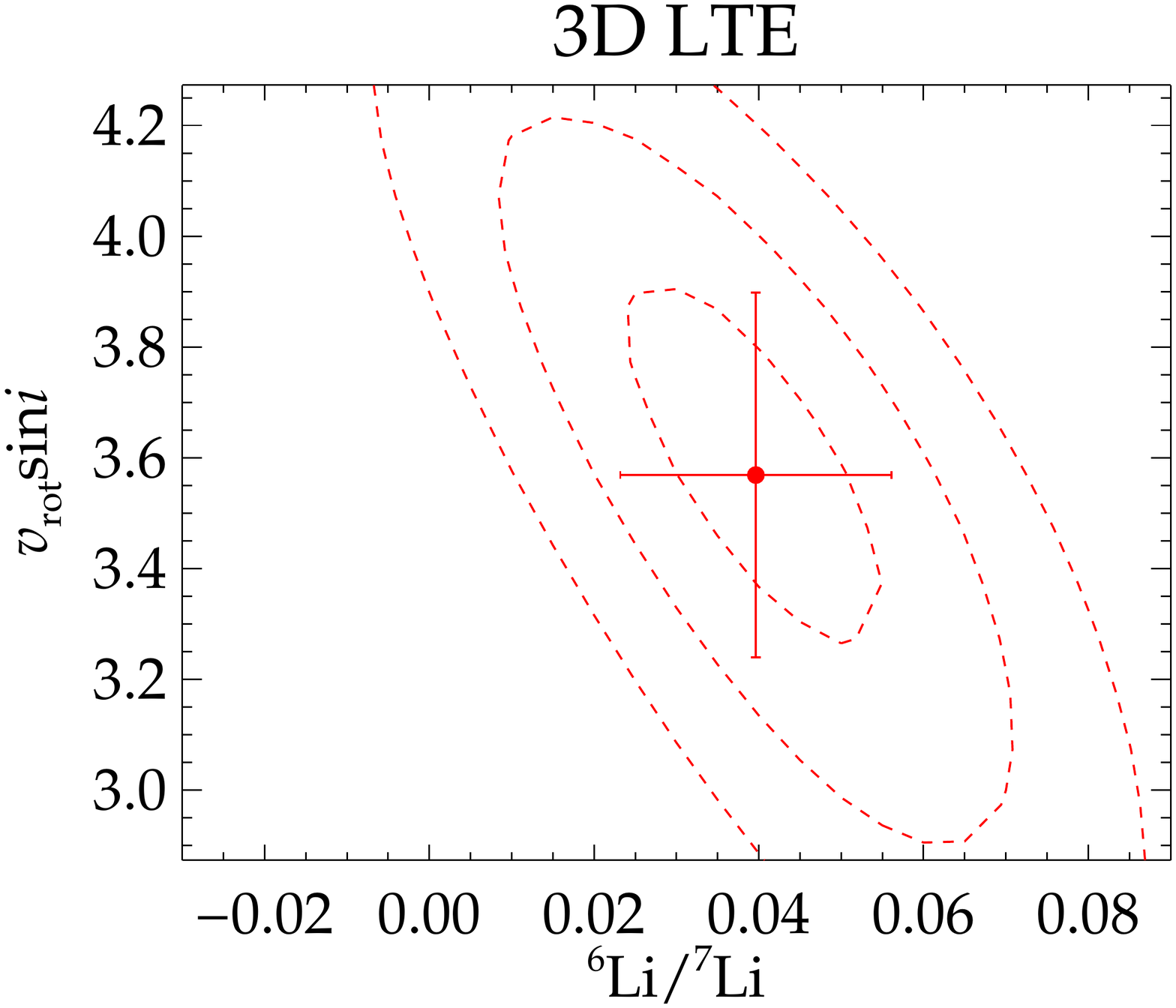}
\end{minipage}
\begin{minipage}[b]{0.495\linewidth}
\centering
\includegraphics[scale=0.19,viewport=5cm 0cm 25cm 20cm]{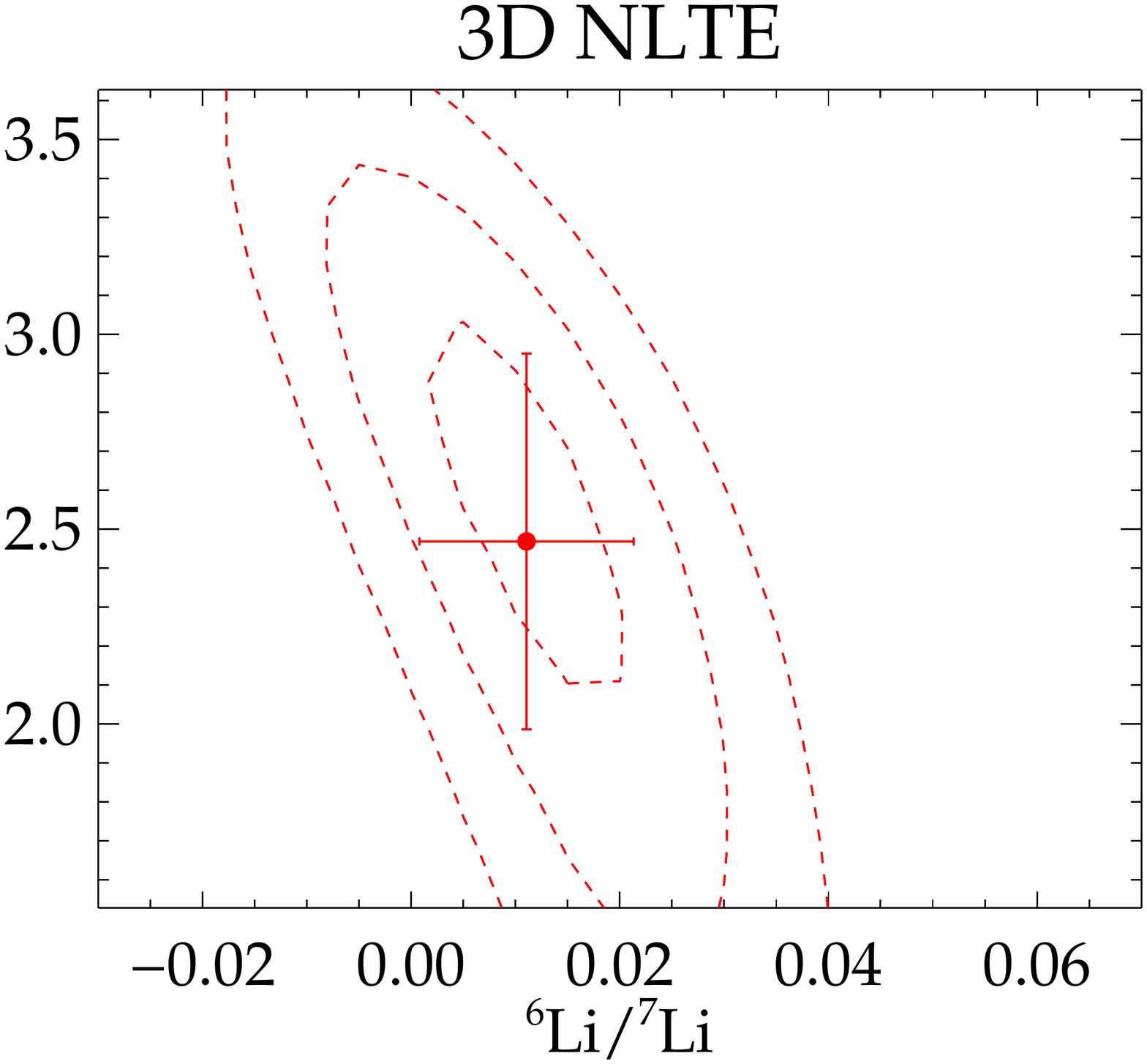}
\end{minipage}
\begin{minipage}[b]{0.495\linewidth}
\centering
\includegraphics[scale=0.19,viewport=5cm 0cm 25cm 20cm]{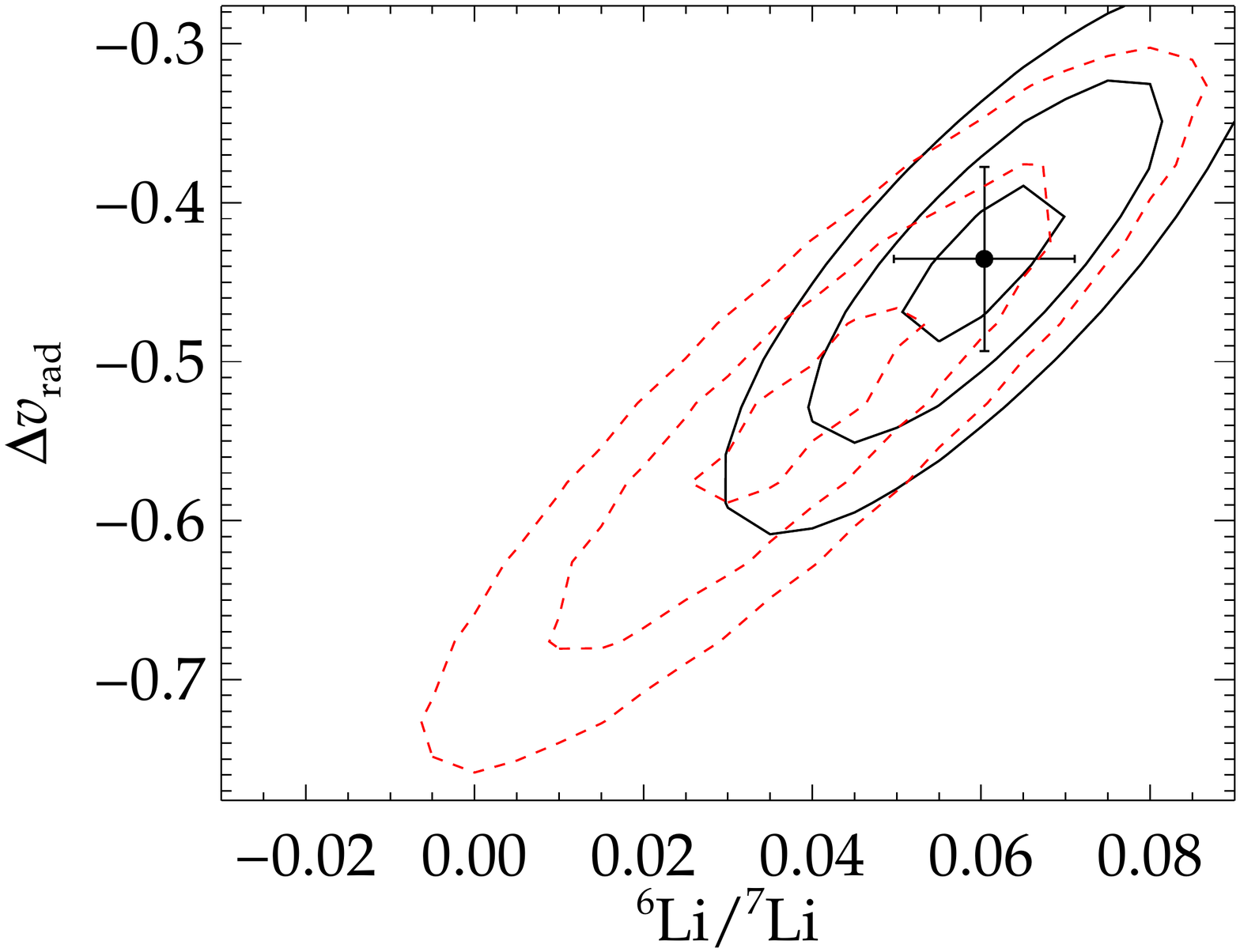}
\end{minipage}
\begin{minipage}[b]{0.495\linewidth}
\centering
\includegraphics[scale=0.19,viewport=5cm 0cm 25cm 20cm]{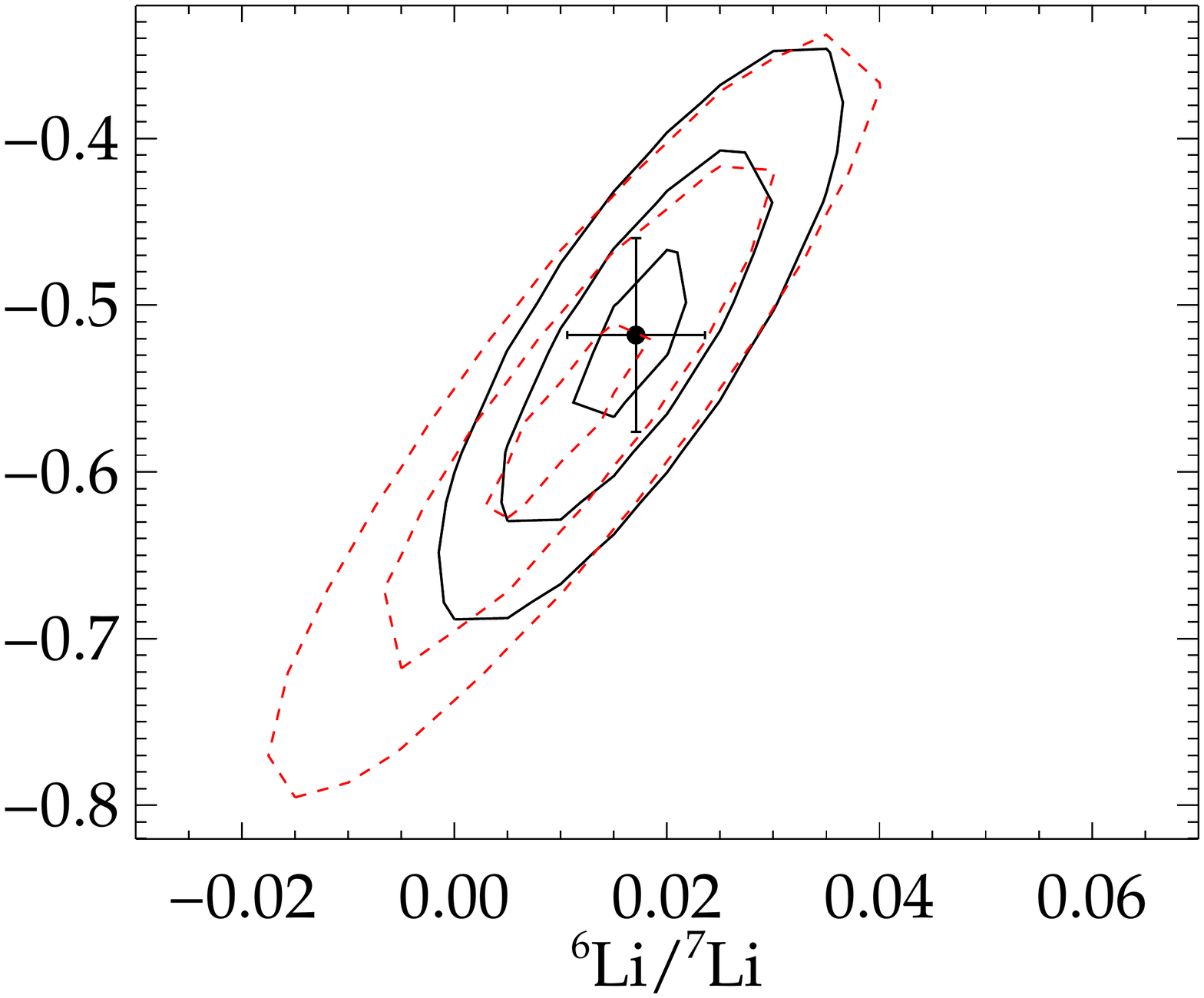}
\end{minipage}
\begin{minipage}[b]{0.495\linewidth}
\centering
\includegraphics[scale=0.19,viewport=5cm 0cm 25cm 20cm]{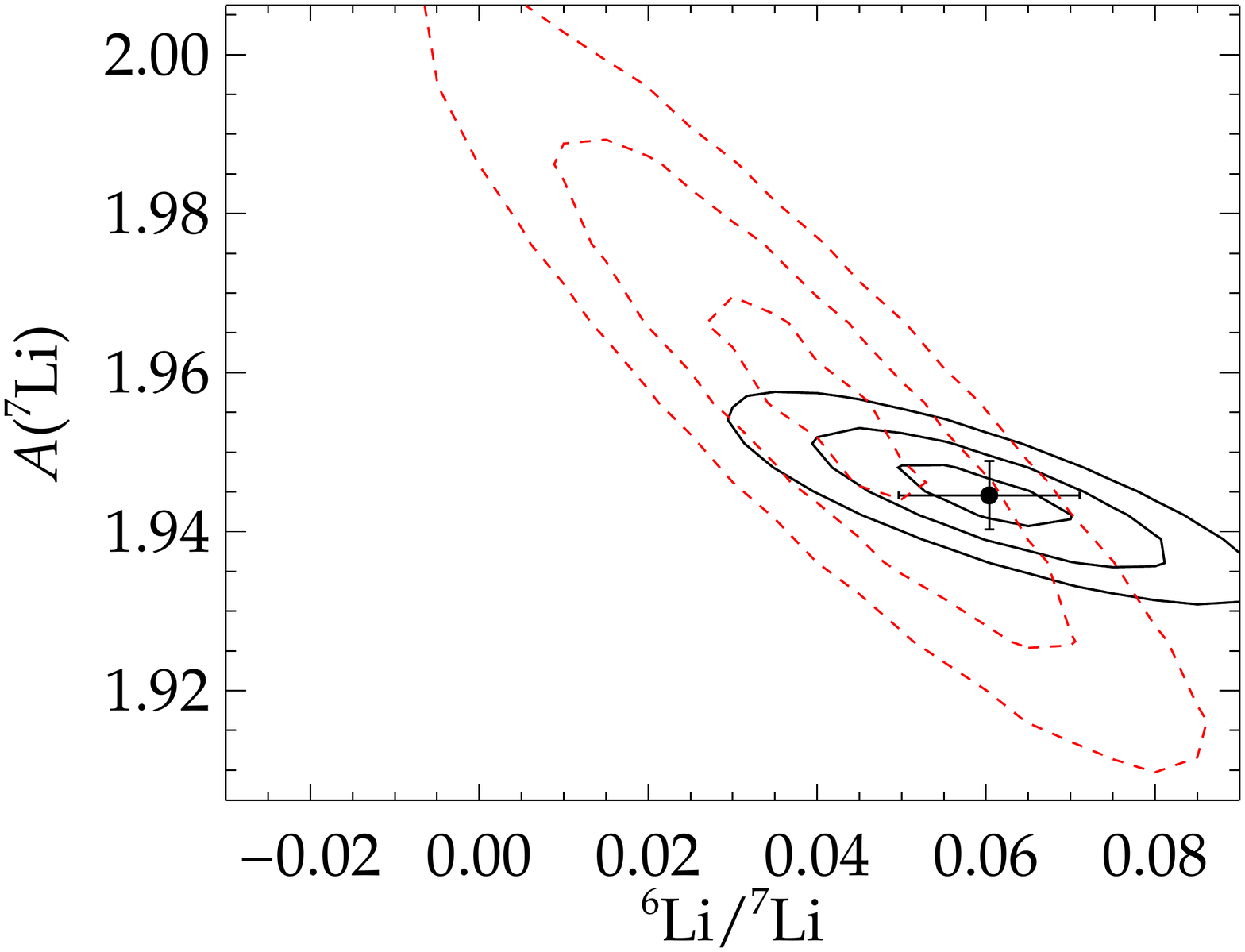}
\end{minipage}
\begin{minipage}[b]{0.495\linewidth}
\centering
\includegraphics[scale=0.19,viewport=5cm 0cm 25cm 20cm]{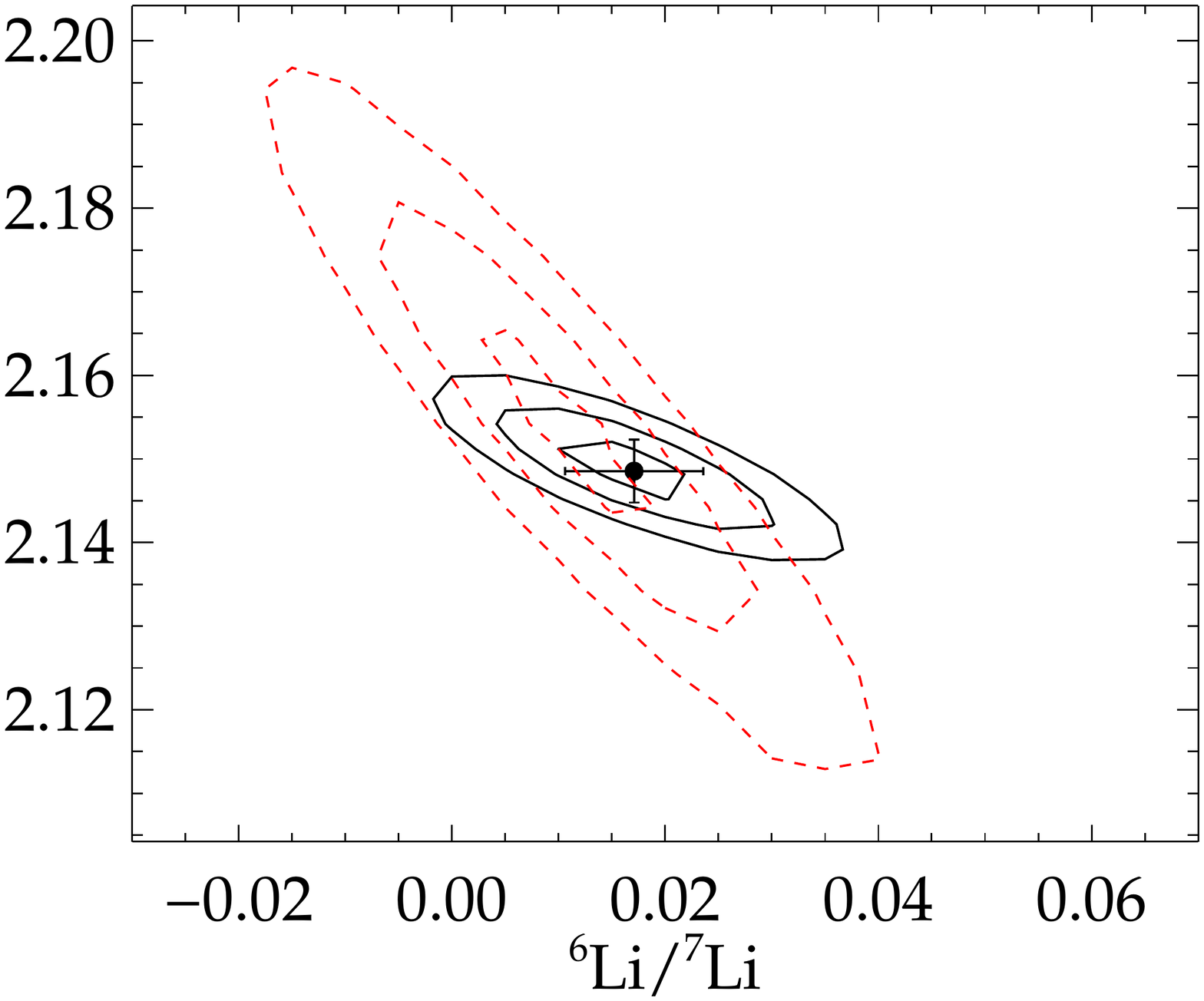}
\end{minipage}
\caption{Same as Fig.\ref{fig:chisquare1} but for HD84937.}
\label{fig:chisquare2}
\end{figure}

The isotopic ratio and its associated error ($\sigma_{\rm obs}$ in Table\,\ref{tab:isotope}) were found by analysing several 2D-surfaces in the $\chi^2$-space. These were formed by fixing the isotopic ratio and one free parameter at the time to a grid of values, while optimising all other free parameters. Examples of the resulting $\chi^2$-contours, which reveal the extent of the parameter degeneracies, are shown in Figs. \ref{fig:chisquare1} and \ref{fig:chisquare2}. The best-fit values and corresponding errors of the fitting parameters were found by parabolic fits to the $\chi^2$ data along the lines of maximum degeneracy. As seen in the figures, the full parameter space defined by the $1\sigma$-contours ($\chi^2=\chi^2_{\rm min}+1$) is adequately covered by the error bars. When $v_{\rm rot}\sin i$ was determined from calibration lines, the associated uncertainty was propagated and added to $\sigma_{\rm obs}$. 

Finally, we have estimated the errors inherent in the synthetic profiles due to uncertainties in stellar parameters and to the limited sampling of the NLTE/LTE profile ratio. We refer to this error as $\sigma_{\rm model}$. We assumed that only the error in effective temperature plays a significant role for the line formation of neutral species, and hence in the determination of the isotopic ratio, and adopted 100K as a reasonable error bar (see Table \ref{tab:param} ). As was pointed out by e.g. \citet{Asplund06}, errors of this magnitude do not contribute significantly to the error in the $\rm^6Li/^7Li$-ratio in a 1D analysis, but in 3D we must account for a non-negligible effect on the shape of the line profile. We have estimated this contribution by repeating the 3D, LTE analysis for a 140K hotter model of HD84937, and adopt 0.009 and 0.004 as reasonable estimates of $\sigma_{\rm model}$ when $N_{\rm free}=5$ and $N_{\rm free}=4$, respectively. The 3D, NLTE analysis was not repeated, however, since the convective motions of the higher temperature model are slightly too high, leading to negative $v_{\rm rot}\sin{i}$. This indicates that the star is indeed a very slow rotator, as expected for an old halo star, and that our 3D model is realistic in terms of predicting the intrinsic line broadening from convective motions. Instead, we adopted the same errors as for 3D, LTE and added to that an estimate of the influence of the limited number of snapshots used to sample the NLTE/LTE profile ratio. For all stars, and both methods, this error on $\rm^6Li/^7Li$ is equal to 0.002.

The isotopic ratios and associated errors due to random and systematic uncertainties should thus be read from Table \ref{tab:isotope} as $\rm^6Li/^7Li\pm\sigma_{\rm obs}\pm\sigma_{\rm model}$.

\subsection{Calibration lines}

Following the same reasoning as detailed in previous studies \citep{Smith98,Asplund06}, simultaneous modelling of lines of other neutral species is important in order to constrain any intrinsic line broadening and thereby reduce the error bar on the isotopic ratio. The broadening due to non-thermal gas motions in the atmosphere is here assumed to be realistically captured by accounting for the velocity field of the hydrodynamical simulations. Hence any remaining line broadening is ascribed to the unknown rotation of the stars and therefore the projected rotational velocity, $v_{\rm rot}\sin{i}$, is treated as a free parameter, assuming an Uns\"old profile  \citep[e.g.][]{Unsoeld38}. While the ability of hydrodynamical models to reproduce observed line profiles has been successfully demonstrated \citep[e.g.][]{Asplund00}, there is a possibility that the true $v_{\rm rot}\sin{i}$ becomes over- or underestimated with this assumption due to inadequate modelling of the convective motions, e.g. due to erroneous stellar parameters. Thereby, $v_{\rm rot}\sin{i}$ may here also compensate to some extent for missing non-rotational broadening. However, we have verified that the Li isotopic ratios obtained using a rotational or Gaussian velocity profile are the same. The uncertainty introduced in the shape of the convolving function used for the unknown line broadening is thus negligible. 

The best-fit values for the free parameters (continuum normalisation, radial velocity, and Ca abundance) were determined from the Ca lines using the same $\chi^2$-minisation technique as described for Li. However, in the determination of the optimal  $v_{\rm rot}\sin{i}$, we considered all the calibration lines simultaneously, i.e. we enforced a single value, rather than computed a mean value. Not all Ca lines that are detectable in the spectra were used as calibration lines, but only those with $W_\lambda<50$m\AA\ . The reason is that we do not wish the formation layers of the calibration lines to deviate too far from that of the Li line, having a total line strength of $23-46$\,m\AA . In addition, stronger lines become increasingly less sensitive to rotational broadening and are therefore less optimal as calibrators (see Appendix). To get a robust constraint on $v_{\rm rot}\sin{i}$ in the most metal-poor star in our sample, G64-12, we complemented the two detectable Ca lines with the Na\,D lines at $\sim$5890\AA , which are of suitable strength ($20-30$\,m\AA\,); at higher [Fe/H] the lines are too strong to be useful. The lines used for calibration in each star are marked in Table \ref{tab:abund}. 

In principle, also the central position of the Li line can be calibrated using other neutral lines, removing yet another free parameter from the determination of the isotopic ratio. However, when investigating this possibility we found that the central radial velocity shifts determined from different lines are not compatible with one another within the errors. For each individual line, the accuracy in $v_{\rm rad}$ implied by the $\chi^2$-analysis is typically $<0.05\,$km/s, while the line-to-line dispersion is closer to $0.1$\,km/s. This may be caused by errors in the adopted laboratory wavelengths or by imperfect modelling of the distribution of convective velocities at different heights of formation. We therefore chose to determine $v_{\rm rad}$ from the Li line itself. We note that the relative central wavelengths of $^6$Li and $^7$Li line components have been measured to a very high accuracy \citep{Sansonetti95}.  

\begin{figure}
\begin{minipage}[b]{0.495\linewidth}
\centering
\includegraphics[scale=0.19,viewport=5cm 0cm 25cm 20cm]{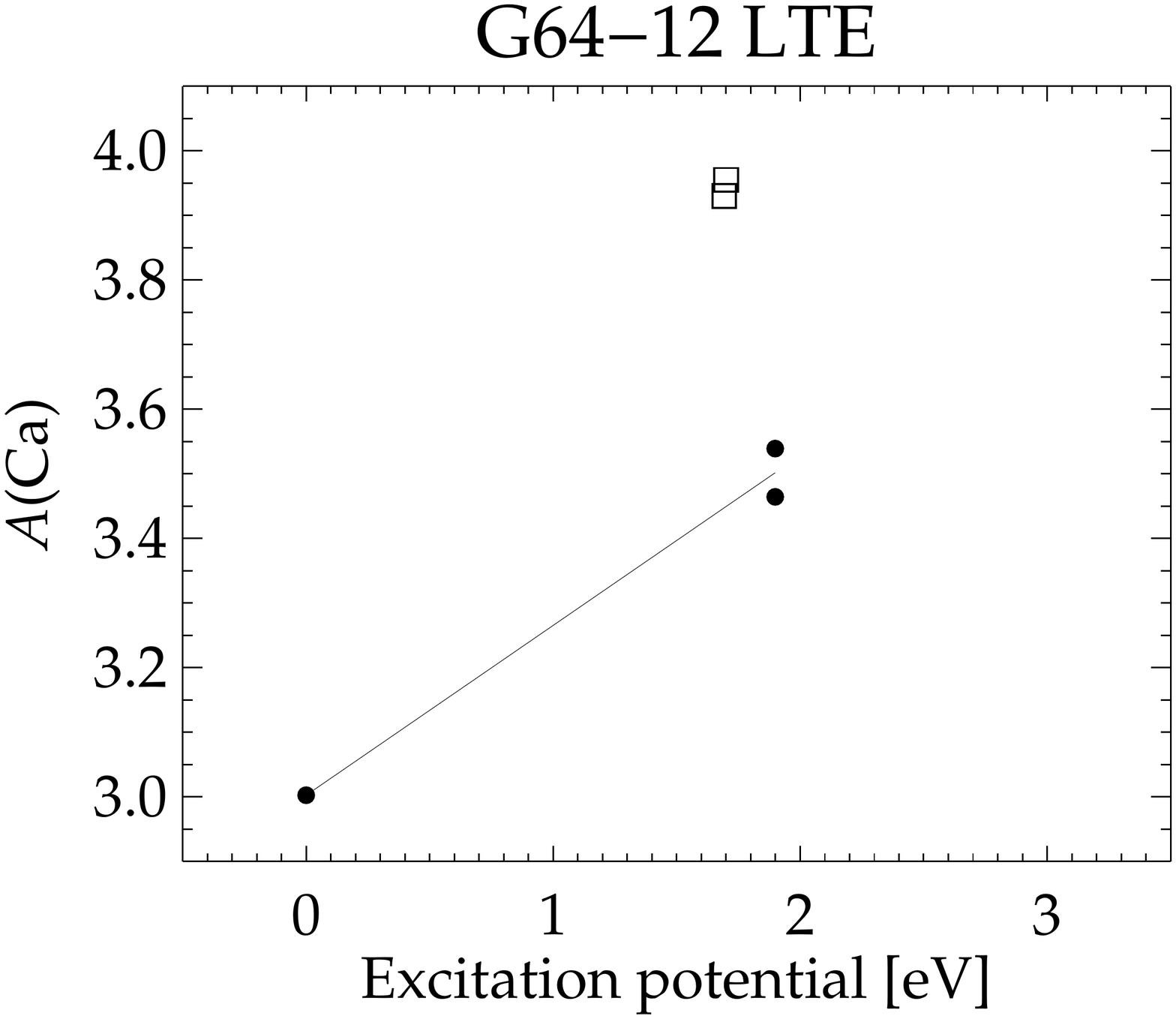}
\end{minipage}
\begin{minipage}[b]{0.495\linewidth}
\centering
\includegraphics[scale=0.19,viewport=5cm 0cm 25cm 20cm]{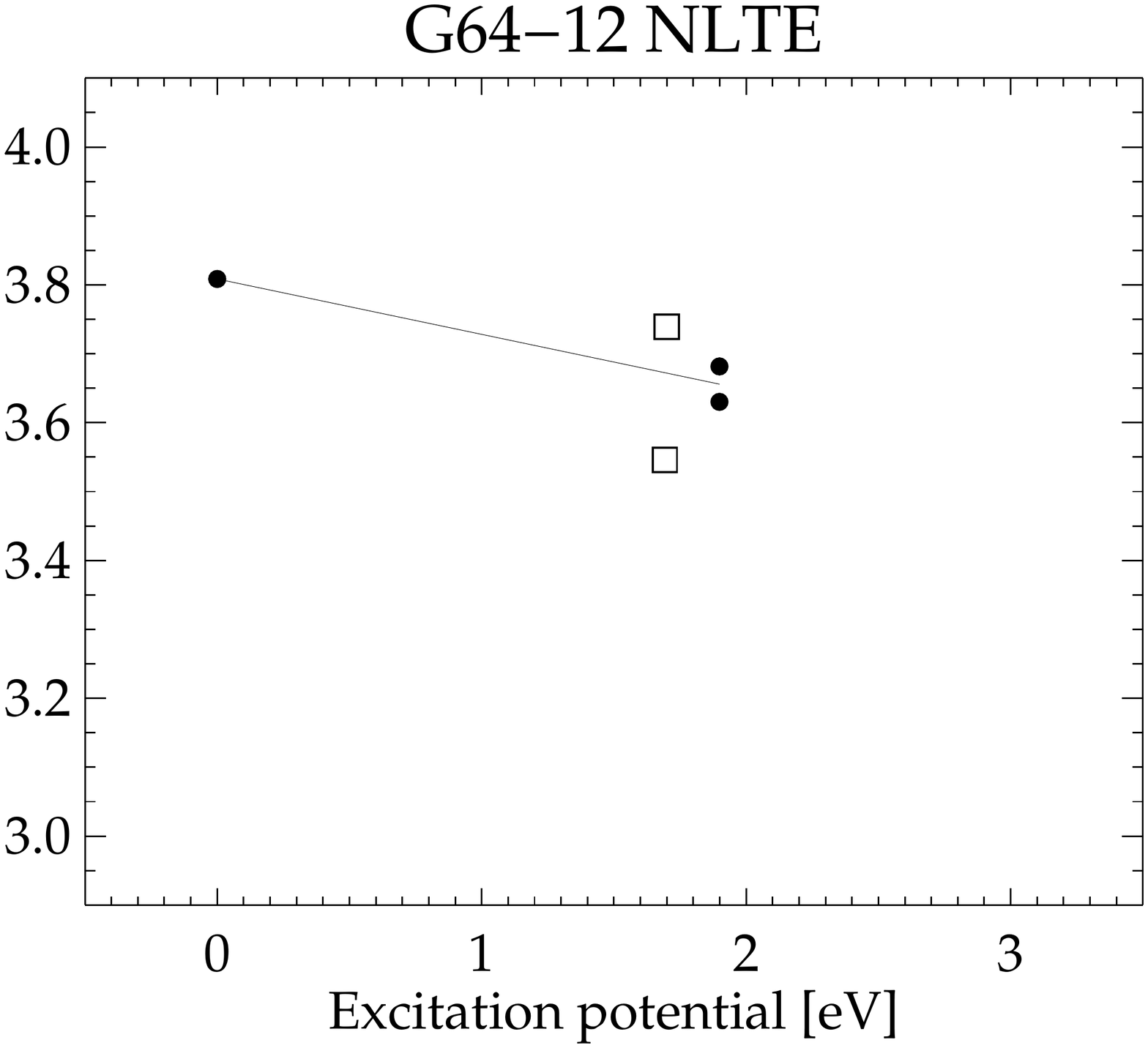}
\end{minipage}
\begin{minipage}[b]{0.495\linewidth}
\centering
\includegraphics[scale=0.19,viewport=5cm 0cm 25cm 20cm]{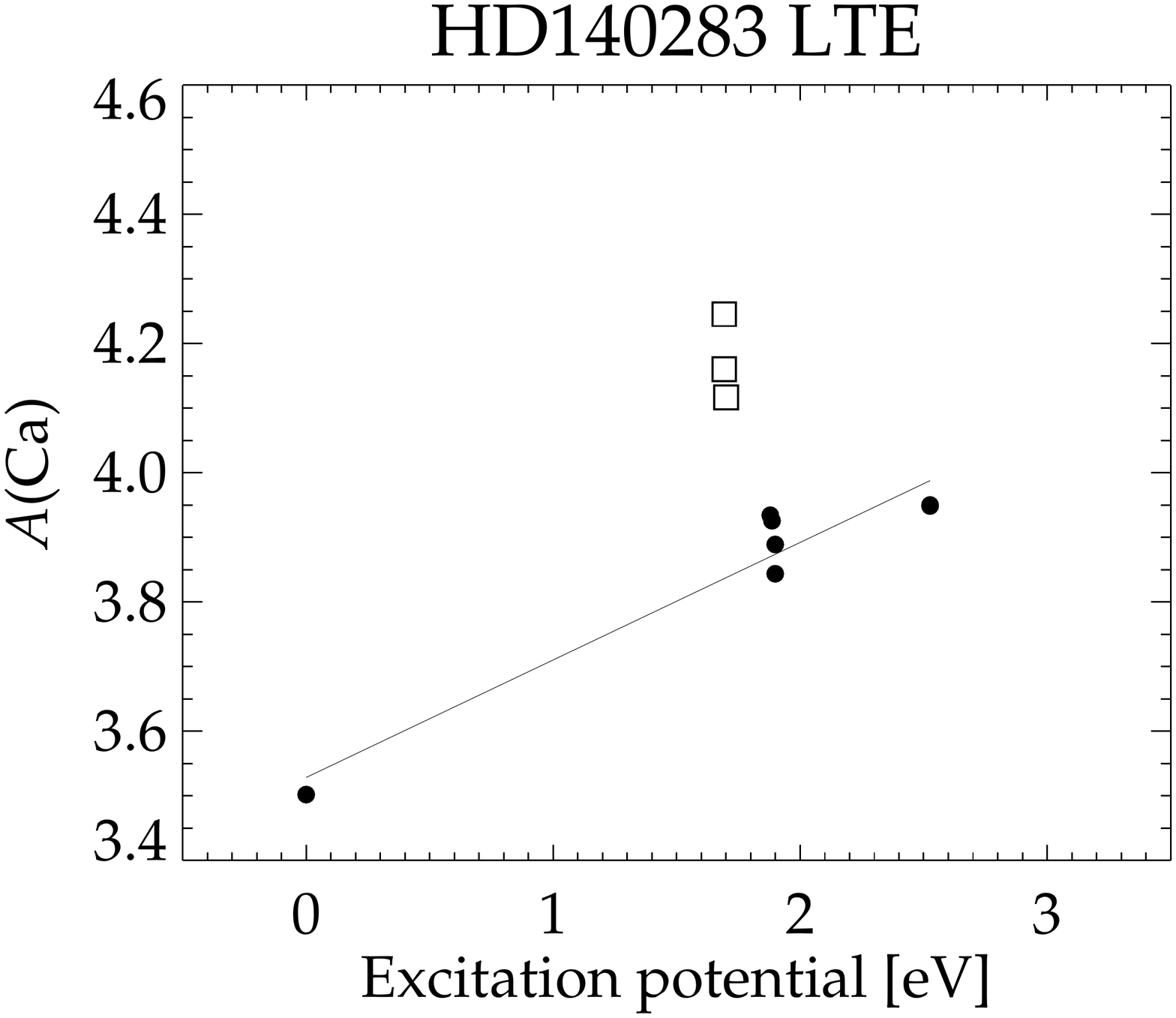}
\end{minipage}
\begin{minipage}[b]{0.495\linewidth}
\centering
\includegraphics[scale=0.19,viewport=5cm 0cm 25cm 20cm]{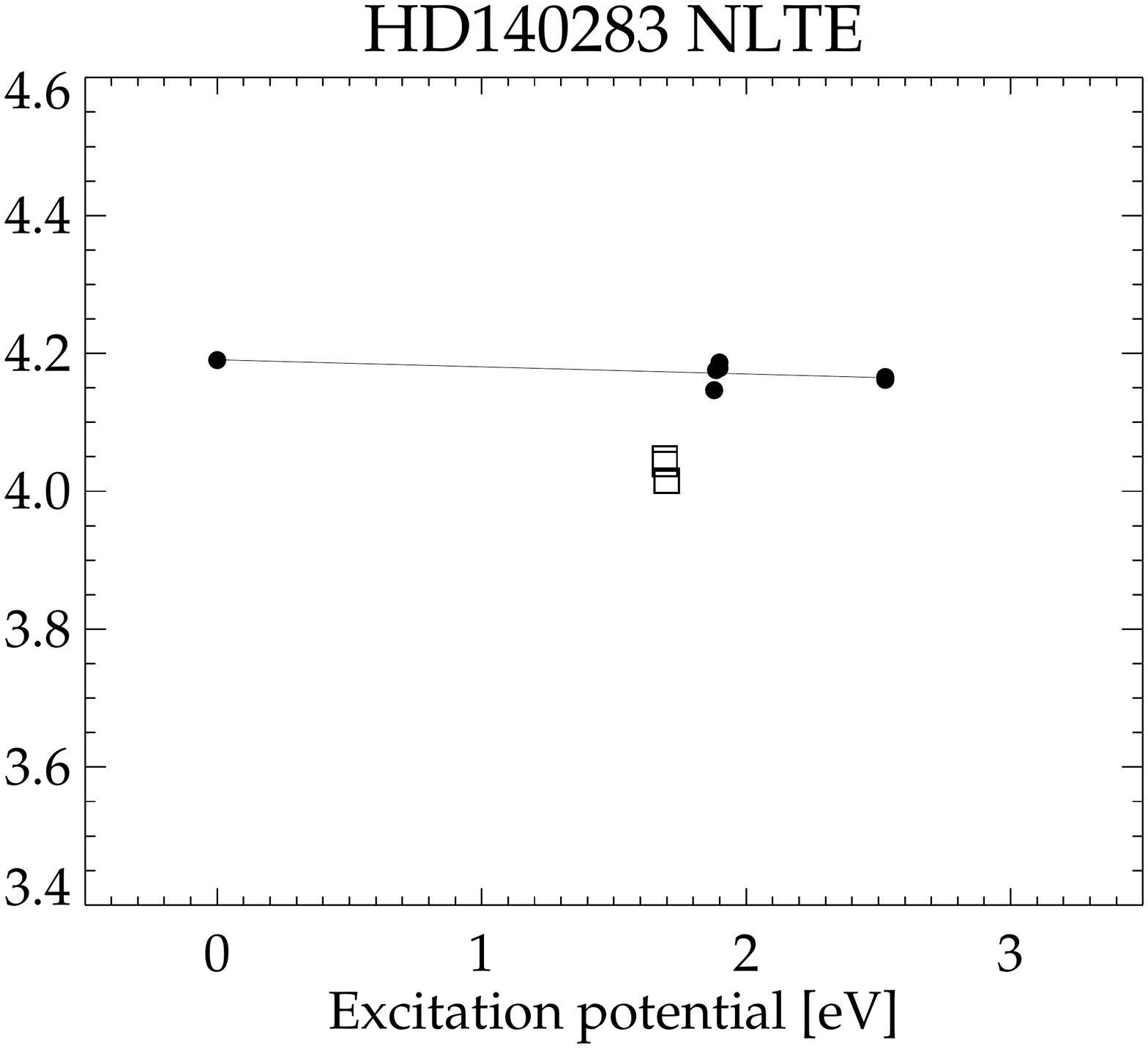}
\end{minipage}
\begin{minipage}[b]{0.495\linewidth}
\centering
\includegraphics[scale=0.19,viewport=5cm 0cm 25cm 20cm]{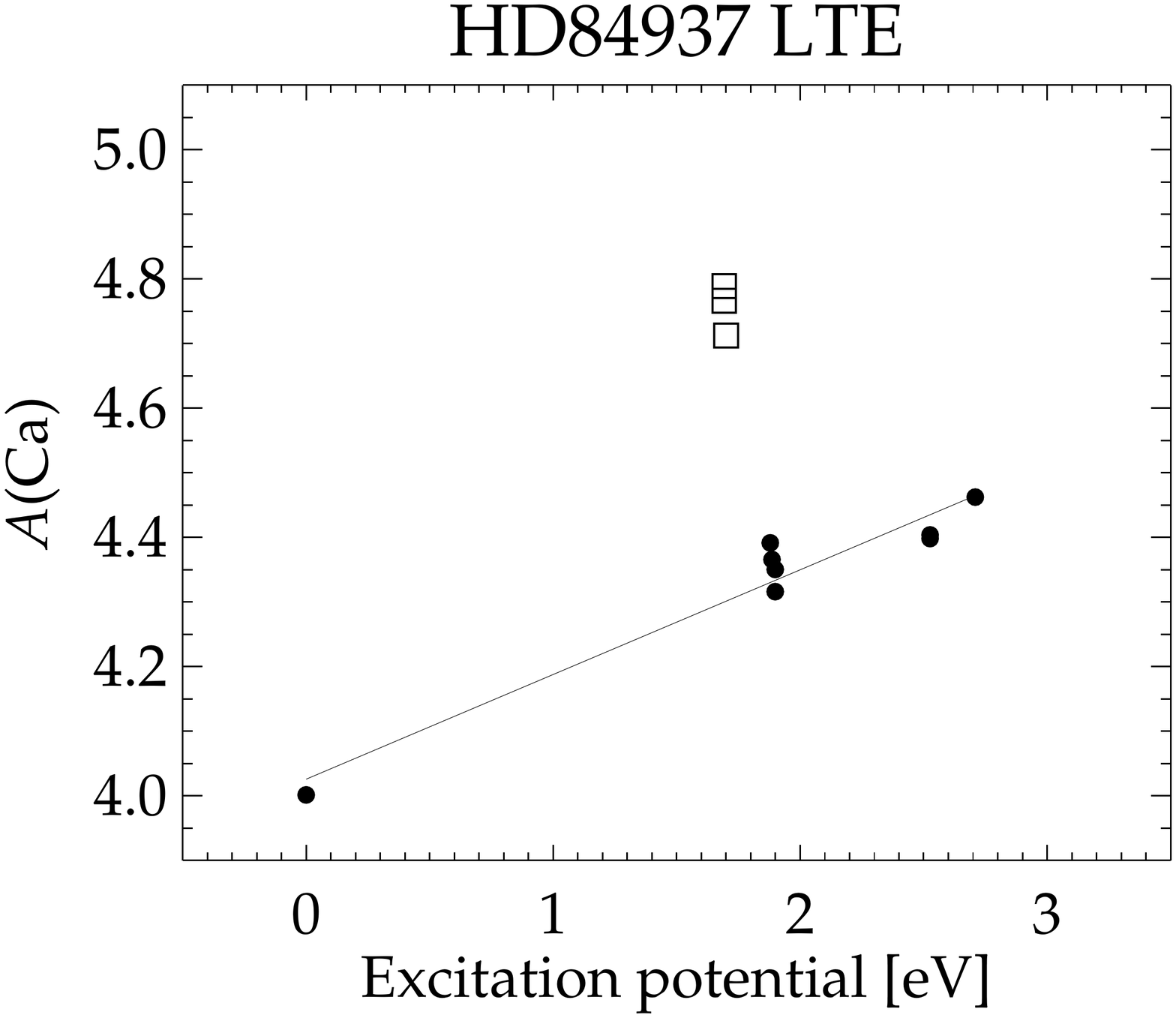}
\end{minipage}
\begin{minipage}[b]{0.495\linewidth}
\centering
\includegraphics[scale=0.19,viewport=5cm 0cm 25cm 20cm]{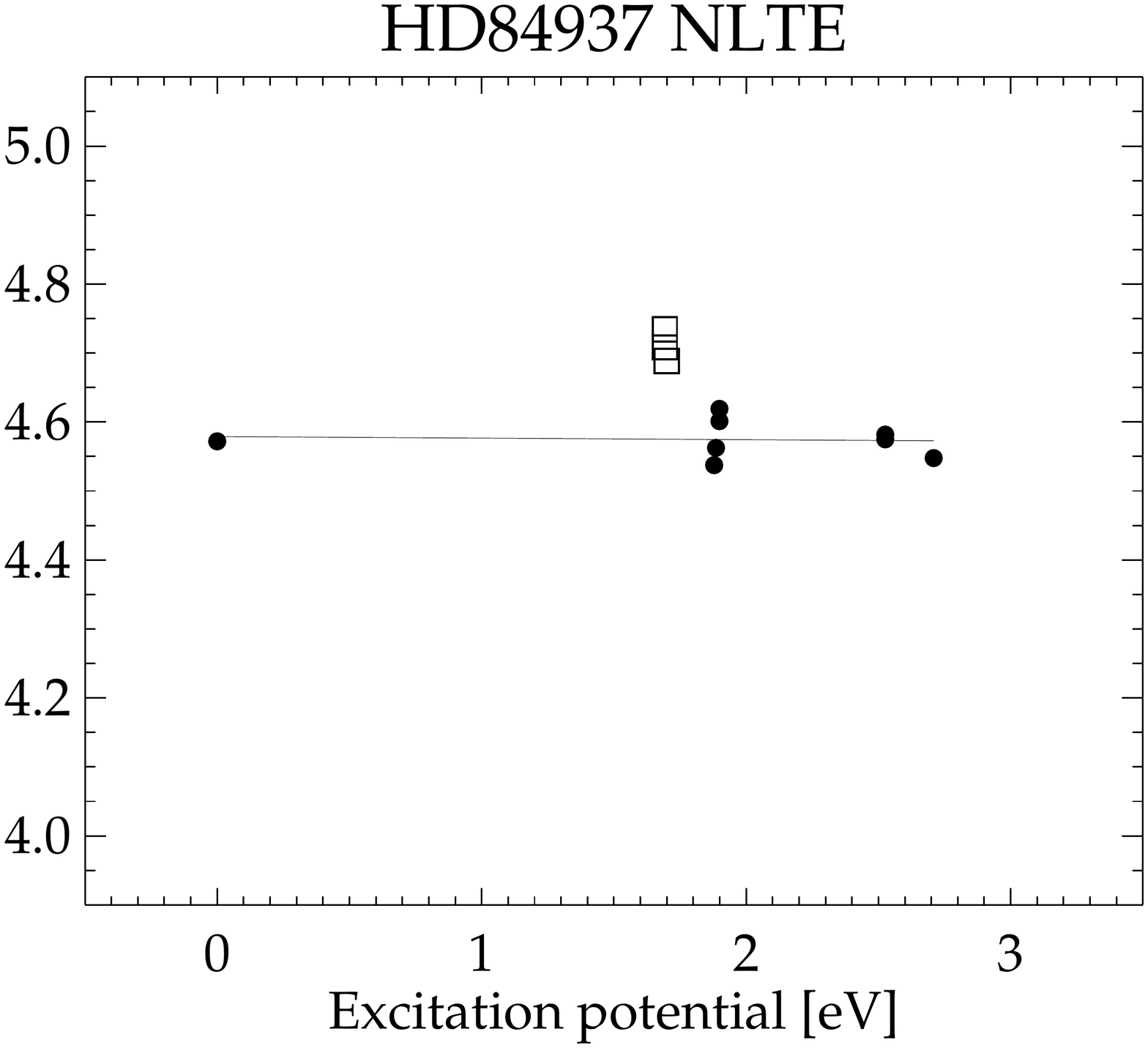}
\end{minipage}
\begin{minipage}[b]{0.495\linewidth}
\centering
\includegraphics[scale=0.19,viewport=5cm 0cm 25cm 20cm]{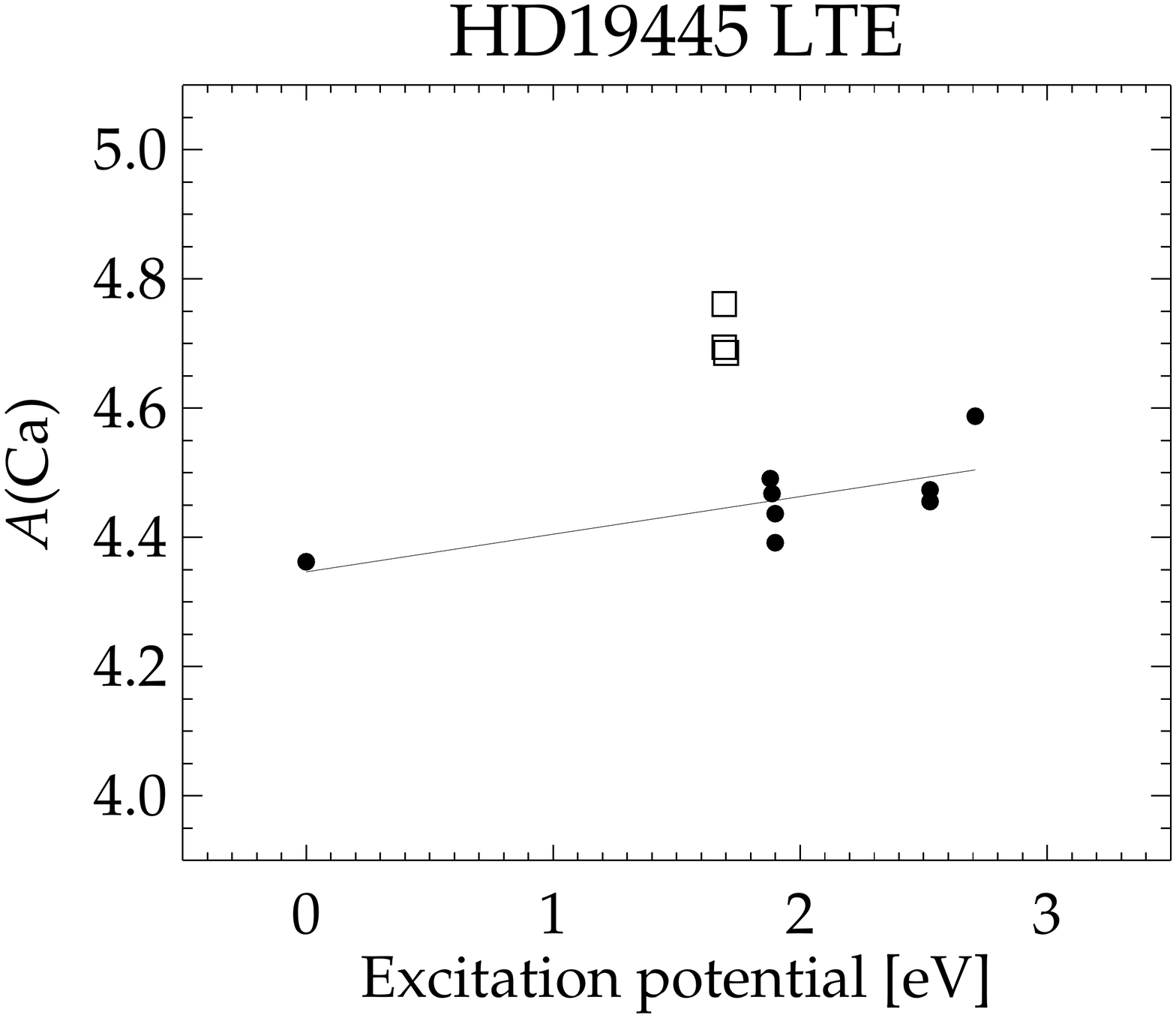}
\end{minipage}
\begin{minipage}[b]{0.495\linewidth}
\centering
\includegraphics[scale=0.19,viewport=5cm 0cm 25cm 20cm]{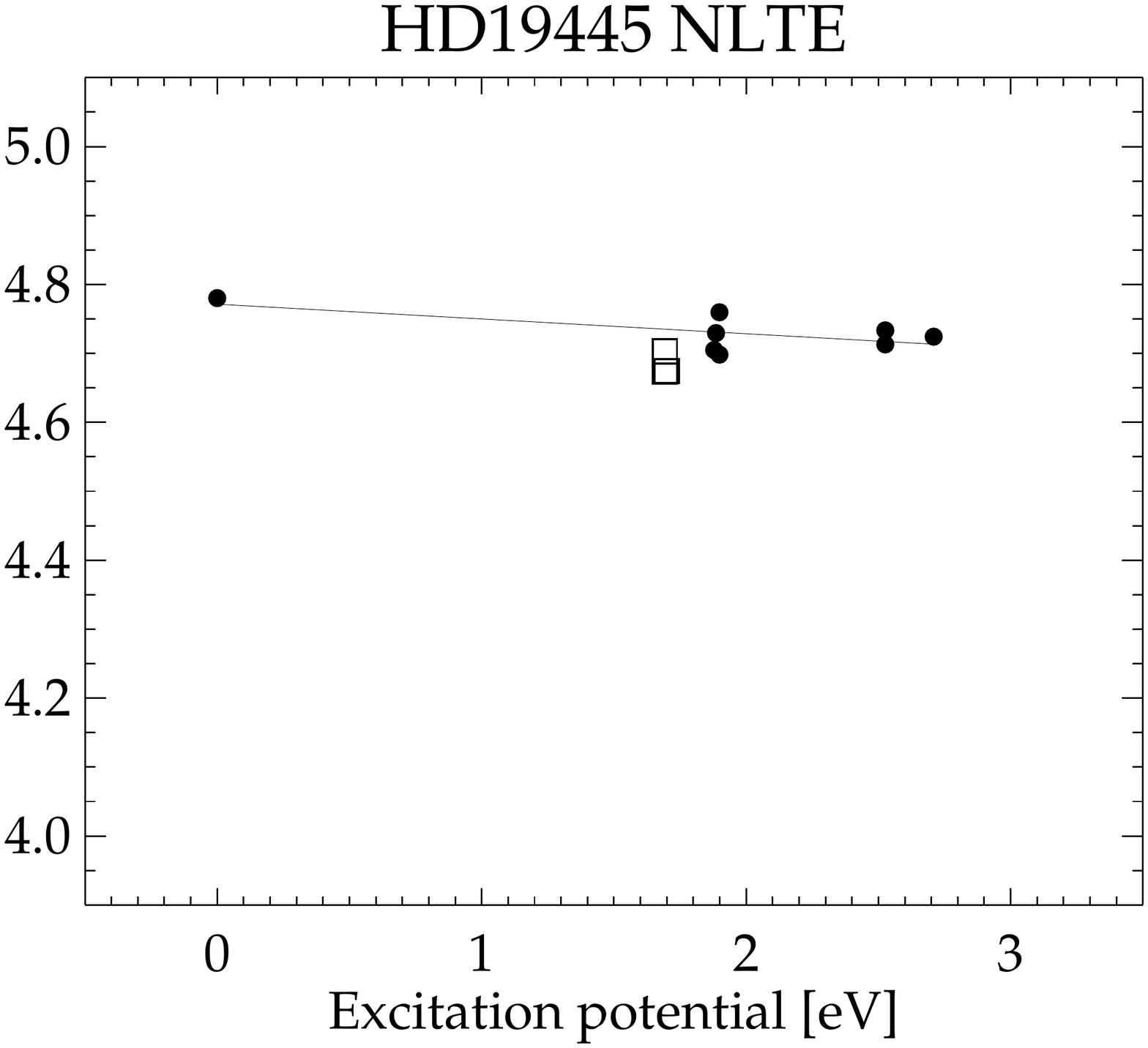}
\end{minipage}
\caption{Ca abundances plotted against lower level excitation potential, for Ca\,I (\textit{bullets}) and Ca\,II lines (\textit{squares}).}
\label{fig:calcium}
\end{figure}

The full set of Ca lines can be used to assess the realism of our stellar parameters and 3D modelling technique. By inclusion of the resonance line 4226\AA\,, the neutral Ca lines span up to 2.7\,eV in lower level excitation potential, which is sufficient to inspect abundance trends with this parameter (see Fig.\,\ref{fig:calcium}). In addition to this verification of the excitation balance, the near-infrared Ca\,II triplet lines provide us with information about the ionisation balance.

\begin{table*}
      \caption{Derived line-by-line $\rm^7Li$ (using $N_{\rm free}=4$), Na, and Ca abundances. Numbers in parentheses represent observational error bars in unit of 0.01\,dex, i.e. 2.14(2) should be read as $2.14\pm0.02$. When not explicitly given, this error is estimated to be $<0.01$\,dex. } 
         \label{tab:abund}
         \centering
         \begin{tabular}{lrrrrllllllll}
                \hline\hline
           Species & $\lambda$ & $\epsilon_{\rm exc}$ & $\log(gf)$ & vdW$^a$   &    \multicolumn{2}{c}{G64-12}   &    \multicolumn{2}{c}{HD140283}     &    \multicolumn{2}{c}{HD84937}    &   \multicolumn{2}{c}{HD19445}  \\

          \hline
         \noalign{\smallskip}	
                        & & & &             & LTE & NLTE & LTE & NLTE & LTE & NLTE & LTE & NLTE \\
	\noalign{\smallskip}	
	 Li\,I   & 6103.6$^b$  & 1.848 & 0.583  &837.274&  2.33(5)& 2.35(6) & 2.10(2) & 2.14(2) & 2.23(4) & 2.26(4) & 2.17(4) & 2.23(5) \\	
	 Li\,I   & 6707.8$^b$ & 0.000 & 0.174 &346.236&  2.12& 2.23 & 1.82 & 2.12 & 1.94 & 2.15 & 1.99 & 2.25 \\
	 Na\,I & 5889.951& 0.000 & 0.117  & 407.273&  2.73$^*$ & 2.78$^*$ & ... & ... & ... & ... & ... & ... \\
	 Na\,I & 5895.924& 0.000 & -0.184 & 407.273&  2.75$^*$ & 2.79$^*$ & ... & ... & ... & ... & ... & ... \\
          Ca\,I & 4226.728& 0.000 & 0.243 & 372.238&   3.00 & 3.81 &  3.57 & 4.19  & 4.16 & 4.57  & 4.36 & 4.78 \\
          Ca\,I & 4544.879& 1.899 & 0.318 & 949.274&   3.46$^*$ & 3.63$^*$  &  3.84$^*$ &  4.18$^*$ & 4.32$^*$ & 4.62$^*$  & 4.39 & 4.70  \\
          Ca\,I & 5588.749& 2.526 & 0.358 & 400.282&   ...      & ...       &  3.95$^*$ &  4.17$^*$ &  4.40$^*$ &  4.57$^*$ & 4.47$^*$ & 4.71$^*$ \\
          Ca\,I & 6102.723& 1.879 & -0.793& 876.233&   ...      & ...       &  3.93$^*$ &  4.15$^*$ &  4.39$^*$ &  4.54$^*$ & 4.49$^*$& 4.70$^*$ \\
          Ca\,I & 6122.217& 1.886 & -0.316& 876.234&   ...      & ...       &  3.93$^*$ &  4.18$^*$ &  4.37$^*$ &  4.56$^*$ & 4.47$^*$ & 4.73$^*$ \\ 
          Ca\,I & 6162.173& 1.899 & -0.090& 876.234&   3.54$^*$ &  3.68$^*$ &  3.89$^*$ &  4.19$^*$  & 4.35$^*$ & 4.60$^*$ &  4.44 & 4.76 \\ 
          Ca\,I & 6439.075& 2.526 & 0.390 & 366.242&   ...      &  ...      &  3.95$^*$ &  4.16$^*$  &  4.40$^*$ & 4.58$^*$ &  4.46$^*$ & 4.73$^*$ \\ 
          Ca\,I  & 6717.681& 2.709 &-0.524& 992.255&   ...      &  ...      &  ...      &  ...       &  4.46$^*$ & 4.55$^*$ &  4.59$^*$ & 4.72$^*$ \\  
          Ca\,II & 8498.023 & 1.692&-1.496& 291.275&   3.93 &  3.55&  4.25 &  4.05  &  4.79 &  4.71 &  4.76 & 4.70\\
          Ca\,II & 8542.091 & 1.700&-0.514& 291.275 &   3.95  &  3.74 &  4.12 &  4.02 &  4.71 &  4.69 &  4.69& 4.67\\
          Ca\,II & 8662.141 & 1.692&-0.770& 291.275&   ...       &  ...      &  4.16 &  4.04 &  4.77 &  4.73 &  4.69 & 4.67\\       
          \noalign{\smallskip}	
                   \hline
          \noalign{\smallskip}	
         \multicolumn{12}{l}{$^a$ The broadening parameters $\alpha$ and $\sigma$ are given in the compressed notation $\rm int(\alpha)+\sigma$ \citep{Barklem00b}}.  \\
         \multicolumn{12}{l}{$^b$ Fine and hyper-fine components listed in \citet{Lind09a}.}  \\
         \multicolumn{12}{l}{$^*$ Used for calibration of $v_{\rm rot}\sin{i}$  }        
         \end{tabular}
\end{table*}

\section{Results}

\subsection{Ca abundances}

Our NLTE modelling predicts very large positive abundance increase for neutral lines in all four stars compared to the LTE case. To investigate whether these effects are realistic we inspect the excitation and ionisation balance of Ca. The results are shown in Fig.\,\ref{fig:calcium}.  Evidently, 3D LTE modelling predicts large positive slopes of line abundance with excitation potential, which would require an increase of $\sim$400K to be removed. In NLTE the slopes are considerably flatter, but a minor over-estimation of the temperatures for G64-12 is implied. The ionisation balance is well established for this star and HD19945 in NLTE, while HD84937 and HD140283 show a small discrepancy ($\sim0.1\rm\,dex$) in opposite directions at these stellar parameters. The positive corrections to neutral lines and negative corrections to singly ionised lines lead to significant improvement of the ionisation balance in NLTE with respect to LTE. 

To perfectly establish agreement between different Ca indicators would require fine-tuning of stellar parameters and the NLTE effects, which are controlled by the unknown efficiency of hydrogen-atom inelastic collisions. This is beyond the scope of this work and will be addressed in a separate publication. We note that the patterns described for the Ca excitation balance are qualitatively similar to those described in the \textless3D\textgreater\,, NLTE analysis of Fe lines in three of these stars in our sample by \citet{Bergemann12}.

\subsection{Li isotopic abundances}

\begin{table*}
      \caption{Lithium isotopic abundances and estimated errors for the two methods with $N_{\rm free}=5$ and 4. The final error associated to $\rm^6Li/^7Li$ has a random ($\sigma_{\rm obs}$) and systematic contribution ( $\sigma_{\rm model}$). }
         \label{tab:isotope}
         \centering
         \begin{tabular}{lrrrrrrrr}
                \hline\hline
               &    \multicolumn{2}{c}{G64-12}   &    \multicolumn{2}{c}{HD140283}     &    \multicolumn{2}{c}{HD84937}    &   \multicolumn{2}{c}{HD19445}  \\
          \hline
	\noalign{\smallskip}	
          & LTE & NLTE & LTE & NLTE & LTE & NLTE & LTE & NLTE \\
	\noalign{\smallskip}	
	$N_{\rm free}=5$      \\
	\noalign{\smallskip}	
	$v_{\rm rot}\sin{i}$ [km/s]              &  3.729 &  1.940   & 3.739 & 2.116 & 3.569 & 2.469 & 3.385 & 1.517\\
	$\sigma$    [km/s]                             &  0.440  &  0.900  & 0.121 & 0.229& 0.329 & 0.482 & 0.220 & 0.560 \\
	$\rm^6Li/^7Li$                        &  0.003 & 0.005 & 0.022 & 0.000 & 0.040 & 0.011 & 0.028 & 0.003\\
         $\sigma_{\rm obs}$                &  0.025 & 0.023  & 0.007 & 0.004  &0.016 & 0.010  & 0.013 & 0.006     \\  
         $\sigma_{\rm model}$            &  0.009 & 0.011 & 0.009 & 0.011 &  0.009 & 0.011 &  0.009 & 0.011     \\         
         $\chi^2_{\rm red}$                &  1.065  &  1.030 &  1.176 & 1.074 & 1.186 & 1.201 & 0.910 & 0.862 \\
 	\noalign{\smallskip}	
         \hline
	\noalign{\smallskip}	
	$N_{\rm free}=4$      \\
	\noalign{\smallskip}	
	$v_{\rm rot}\sin{i}$    [km/s]            &  3.189 &  2.447  & 2.830 & 1.467 & 2.976 & 2.153 & 2.706 & 1.436\\
	$\sigma$            [km/s]                     &  0.067  &  0.093  & 0.025 & 0.048& 0.023 & 0.032 & 0.027 & 0.049 \\
	$\rm^6Li/^7Li$                     &   0.026 & -0.002 & 0.052 & 0.007 & 0.060 & 0.017 & 0.051 & 0.005 \\
         $\sigma_{\rm obs}$             &  0.016 & 0.013  & 0.006 & 0.003  & 0.011 & 0.007  & 0.010 & 0.005     \\ 
         $\sigma_{\rm model}$         &  0.004 & 0.006 & 0.004 & 0.006 &  0.004 & 0.006 &  0.004 & 0.006     \\
         $\chi^2_{\rm red}$             & 1.069 & 1.022  & 1.632 & 1.148 & 1.204 & 1.198 & 0.988 & 0.855 \\
	\noalign{\smallskip}	
                   \hline

         \end{tabular}
\end{table*}

\begin{figure*}[htb]
\begin{minipage}[b]{0.5\linewidth}
\centering
\includegraphics[scale=0.35,viewport=5cm 0cm 25cm 21cm]{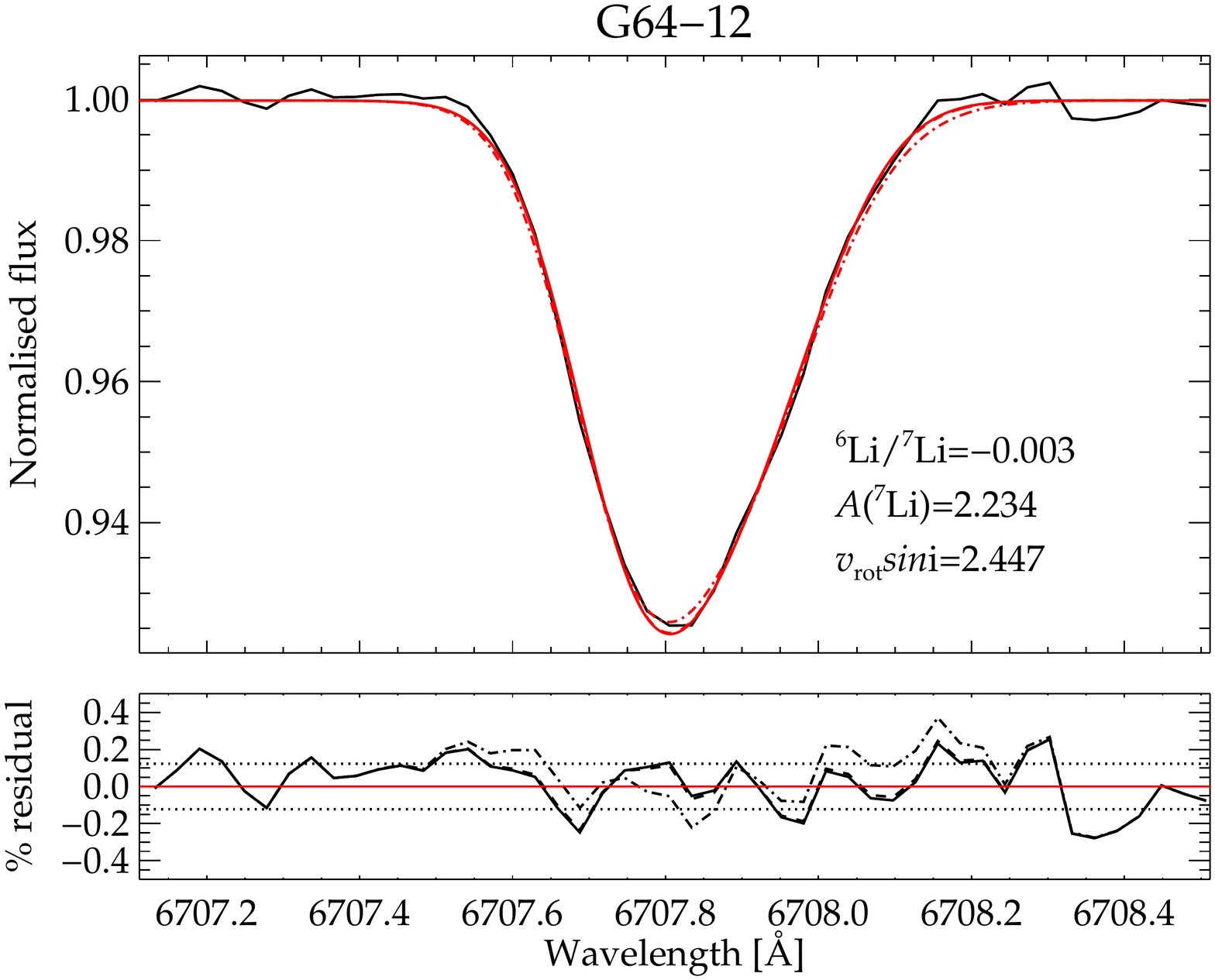}
\end{minipage}
\begin{minipage}[b]{0.5\linewidth}
\centering
\includegraphics[scale=0.35,viewport=5cm 0cm 25cm 21cm]{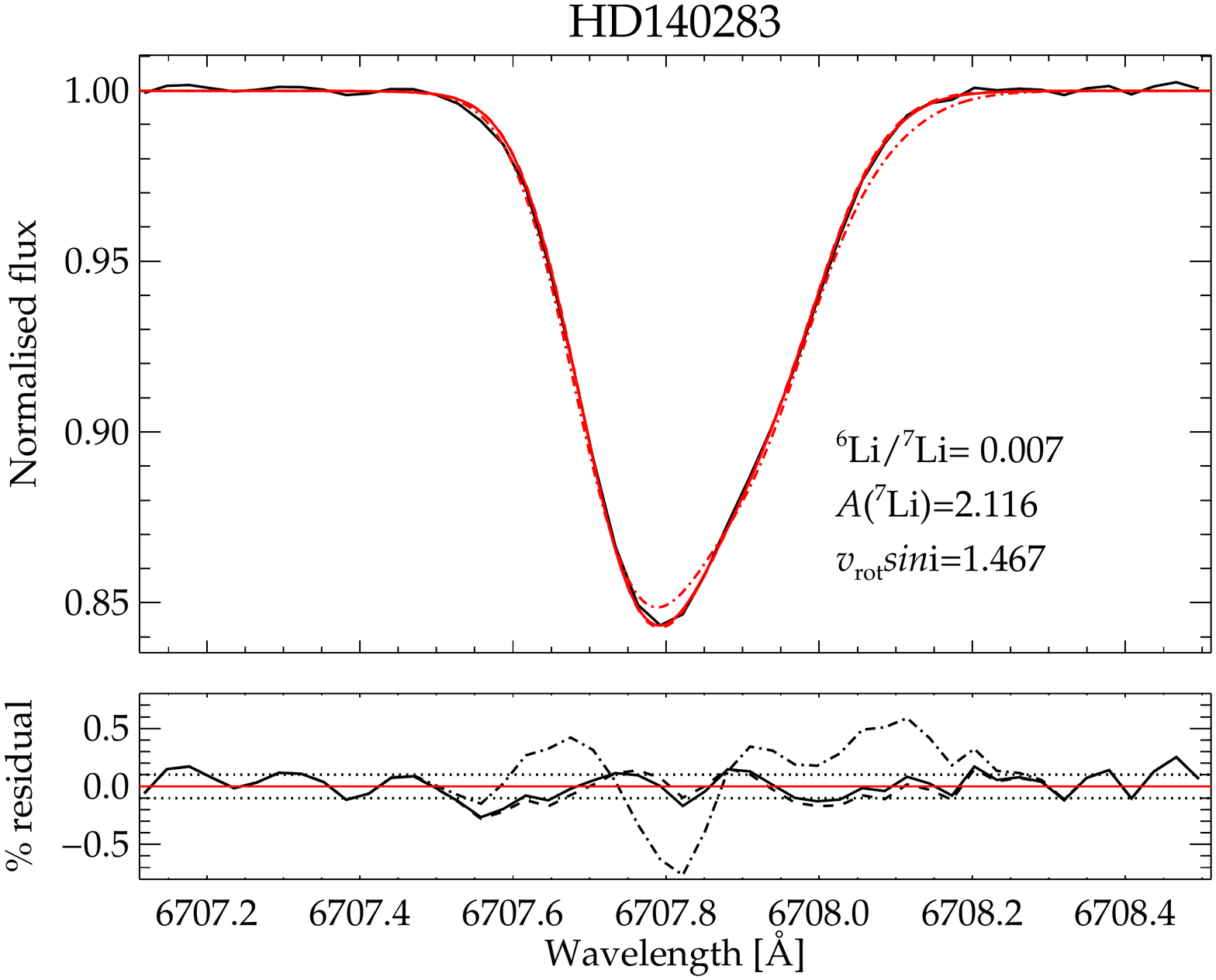}
\end{minipage}
\begin{minipage}[b]{0.5\linewidth}
\centering
\includegraphics[scale=0.35,viewport=5cm 0cm 25cm 21cm]{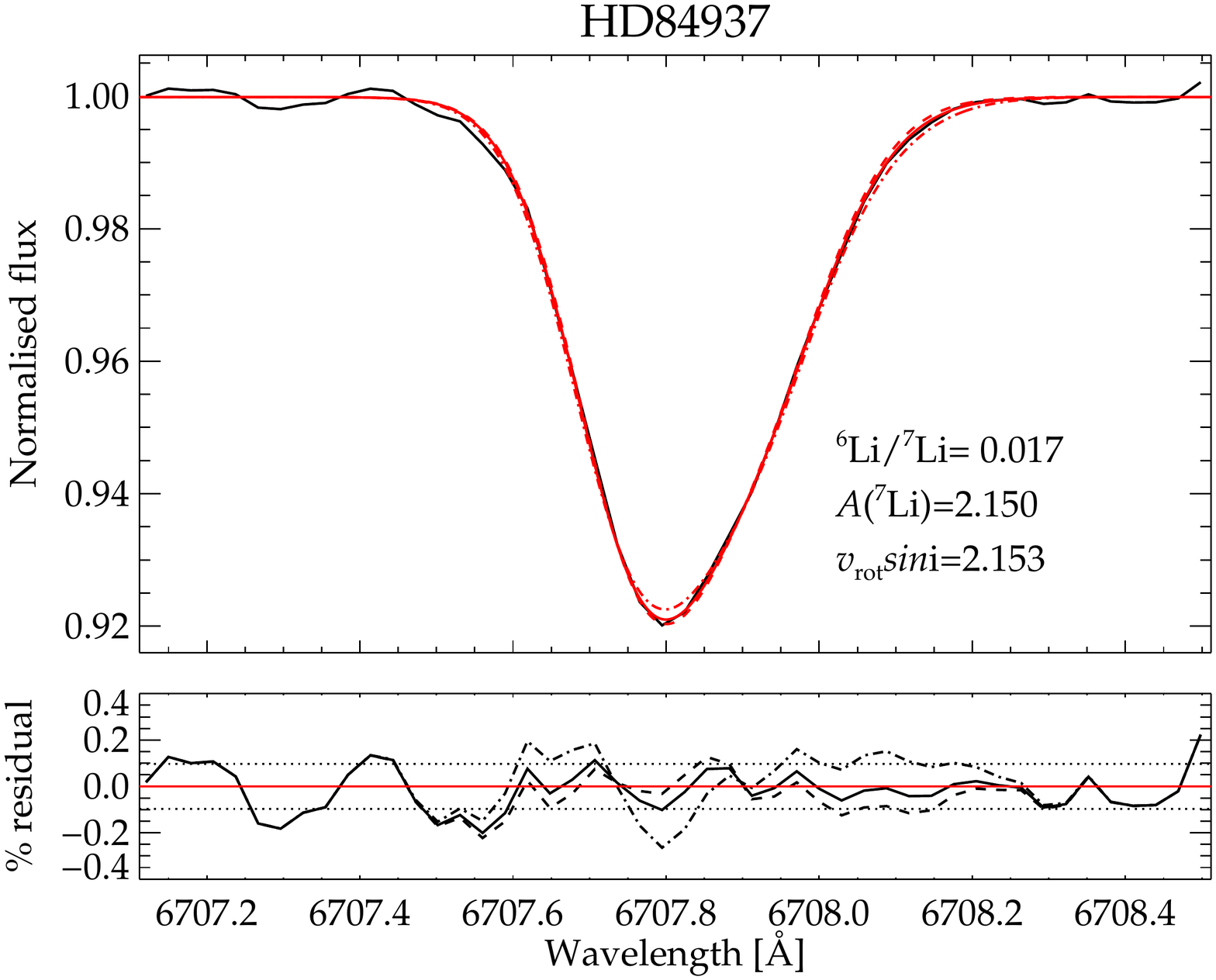}
\end{minipage}
\begin{minipage}[b]{0.5\linewidth}
\centering
\includegraphics[scale=0.35,viewport=5cm 0cm 25cm 21cm]{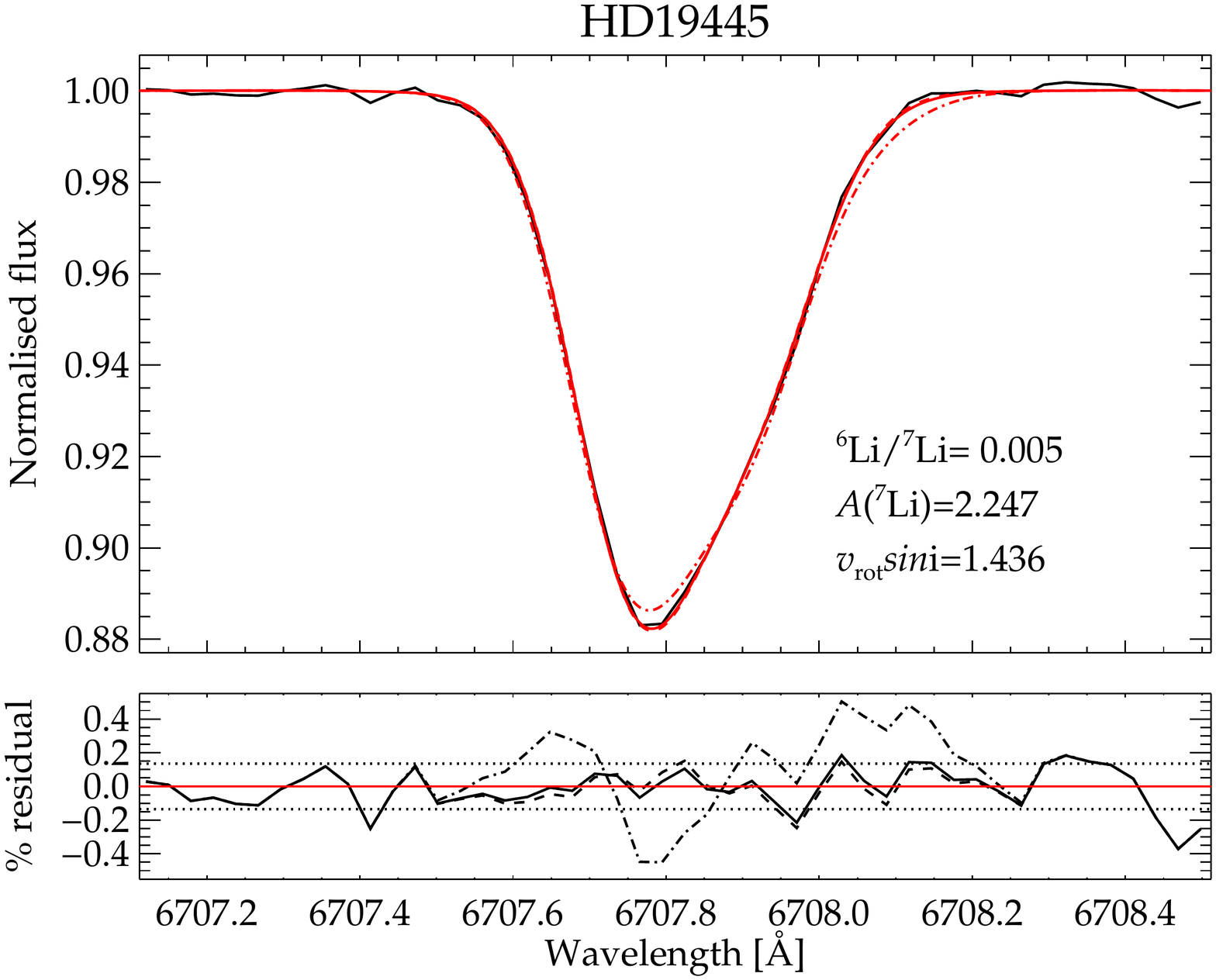}
\end{minipage}
\caption{Observed line profiles \textit{(black solid lines)} and best-fit 3D, NLTE synthetic profiles. The projected rotational velocity has been constrained by calibration lines. Below each panel the flux residual (observed-synthetic) in per cent is shown. The three line profiles correspond to the best-fit value indicated in the plot \textit{(solid line)}, $\rm^6Li/^7Li=0.0$ \textit{(dashed line)} and $\rm^6Li/^7Li=0.05$ \textit{(dotted-dashed lines)}. The   estimated observational uncertainty per pixel is also indicated \textit{(dotted horisontal lines)}. }
\label{fig:liprof}
\end{figure*}

The Li isotopic abundances were derived from the 6707\AA\ feature using either the line itself or other neutral lines to constrain the rotational line broadening, under the assumptions of LTE and NLTE. The four sets of results for $\rm^6Li/^7Li$, $v_{\rm rot}\sin{i}$, and associated $\chi_{\rm red}^2$ are summarised in Table \ref{tab:isotope}. The best-fit 3D, NLTE Li synthetic profiles obtained when using the calibration lines are shown in Fig.\,\ref{fig:liprof}.

The shapes and sizes of the probability contours displayed in Fig.\,\ref{fig:chisquare1} and \ref{fig:chisquare2} determine the sizes of the error bars. It is evident that increasing Li isotopic ratio is partly degenerate with decreasing projected rotational velocity, increasing central radial-velocity shift, and decreasing $^7$Li-abundance. The degeneracy with continuum normalisation is very small in comparison and therefore not shown. A striking result is how much the definition of the minimum benefits from the use of calibration lines, which can be appreciated by comparing the two sets of contours in each panel in Fig.\,\ref{fig:chisquare1} and \ref{fig:chisquare2}. The random error component, $\sigma_{\rm obs}$, decreases by up to a factor of two. Indeed, for G64-12 in particular, the use of calibration lines is necessary to obtain a total random error in the isotopic ratio below 2.5 percentage points. This is partly a reflection of the weakness of the Li line in this hot star.  However, the minimum $\chi_{\rm red}^2$-value is either decreased or increased when using calibration lines (see Table\,\ref{tab:isotope}), indicating either a better or worse goodness-of-fit. The choice of which method is better is thus not straightforward, but if the modelling is sound the two should agree on the location of the minimum value within the error bars. Inspecting Table \ref{tab:isotope}, the agreement is substantially better in NLTE compared to LTE, with HD140283 being the most prominent example. For this star, the $v_{\rm rot}\sin{i}$-value derived from the Li line in LTE is almost $1$\,km/s greater than that from the calibration lines, which propagates into a significant difference in the Li isotopic ratio.

The great importance of accounting for NLTE effects in the line formation is evident from the Li isotopic ratios displayed in Table \ref{tab:isotope}. The best-fit ratios decrease by up to five percentage points in NLTE compared to LTE, and it is clear that the latter assumption can lead to spurious detections. In NLTE, no star has a $2\sigma$-detection of $^6\rm Li$ and the reported values should be regarded as upper limits. For HD84937, the non-detection found using only the line itself ($0.011\pm0.010\pm0.011$) challenges the significance of the $1\sigma$-detection when calibration lines are used ($0.017\pm0.007\pm0.006$). Extending the analysis to include more calibration lines would be desirable to further decrease the error bar.

Finally, we emphasize that the small difference in profile shape in NLTE with respect to LTE acts to increase the error in $v_{\rm rot}\sin{i}$, but decrease the error in $\rm^6Li/^7Li$. This is not surprising since with the more realistic NLTE modelling, the convective broadening becomes relatively more influential, which forces a larger relative error on the rotational velocity.

\section{Discussion}   

It has only recently become numerically feasible to perform the complex 3D, NLTE calculations outlined in this study. Therefore, until a few years ago, all quantitative analyses of the lithium isotopic ratio were performed under the 1D, LTE assumptions. \citet{Asplund06} were the first to investigate the influence of 3D, LTE modelling on the abundances and found minor differences with respect to 1D, with no large systematic influence. Due to the stronger departures from LTE expected in 3D, the authors put more trust in the 1D, LTE results. Including also the follow-up paper by \citet{Asplund08}, a total of 27 stars were analysed in the two studies, 11 of which were found to have a significant $2\sigma$-detection of $^6$Li. The studied sample contains all four of our stars and significant detections were then reported for HD84937 ($0.051\pm0.015$) and G64-12 ($0.059\pm0.021$), while the other two stars were found compatible with a vanishing $^6$Li-signature. In addition to the 1D, LTE modelling technique, their analysis differs from our new study in the choice of calibration lines; the studies by \citet{Asplund06} and \citet{Asplund08} used a larger set of 14--35 lines of NaI\,, Mg\,I, Ca\,I, Sc\,II, Ti\,II, Cr\,I, Fe\,I, and Fe\,II, while we are here limited to 4--11 lines of Na\,I and Ca\,I, however in our case analysed in 3D, NLTE for the first time. 

We have confirmed that the conclusions of the two previous studies are verified using 1D, LTE calculations and the same set of calibration lines, but with our new $\chi^2$-analysis routines. However, trusting only the smaller set of calibration lines used in the present study, the detection vanishes for G64-12 and is thus in agreement with our 3D, LTE and 3D, NLTE results. The other stars are not much affected. We believe that an important reason why the results for G64-12 are less model dependent compared to the other stars in our analysis (see Table\,\ref{tab:isotope}), is the strong similarity in the line formation of Li\,I and Na\,I resonance lines (see Fig.\,\ref{fig:hist}), which here have similar strengths. The use of appropriate calibration lines thus effectively cancel out systematic uncertainties in the determination of the isotopic ratio. Of course, one should remember that the Na D lines may in general be less suitable for other reasons, e.g. interstellar absorption. From our results for the other stars, for which only Ca\,I lines have been used, it seems that the cancellation of 3D and NLTE effects work less well between Li and Ca. 3D, NLTE modelling of both calibration lines and the Li line is required to not make false detections of $\rm^6Li$. We emphasize that it would be desirable to have 3D, NLTE modelling of Fe and other elements to increase the number of available calibration lines.

\citet{Cayrel07} demonstrated with a mostly theoretical analysis how the neglected Li line asymmetry in 1D, LTE, with respect to 3D, NLTE, may lead to overestimation of the lithium isotopic ratio. Following this line of arguments, \citet{Steffen10a,Steffen10b} and \citet{Steffen12} applied theoretical corrections to the 1D, LTE results of  \citet{Asplund06} and \citet{Asplund08}, based on modelling of only the Li line in 1D, LTE and 3D, NLTE.  By repeating our isotopic analysis in 1D, LTE without use of calibration lines, we confirm that the typical sizes of such corrections would be of order $1-2\%$. However, as \citet{Steffen12} also points out, this post-correction procedure is complicated by the fact that the results of  \citet{Asplund06} and \citet{Asplund08} were based on simultaneous modelling of calibration lines. As is evident from the results presented in Table \ref{tab:isotope}, the lithium isotopic ratios are  systematically amplified by the use of calibration lines in LTE, which is also in agreement with the findings of \citet{Steffen10b}. In 3D, NLTE, the calibration lines serve their intended purpose to decrease the observational error. 

The main conclusions outlined in this paper agree with those presented in \citet{Lind12c}, where we first tested our new 3D, NLTE modelling technique. Quantitative differences in isotopic ratio lie within the error bars and are most notable for G64-12. They arise mainly because our previous study used observed spectra that had not yet been optimally processed on a line-by-line basis and a different set of calibration lines. 

Further, we note that our 3D, NLTE result for the metal-poor turn-off star HD84937 ($0.011\pm0.010\pm0.011$ without the use of calibration lines) is barely in agreement with the corresponding result by \citet{Steffen12}, who find $0.051\pm0.023$, based on a direct comparison to observed spectra. However, the difference is critical because it leads to different conclusions whether the lighter isotope has been detected or not. We are confident that our upper limit is more realistic, based on our superior spectrum quality and important verification using calibration lines. 

In this study, we have demonstrated that lithium isotopic abundances in metal-poor halo stars are prone to systematic uncertainties due to the common simplifying assumptions of 1D and LTE and that these uncertainties do not necessarily cancel out using calibration lines. A full understanding of 3D, NLTE line formation is necessary to make correct measurements of the level of $^6\rm Li$. We conclude from our study that only upper limits can be derived on the isotopic ratios in our studied stars; there is thus currently no empirical evidence for a high $^6\rm Li$ content in the early Galaxy that could signal a cosmological production, perhaps stemming from non-standard Big Bang nucleosynthesis. It will be of great value to extend our study to higher metallicity stars, in order to pinpoint when Galactic production of the lighter isotope starts to significantly influence spectral line profiles and thereby testing our understanding of the Galactic chemical evolution of the Li isotopic ratio.

\begin{acknowledgements}
We acknowledge W. Hayek for his help with 3D line formation calculations and M. Bergemann, T. Gehren. Klaus Fuhrmann, and Michael Pfeiffer for providing the FOCES spectra. The data presented herein were obtained at the W.M. Keck Observatory, which is operated as a scientific partnership among the California Institute of Technology, the University of California and the National Aeronautics and Space Administration. The Observatory was made possible by the generous financial support of the W.\,M.\,Keck Foundation.
\end{acknowledgements}

\bibliographystyle{aa}

\begin{thebibliography}{51}
\expandafter\ifx\csname natexlab\endcsname\relax\def\natexlab#1{#1}\fi

\bibitem[{{Asplund}(2005)}]{Asplund05}
{Asplund}, M. 2005, \araa, 43, 481

\bibitem[{{Asplund} {et~al.}(2003){Asplund}, {Carlsson}, \&
  {Botnen}}]{Asplund03}
{Asplund}, M., {Carlsson}, M., \& {Botnen}, A.~V. 2003, \aap, 399, L31

\bibitem[{{Asplund} {et~al.}(2006){Asplund}, {Lambert}, {Nissen}, {Primas}, \&
  {Smith}}]{Asplund06}
{Asplund}, M., {Lambert}, D.~L., {Nissen}, P.~E., {Primas}, F., \& {Smith},
  V.~V. 2006, \apj, 644, 229

\bibitem[{{Asplund} \& {Mel{\'e}ndez}(2008)}]{Asplund08}
{Asplund}, M. \& {Mel{\'e}ndez}, J. 2008, in American Institute of Physics
  Conference Series, Vol. 990, First Stars III, ed. B.~W. {O'Shea} \&
  A.~{Heger}, 342--346

\bibitem[{{Asplund} {et~al.}(2000){Asplund}, {Nordlund}, {Trampedach}, {Allende
  Prieto}, \& {Stein}}]{Asplund00}
{Asplund}, M., {Nordlund}, {\AA}., {Trampedach}, R., {Allende Prieto}, C., \&
  {Stein}, R.~F. 2000, \aap, 359, 729

\bibitem[{{Asplund} {et~al.}(1999){Asplund}, {Nordlund}, {Trampedach}, \&
  {Stein}}]{Asplund99}
{Asplund}, M., {Nordlund}, {\AA}., {Trampedach}, R., \& {Stein}, R.~F. 1999,
  \aap, 346, L17

\bibitem[{{Barklem} {et~al.}(2010){Barklem}, {Belyaev}, {Dickinson}, \&
  {Gad{\'e}a}}]{Barklem10}
{Barklem}, P.~S., {Belyaev}, A.~K., {Dickinson}, A.~S., \& {Gad{\'e}a}, F.~X.
  2010, \aap, 519, A20+

\bibitem[{{Barklem} {et~al.}(2000){Barklem}, {Piskunov}, \&
  {O'Mara}}]{Barklem00b}
{Barklem}, P.~S., {Piskunov}, N., \& {O'Mara}, B.~J. 2000, \aaps, 142, 467

\bibitem[{{Bergemann} {et~al.}(2012){Bergemann}, {Lind}, {Collet}, {Magic}, \&
  {Asplund}}]{Bergemann12}
{Bergemann}, M., {Lind}, K., {Collet}, R., {Magic}, Z., \& {Asplund}, M. 2012,
  \mnras, 427, 27

\bibitem[{{Botnen}(1997)}]{Botnen97}
{Botnen}, A. 1997, Master's thesis, Master's thesis,
  Inst.~Theor.~Astrophys.~Oslo

\bibitem[{{Casagrande} {et~al.}(2010){Casagrande}, {Ram{\'{\i}}rez},
  {Mel{\'e}ndez}, {Bessell}, \& {Asplund}}]{Casagrande10}
{Casagrande}, L., {Ram{\'{\i}}rez}, I., {Mel{\'e}ndez}, J., {Bessell}, M., \&
  {Asplund}, M. 2010, \aap, 512, A54+

\bibitem[{{Cayrel} {et~al.}(1999){Cayrel}, {Spite}, {Spite}, {Vangioni-Flam},
  {Cass{\'e}}, \& {Audouze}}]{Cayrel99}
{Cayrel}, R., {Spite}, M., {Spite}, F., {et~al.} 1999, \aap, 343, 923

\bibitem[{{Cayrel} {et~al.}(2007){Cayrel}, {Steffen}, {Chand}, {Bonifacio},
  {Spite}, {Spite}, {Petitjean}, {Ludwig}, \& {Caffau}}]{Cayrel07}
{Cayrel}, R., {Steffen}, M., {Chand}, H., {et~al.} 2007, \aap, 473, L37

\bibitem[{{Coc} {et~al.}(2012){Coc}, {Goriely}, {Xu}, {Saimpert}, \&
  {Vangioni}}]{Coc12}
{Coc}, A., {Goriely}, S., {Xu}, Y., {Saimpert}, M., \& {Vangioni}, E. 2012,
  \apj, 744, 158

\bibitem[{{Collet} {et~al.}(2011){Collet}, {Hayek}, {Asplund}, {Nordlund},
  {Trampedach}, \& {Gudiksen}}]{Collet11a}
{Collet}, R., {Hayek}, W., {Asplund}, M., {et~al.} 2011, \aap, 528, A32

\bibitem[{{D'Antona} \& {Ventura}(2010)}]{Dantona10}
{D'Antona}, F. \& {Ventura}, P. 2010, in IAU Symposium, Vol. 268, IAU
  Symposium, ed. C.~{Charbonnel}, M.~{Tosi}, F.~{Primas}, \& C.~{Chiappini},
  395--404

\bibitem[{{Drawin}(1968)}]{Drawin68}
{Drawin}, H.-W. 1968, Zeitschrift f\"ur Physik, 211, 404

\bibitem[{{Feautrier}(1964)}]{Feautrier64}
{Feautrier}, P. 1964, Comptes Rendus Academie des Sciences, 258, 3189

\bibitem[{{Fields}(2011)}]{Fields11}
{Fields}, B.~D. 2011, Annual Review of Nuclear and Particle Science, 61, 47

\bibitem[{{Hayek} {et~al.}(2011){Hayek}, {Asplund}, {Collet}, \&
  {Nordlund}}]{Hayek11}
{Hayek}, W., {Asplund}, M., {Collet}, R., \& {Nordlund}, {\AA}. 2011, \aap,
  529, A158

\bibitem[{{Hobbs} \& {Thorburn}(1994)}]{Hobbs94}
{Hobbs}, L.~M. \& {Thorburn}, J.~A. 1994, \apjl, 428, L25

\bibitem[{{Jedamzik} \& {Pospelov}(2009)}]{Jedamzik09}
{Jedamzik}, K. \& {Pospelov}, M. 2009, New Journal of Physics, 11, 105028

\bibitem[{{Komatsu} {et~al.}(2011){Komatsu}, {Smith}, {Dunkley}, {Bennett},
  {Gold}, {Hinshaw}, {Jarosik}, {Larson}, {Nolta}, {Page}, {Spergel},
  {Halpern}, {Hill}, {Kogut}, {Limon}, {Meyer}, {Odegard}, {Tucker}, {Weiland},
  {Wollack}, \& {Wright}}]{Komatsu11}
{Komatsu}, E., {Smith}, K.~M., {Dunkley}, J., {et~al.} 2011, \apjs, 192, 18

\bibitem[{{Korn} {et~al.}(2007){Korn}, {Grundahl}, {Richard}, {Mashonkina},
  {Barklem}, {Collet}, {Gustafsson}, \& {Piskunov}}]{Korn07}
{Korn}, A.~J., {Grundahl}, F., {Richard}, O., {et~al.} 2007, \apj, 671, 402

\bibitem[{{Leenaarts} {et~al.}(2010){Leenaarts}, {Rutten}, {Reardon},
  {Carlsson}, \& {Hansteen}}]{Leenaarts10}
{Leenaarts}, J., {Rutten}, R.~J., {Reardon}, K., {Carlsson}, M., \& {Hansteen},
  V. 2010, \apj, 709, 1362

\bibitem[{{Lind} {et~al.}(2009{\natexlab{a}}){Lind}, {Asplund}, \&
  {Barklem}}]{Lind09a}
{Lind}, K., {Asplund}, M., \& {Barklem}, P.~S. 2009{\natexlab{a}}, \aap, 503,
  541

\bibitem[{{Lind} {et~al.}(2011){Lind}, {Asplund}, {Barklem}, \&
  {Belyaev}}]{Lind11b}
{Lind}, K., {Asplund}, M., {Barklem}, P.~S., \& {Belyaev}, A.~K. 2011, \aap,
  528, A103

\bibitem[{{Lind} {et~al.}(2012){Lind}, {Asplund}, {Collet}, \&
  {Mel{\'e}ndez}}]{Lind12c}
{Lind}, K., {Asplund}, M., {Collet}, R., \& {Mel{\'e}ndez}, J. 2012, Memorie
  della Societa Astronomica Italiana Supplementi, 22, 142

\bibitem[{{Lind} {et~al.}(2009{\natexlab{b}}){Lind}, {Primas}, {Charbonnel},
  {Grundahl}, \& {Asplund}}]{Lind09b}
{Lind}, K., {Primas}, F., {Charbonnel}, C., {Grundahl}, F., \& {Asplund}, M.
  2009{\natexlab{b}}, \aap, 503, 545

\bibitem[{{Magic} {et~al.}(2013){Magic}, {Collet}, {Asplund}, {Trampedach},
  {Hayek}, {Chiavassa}, {Stein}, \& {Nordlund}}]{Magic13}
{Magic}, Z., {Collet}, R., {Asplund}, M., {et~al.} 2013, ArXiv e-prints

\bibitem[{{Mashonkina} {et~al.}(2007){Mashonkina}, {Korn}, \&
  {Przybilla}}]{Mashonkina07}
{Mashonkina}, L., {Korn}, A.~J., \& {Przybilla}, N. 2007, \aap, 461, 261

\bibitem[{{Maurice} {et~al.}(1984){Maurice}, {Spite}, \& {Spite}}]{Maurice84}
{Maurice}, E., {Spite}, F., \& {Spite}, M. 1984, \aap, 132, 278

\bibitem[{{Mel{\'e}ndez} {et~al.}(2010){Mel{\'e}ndez}, {Casagrande},
  {Ram{\'{\i}}rez}, {Asplund}, \& {Schuster}}]{Melendez10}
{Mel{\'e}ndez}, J., {Casagrande}, L., {Ram{\'{\i}}rez}, I., {Asplund}, M., \&
  {Schuster}, W.~J. 2010, \aap, 515, L3+

\bibitem[{{Mihalas} {et~al.}(1988){Mihalas}, {Dappen}, \& {Hummer}}]{Mihalas88}
{Mihalas}, D., {Dappen}, W., \& {Hummer}, D.~G. 1988, \apj, 331, 815

\bibitem[{{Nissen} {et~al.}(2007){Nissen}, {Akerman}, {Asplund}, {Fabbian},
  {Kerber}, {Kaufl}, \& {Pettini}}]{Nissen07}
{Nissen}, P.~E., {Akerman}, C., {Asplund}, M., {et~al.} 2007, \aap, 469, 319

\bibitem[{{Nordlund}(1982)}]{Nordlund82}
{Nordlund}, A. 1982, \aap, 107, 1

\bibitem[{{Pfeiffer} {et~al.}(1998){Pfeiffer}, {Frank}, {Baumueller},
  {Fuhrmann}, \& {Gehren}}]{Pfeiffer98}
{Pfeiffer}, M.~J., {Frank}, C., {Baumueller}, D., {Fuhrmann}, K., \& {Gehren},
  T. 1998, \aaps, 130, 381

\bibitem[{{Pinsonneault}(1997)}]{Pinsonneault97}
{Pinsonneault}, M. 1997, \araa, 35, 557

\bibitem[{{Prantzos}(2012)}]{Prantzos12}
{Prantzos}, N. 2012, \aap, 542, A67

\bibitem[{{Richard}(2005)}]{Richard05b}
{Richard}, O. 2005, in Engineering and Science, Vol.~17, EAS Publications
  Series, ed. G.~{Alecian}, O.~{Richard}, \& S.~{Vauclair}, 43--52

\bibitem[{{Rollinde} {et~al.}(2006){Rollinde}, {Vangioni}, \&
  {Olive}}]{Rollinde06}
{Rollinde}, E., {Vangioni}, E., \& {Olive}, K.~A. 2006, \apj, 651, 658

\bibitem[{{Sansonetti} {et~al.}(1995){Sansonetti}, {Richou}, {Engleman}, \&
  {Radziemski}}]{Sansonetti95}
{Sansonetti}, C.~J., {Richou}, B., {Engleman}, R.~J., \& {Radziemski}, L.~J.
  1995, \pra, 52, 2682

\bibitem[{{Skartlien}(2000)}]{Skartlien00}
{Skartlien}, R. 2000, \apj, 536, 465

\bibitem[{{Smith} {et~al.}(1993){Smith}, {Lambert}, \& {Nissen}}]{Smith93}
{Smith}, V.~V., {Lambert}, D.~L., \& {Nissen}, P.~E. 1993, \apj, 408, 262

\bibitem[{{Smith} {et~al.}(1998){Smith}, {Lambert}, \& {Nissen}}]{Smith98}
{Smith}, V.~V., {Lambert}, D.~L., \& {Nissen}, P.~E. 1998, \apj, 506, 405

\bibitem[{{Steffen} {et~al.}(2010{\natexlab{a}}){Steffen}, {Cayrel},
  {Bonifacio}, {Ludwig}, \& {Caffau}}]{Steffen10b}
{Steffen}, M., {Cayrel}, R., {Bonifacio}, P., {Ludwig}, H., \& {Caffau}, E.
  2010{\natexlab{a}}, in IAU Symposium, Vol. 268, IAU Symposium, ed.
  {C.~Charbonnel, M.~Tosi, F.~Primas, \& C.~Chiappini}, 215--220

\bibitem[{{Steffen} {et~al.}(2010{\natexlab{b}}){Steffen}, {Cayrel},
  {Bonifacio}, {Ludwig}, \& {Caffau}}]{Steffen10a}
{Steffen}, M., {Cayrel}, R., {Bonifacio}, P., {Ludwig}, H.-G., \& {Caffau}, E.
  2010{\natexlab{b}}, in IAU Symposium, Vol. 265, IAU Symposium, ed.
  K.~{Cunha}, M.~{Spite}, \& B.~{Barbuy}, 23--26

\bibitem[{{Steffen} {et~al.}(2012){Steffen}, {Cayrel}, {Caffau}, {Bonifacio},
  {Ludwig}, \& {Spite}}]{Steffen12}
{Steffen}, M., {Cayrel}, R., {Caffau}, E., {et~al.} 2012, Memorie della Societa
  Astronomica Italiana Supplementi, 22, 152

\bibitem[{{Suzuki} \& {Inoue}(2002)}]{Suzuki02}
{Suzuki}, T.~K. \& {Inoue}, S. 2002, \apj, 573, 168

\bibitem[{{Uns\"old}(1938)}]{Unsoeld38}
{Uns\"old}, A. 1938, {Physik der Sternamosph\"aren, MIT besonderer
  Ber\"ucksichtigung der Sonne}

\bibitem[{{Vogt} {et~al.}(1994){Vogt}, {Allen}, {Bigelow}, {Bresee}, {Brown},
  {Cantrall}, {Conrad}, {Couture}, {Delaney}, {Epps}, {Hilyard}, {Hilyard},
  {Horn}, {Jern}, {Kanto}, {Keane}, {Kibrick}, {Lewis}, {Osborne},
  {Pardeilhan}, {Pfister}, {Ricketts}, {Robinson}, {Stover}, {Tucker}, {Ward},
  \& {Wei}}]{Vogt94}
{Vogt}, S.~S., {Allen}, S.~L., {Bigelow}, B.~C., {et~al.} 1994, in Society of
  Photo-Optical Instrumentation Engineers (SPIE) Conference Series, Vol. 2198,
  Society of Photo-Optical Instrumentation Engineers (SPIE) Conference Series,
  ed. D.~L. {Crawford} \& E.~R. {Craine}, 362

\end{thebibliography}

\begin{appendix} 
\section{3D NLTE synthetic line profiles}

\begin{figure*}[htb]
\begin{minipage}[b]{0.33\linewidth}
\centering
\includegraphics[scale=0.23,viewport=5cm 0cm 25cm 21cm]{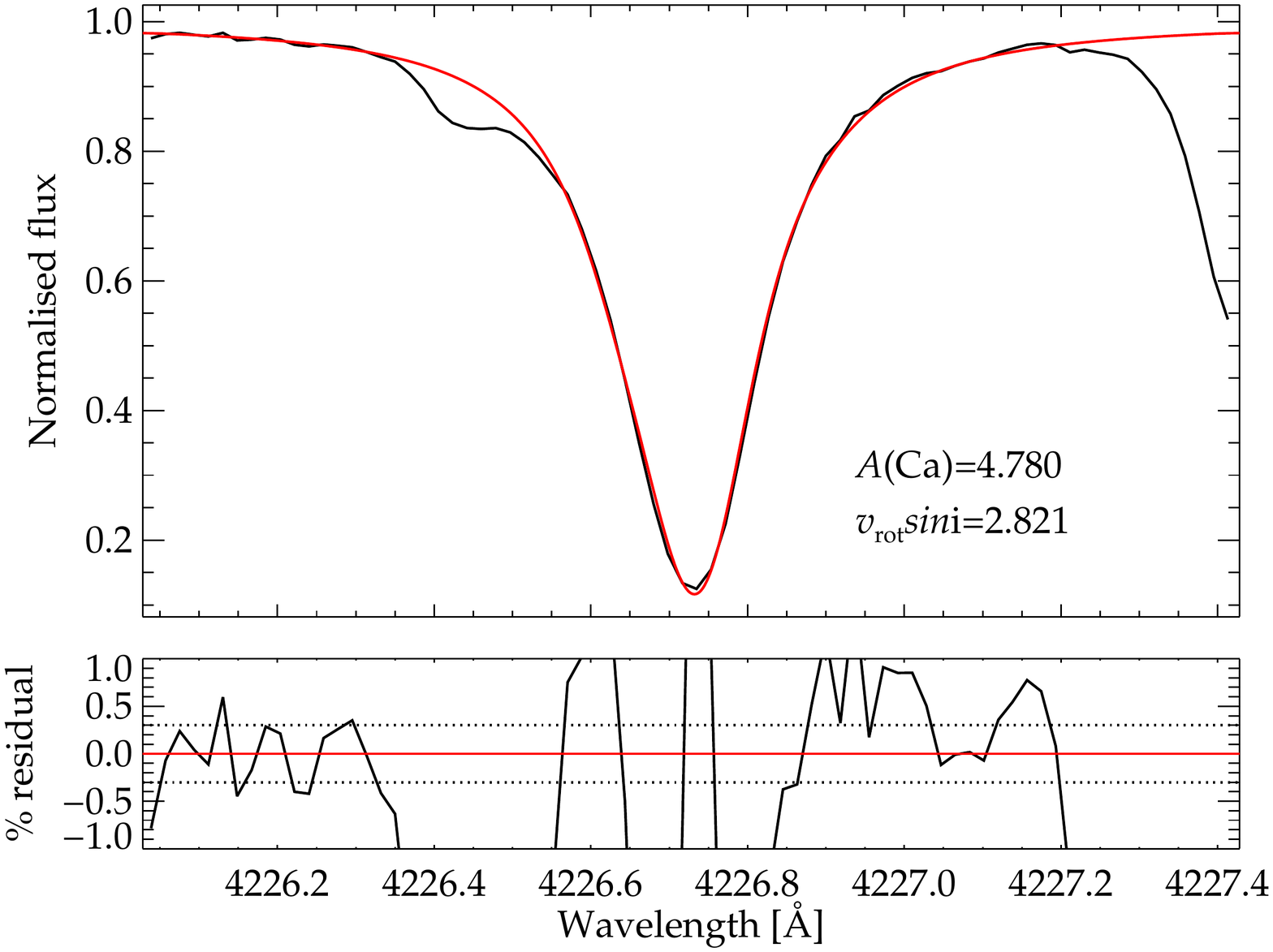}
\end{minipage}
\begin{minipage}[b]{0.33\linewidth}
\centering
\includegraphics[scale=0.23,viewport=5cm 0cm 25cm 21cm]{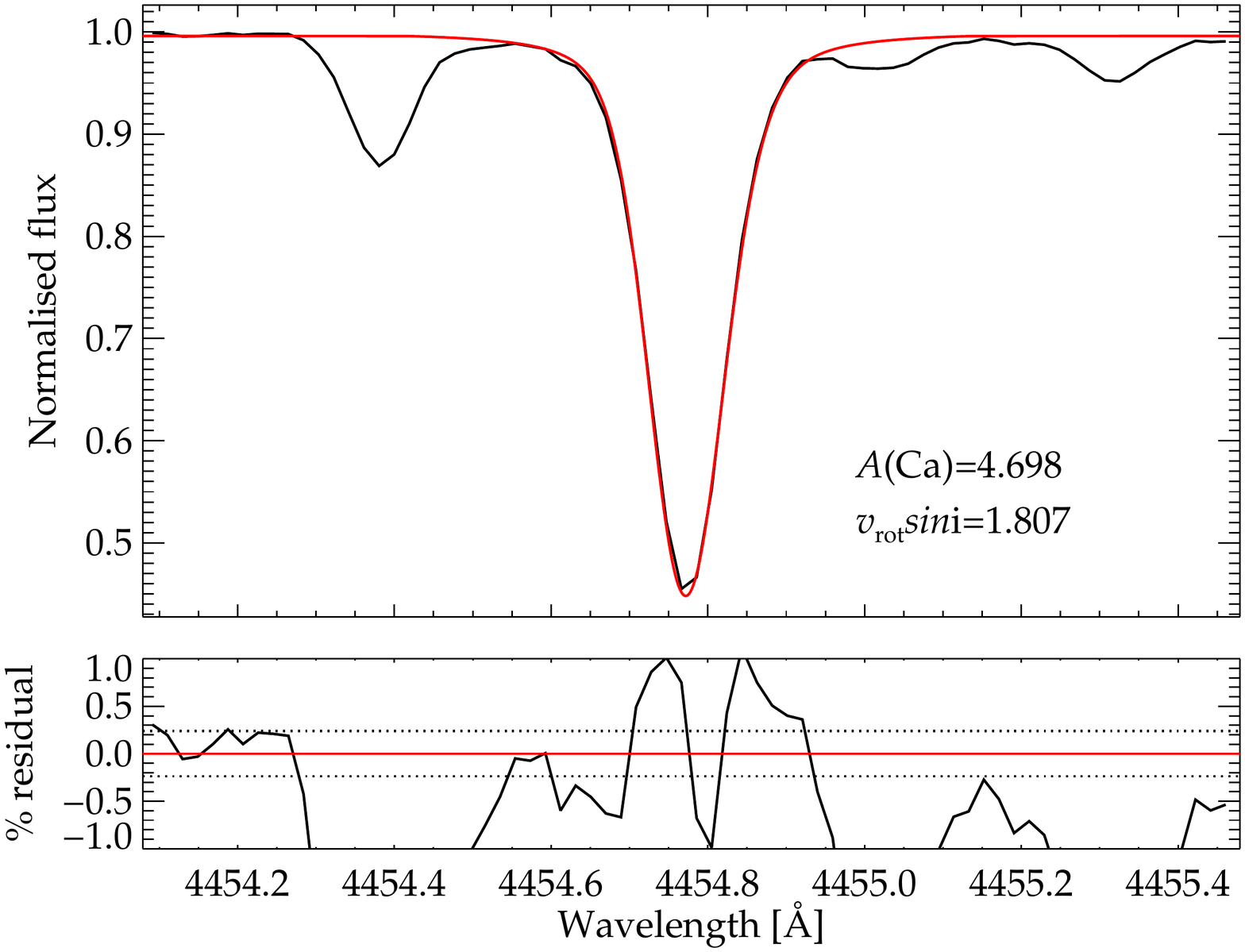}
\end{minipage}
\begin{minipage}[b]{0.33\linewidth}
\centering
\includegraphics[scale=0.23,viewport=5cm 0cm 25cm 21cm]{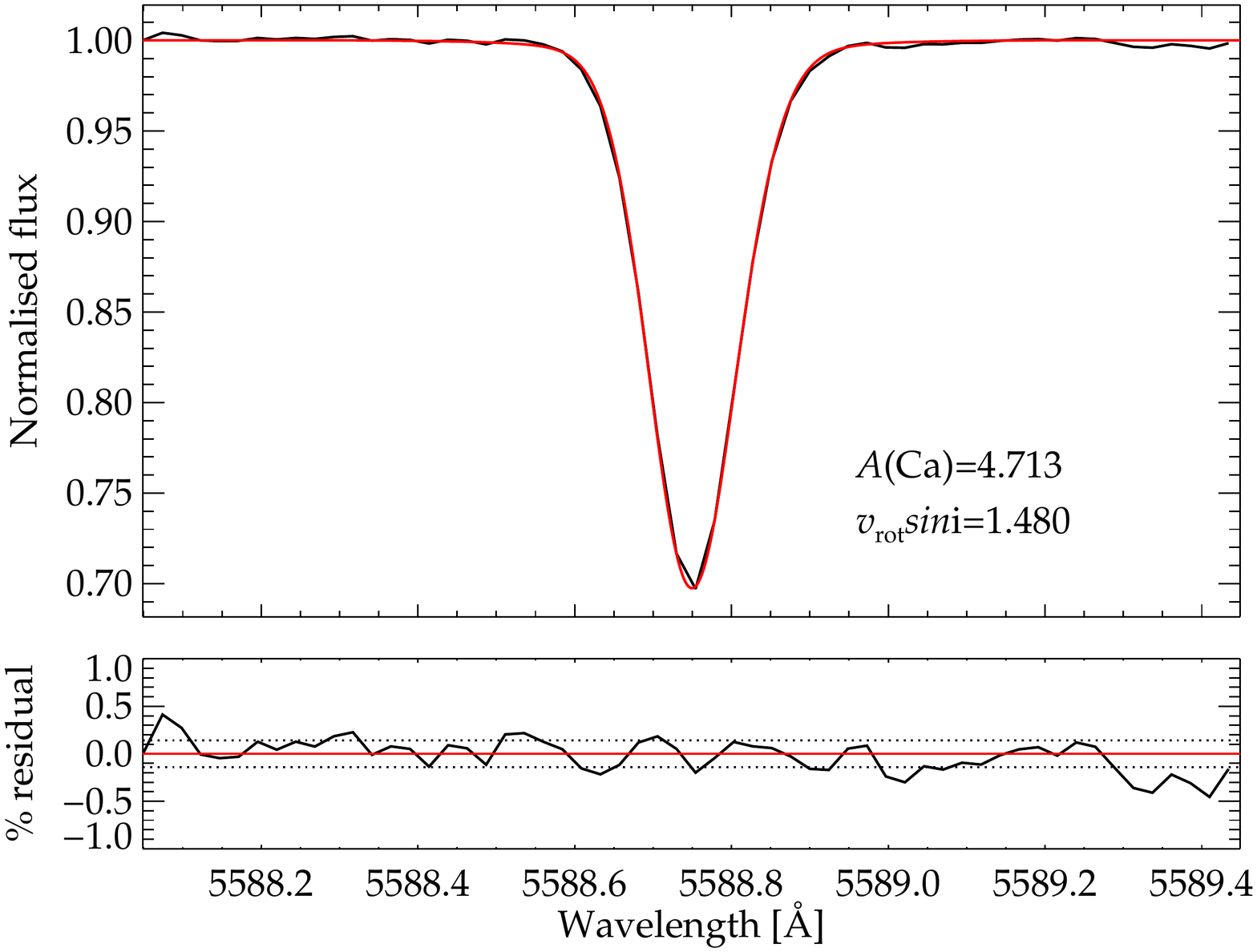}
\end{minipage}
\begin{minipage}[b]{0.33\linewidth}
\centering
\includegraphics[scale=0.23,viewport=5cm 0cm 25cm 21cm]{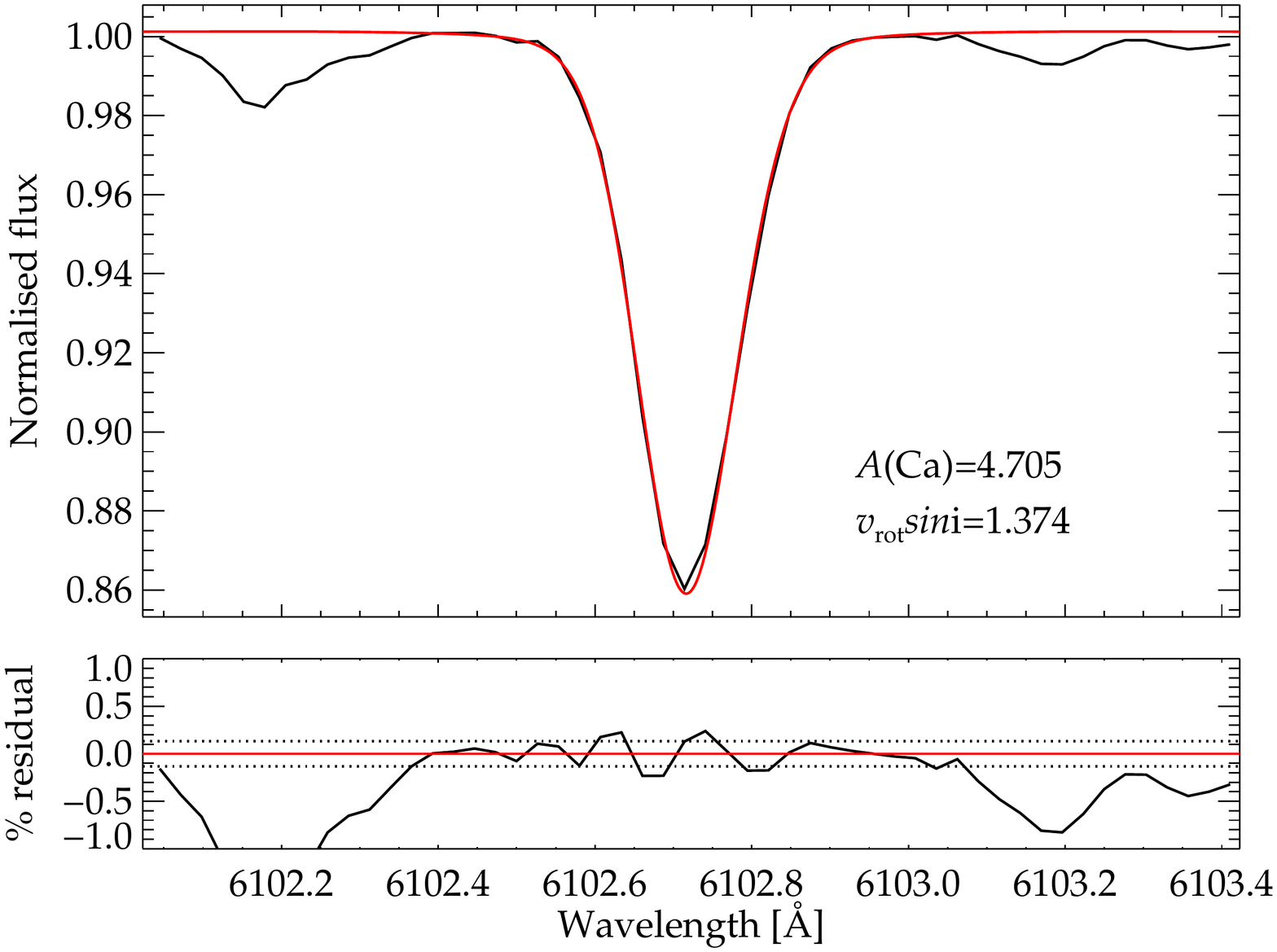}
\end{minipage}
\begin{minipage}[b]{0.33\linewidth}
\centering
\includegraphics[scale=0.23,viewport=5cm 0cm 25cm 21cm]{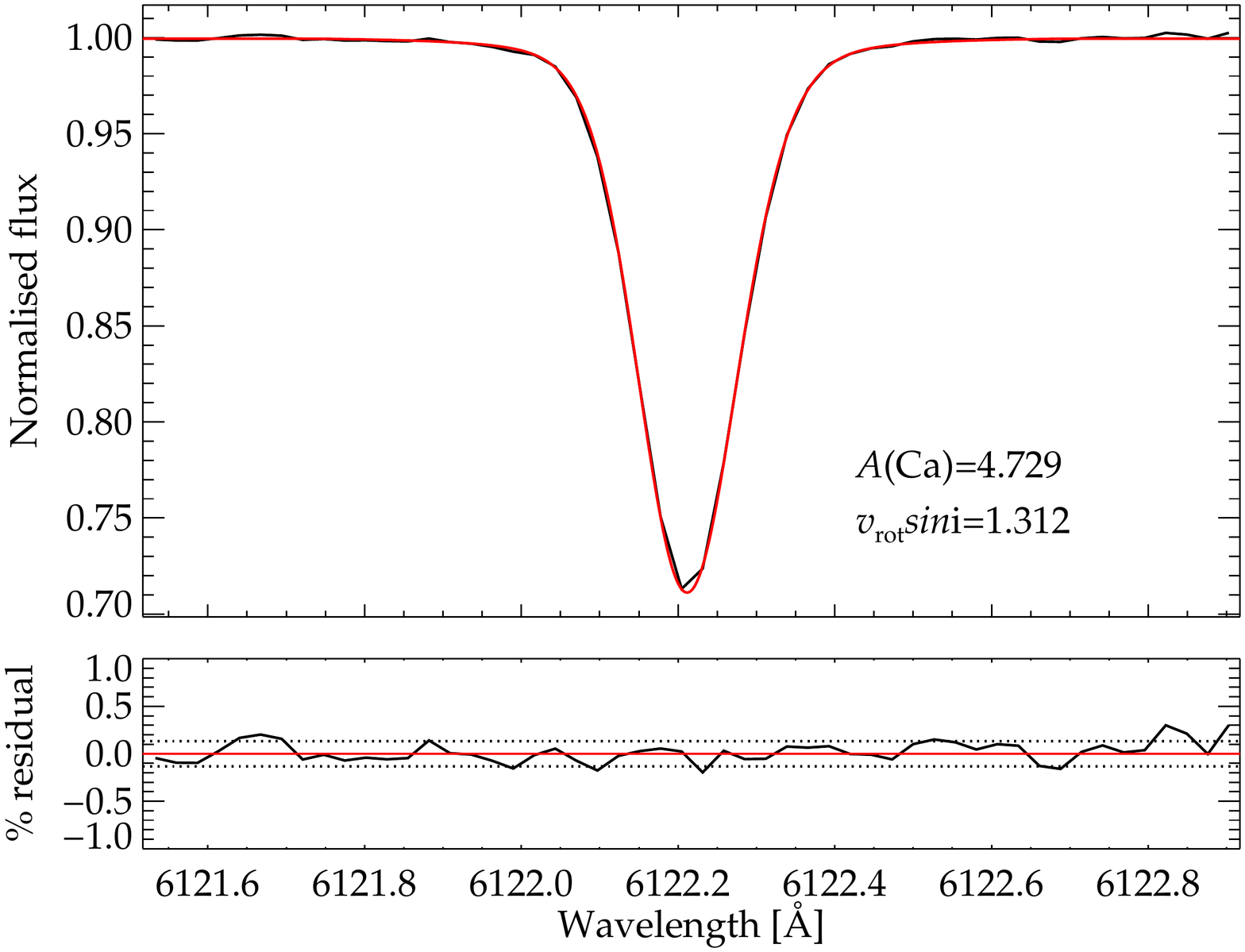}
\end{minipage}
\begin{minipage}[b]{0.33\linewidth}
\centering
\includegraphics[scale=0.23,viewport=5cm 0cm 25cm 21cm]{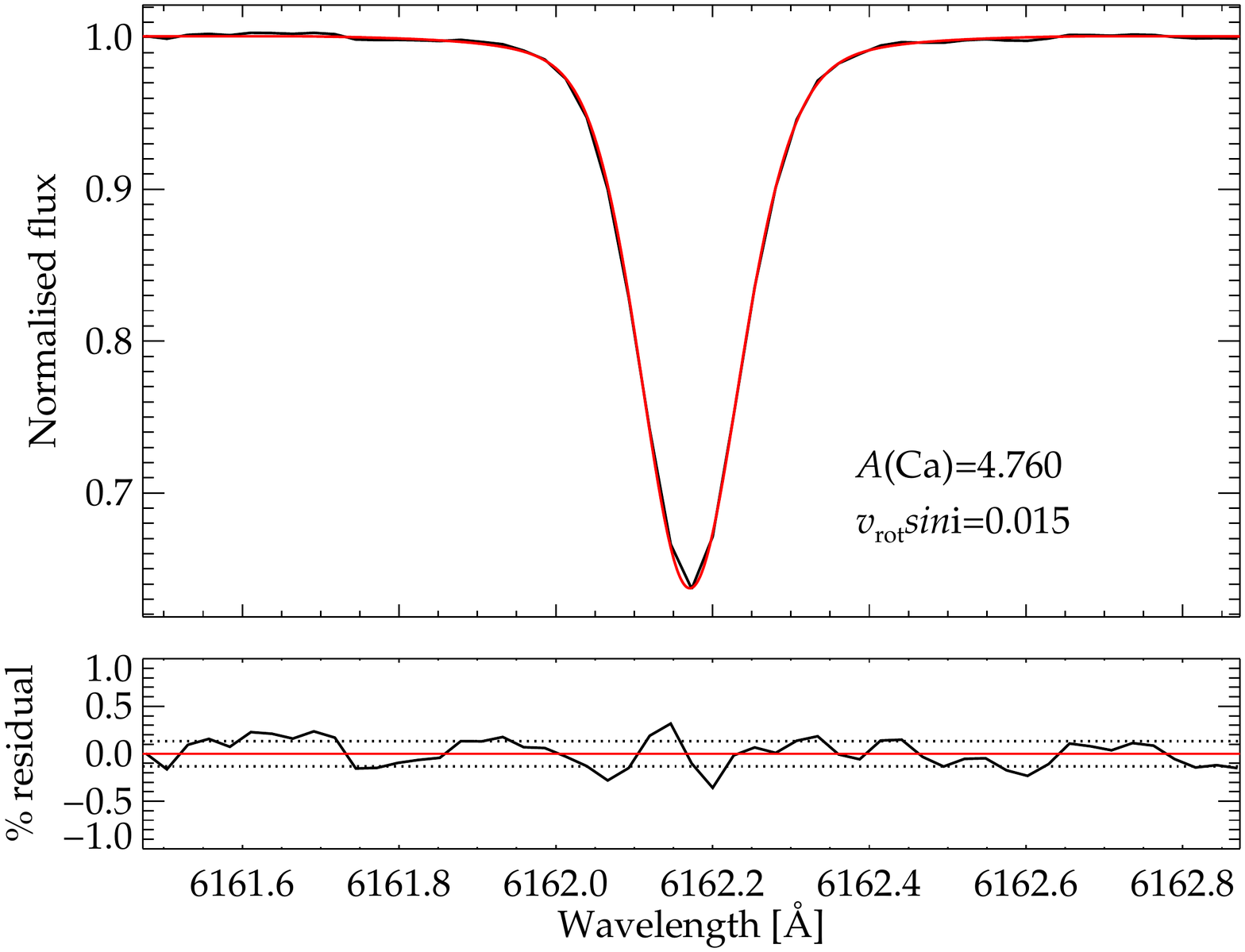}
\end{minipage}
\begin{minipage}[b]{0.33\linewidth}
\centering
\includegraphics[scale=0.23,viewport=5cm 0cm 25cm 21cm]{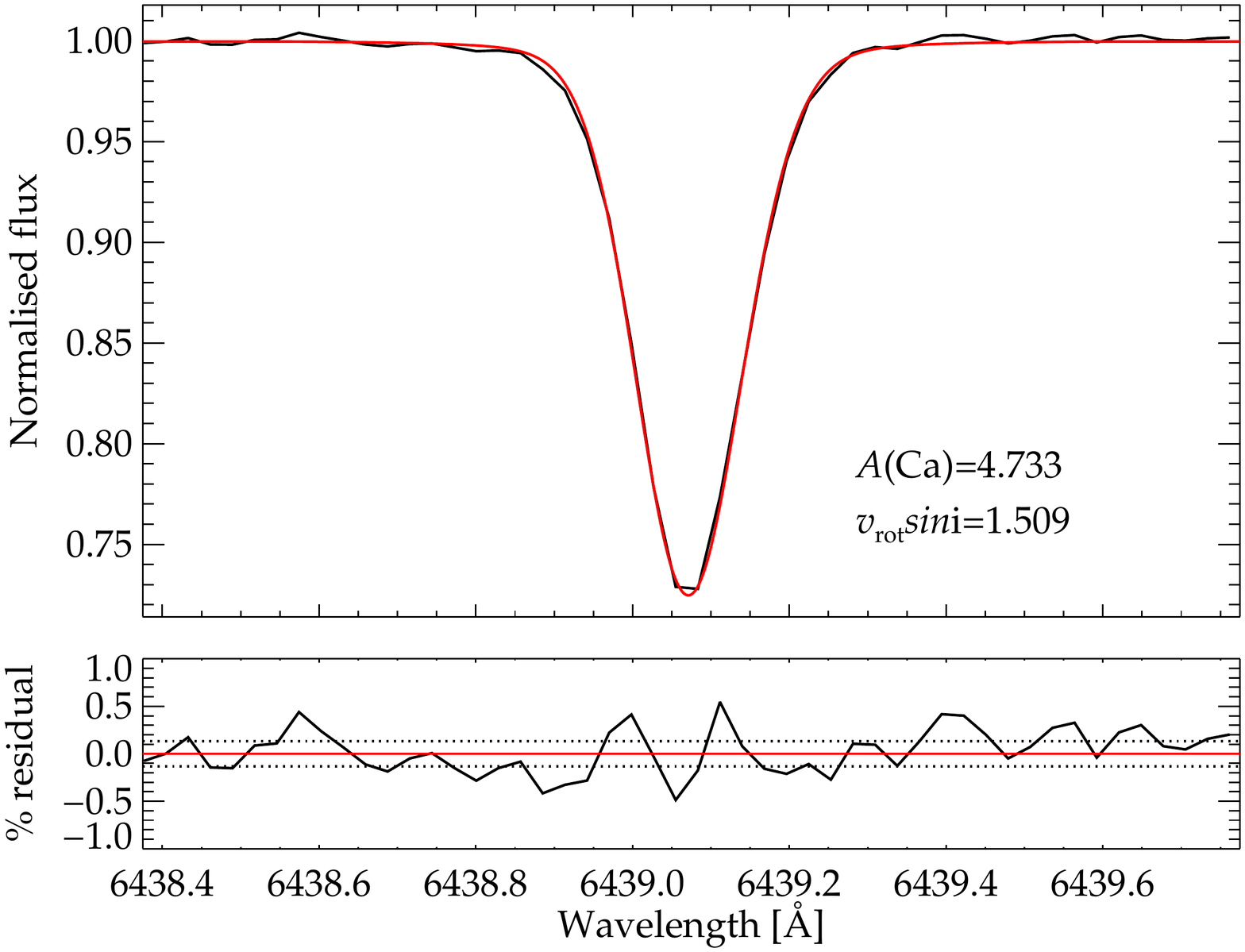}
\end{minipage}
\begin{minipage}[b]{0.33\linewidth}
\centering
\includegraphics[scale=0.23,viewport=5cm 0cm 25cm 21cm]{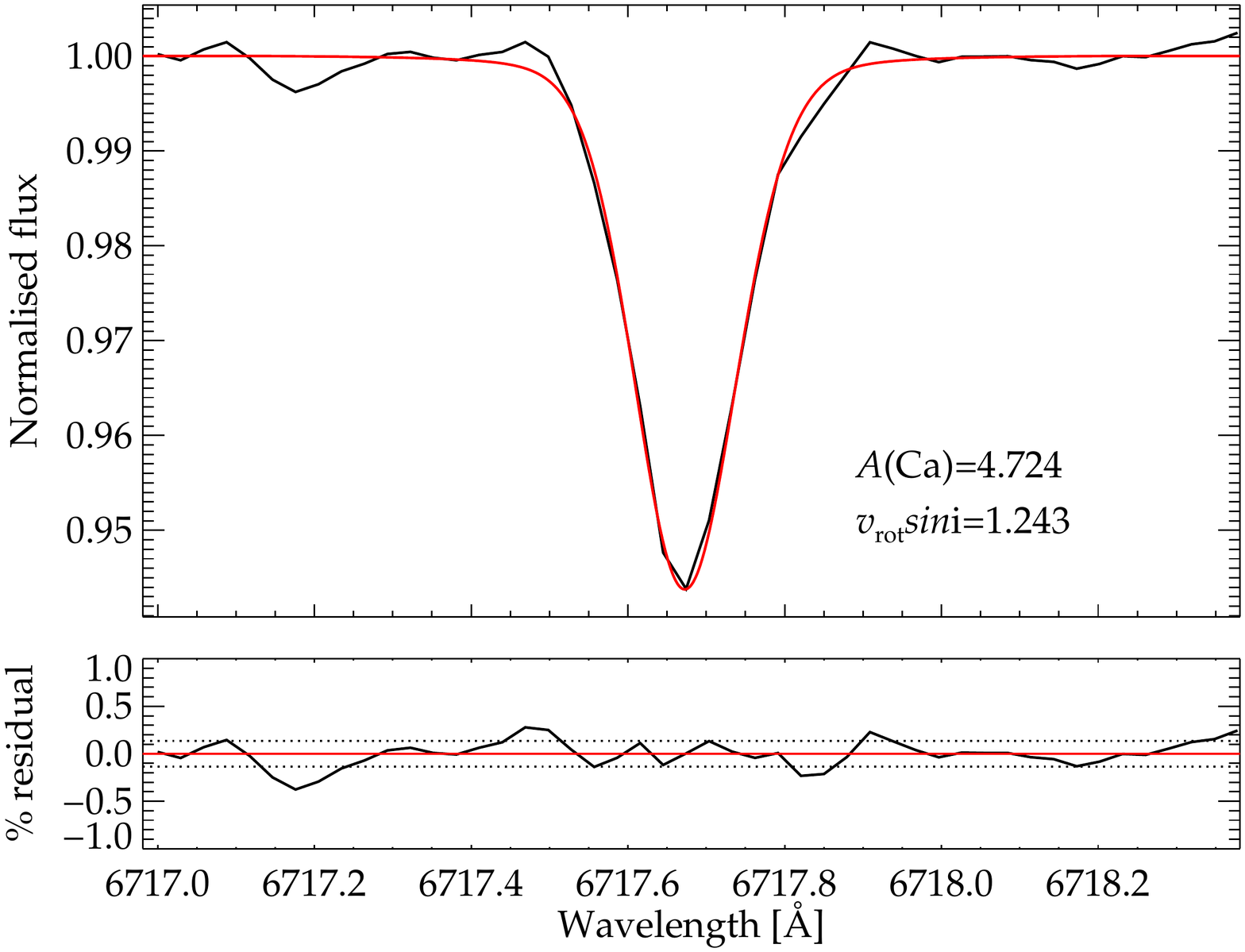}
\end{minipage}
\begin{minipage}[b]{0.33\linewidth}
\centering
\includegraphics[scale=0.23,viewport=5cm 0cm 25cm 21cm]{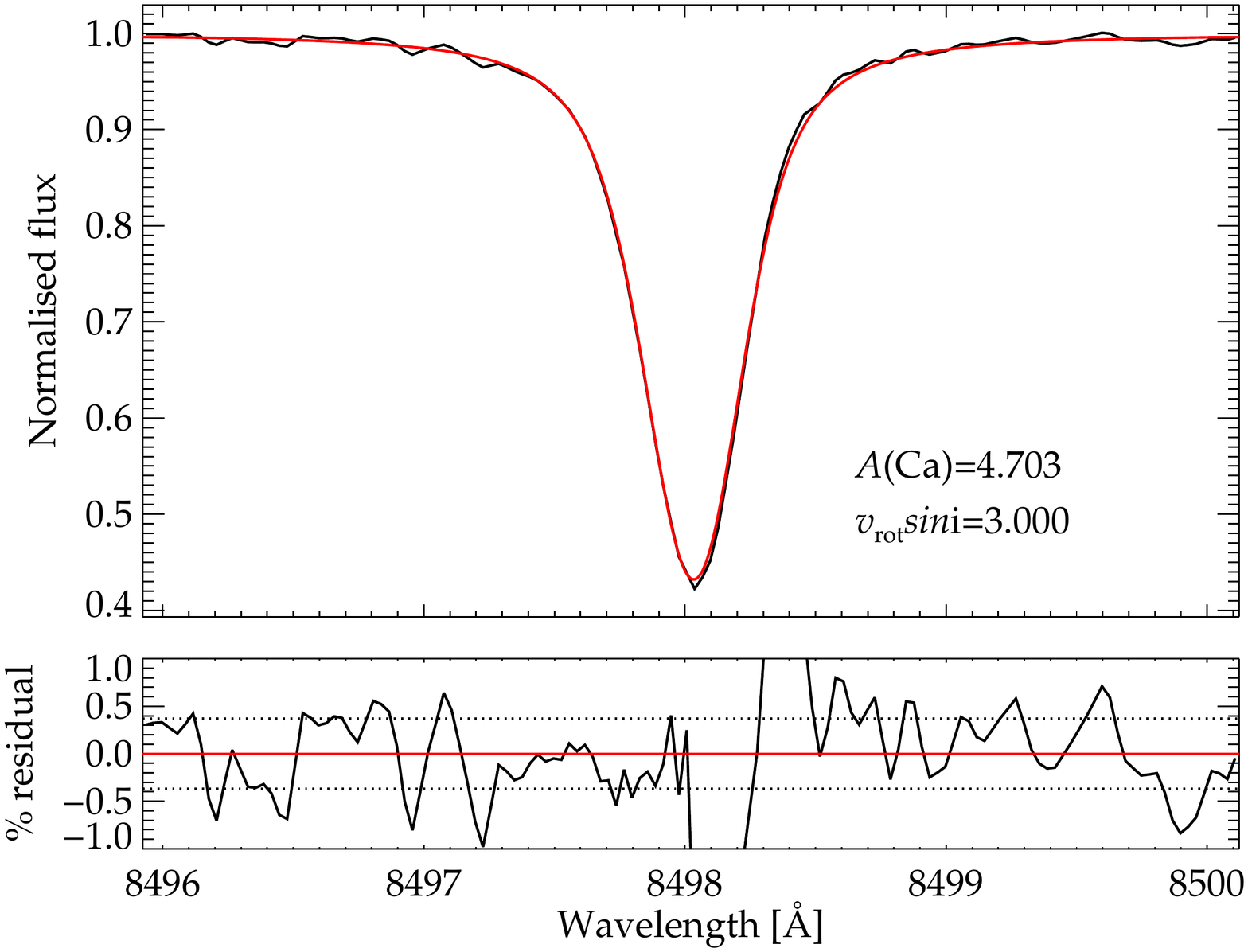}
\end{minipage}
\begin{minipage}[b]{0.33\linewidth}
\centering
\includegraphics[scale=0.23,viewport=5cm 0cm 25cm 21cm]{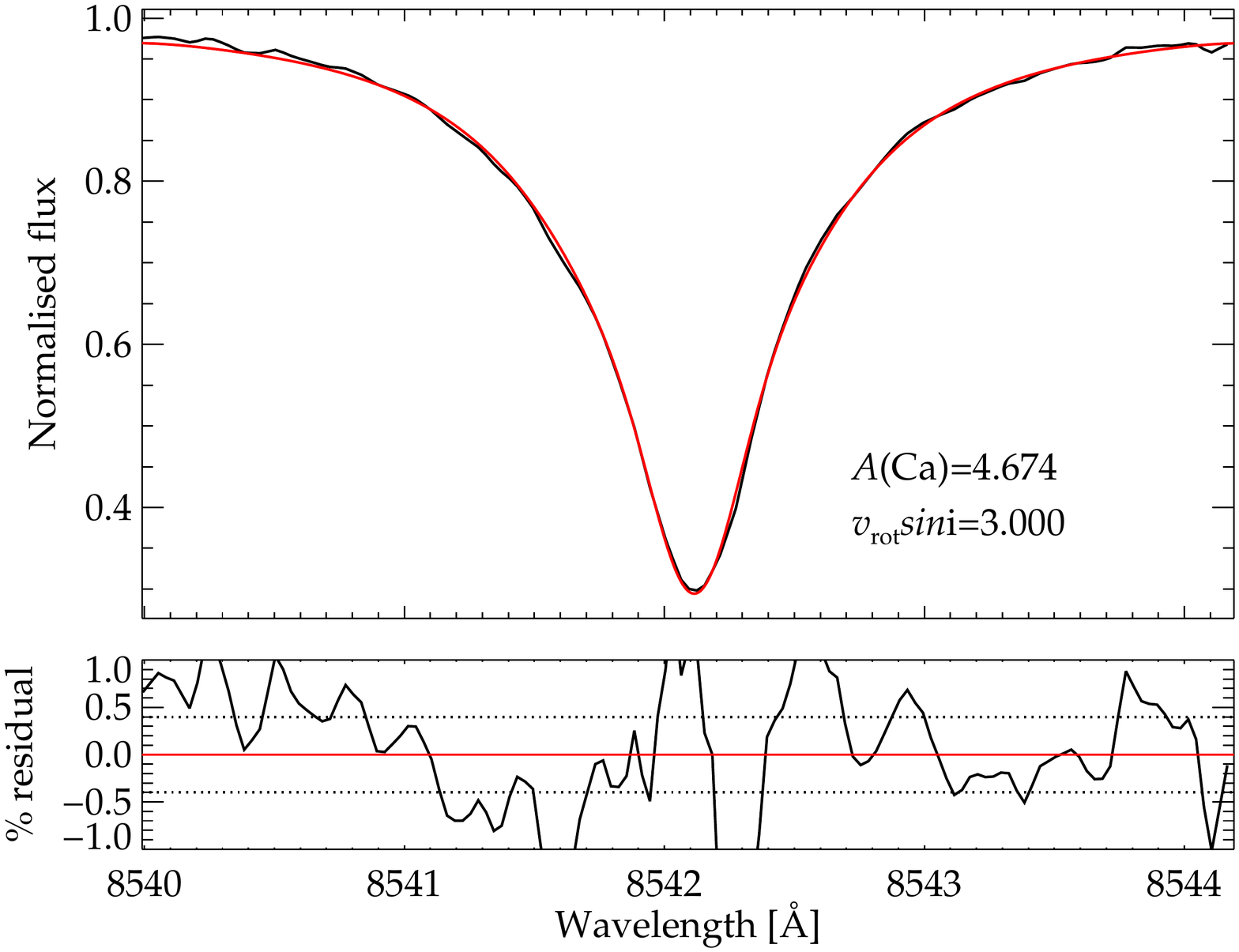}
\end{minipage}
\begin{minipage}[b]{0.33\linewidth}
\centering
\includegraphics[scale=0.23,viewport=5cm 0cm 25cm 21cm]{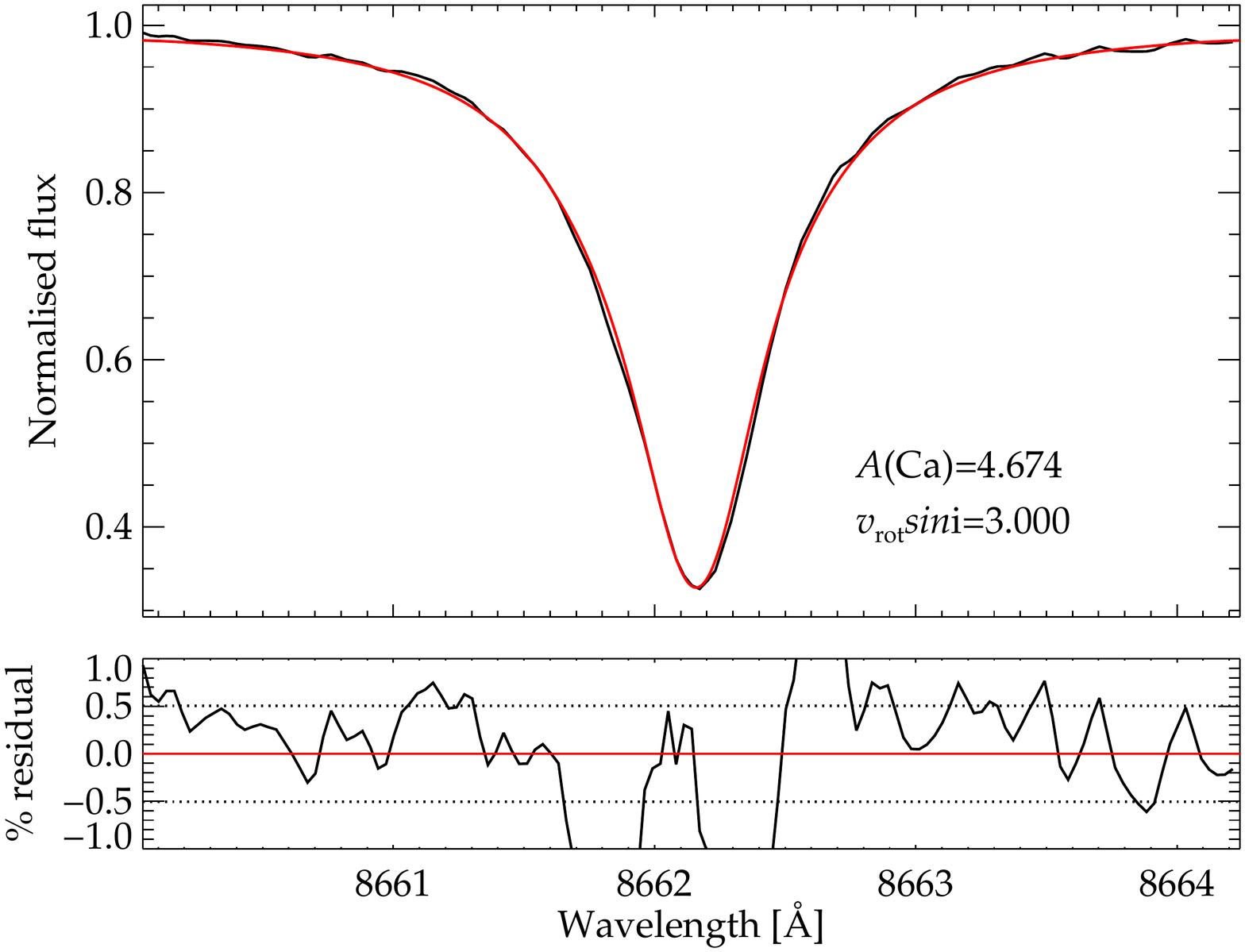}
\end{minipage}
\caption{Observed line profiles \textit{(black solid lines)} and best-fit 3D, NLTE synthetic profiles \textit{(red solid lines)} for HD19445. Below each panel is shown the flux residual (observed-synthetic) in per cent. When $v_{\rm rot}\sin{i}$ has not been solved for, the value has been fixed to $3.0$. Only lines with $W_{\lambda}<50$m\AA\ have been included to estimate the $v_{\rm rot}\sin{i}$-values used for analysis of Li line to determine the Li isotopic ratio.}
\label{fig:hd19445}
\end{figure*}

\begin{figure*}[htb]
\begin{minipage}[b]{0.33\linewidth}
\centering
\includegraphics[scale=0.23,viewport=5cm 0cm 25cm 21cm]{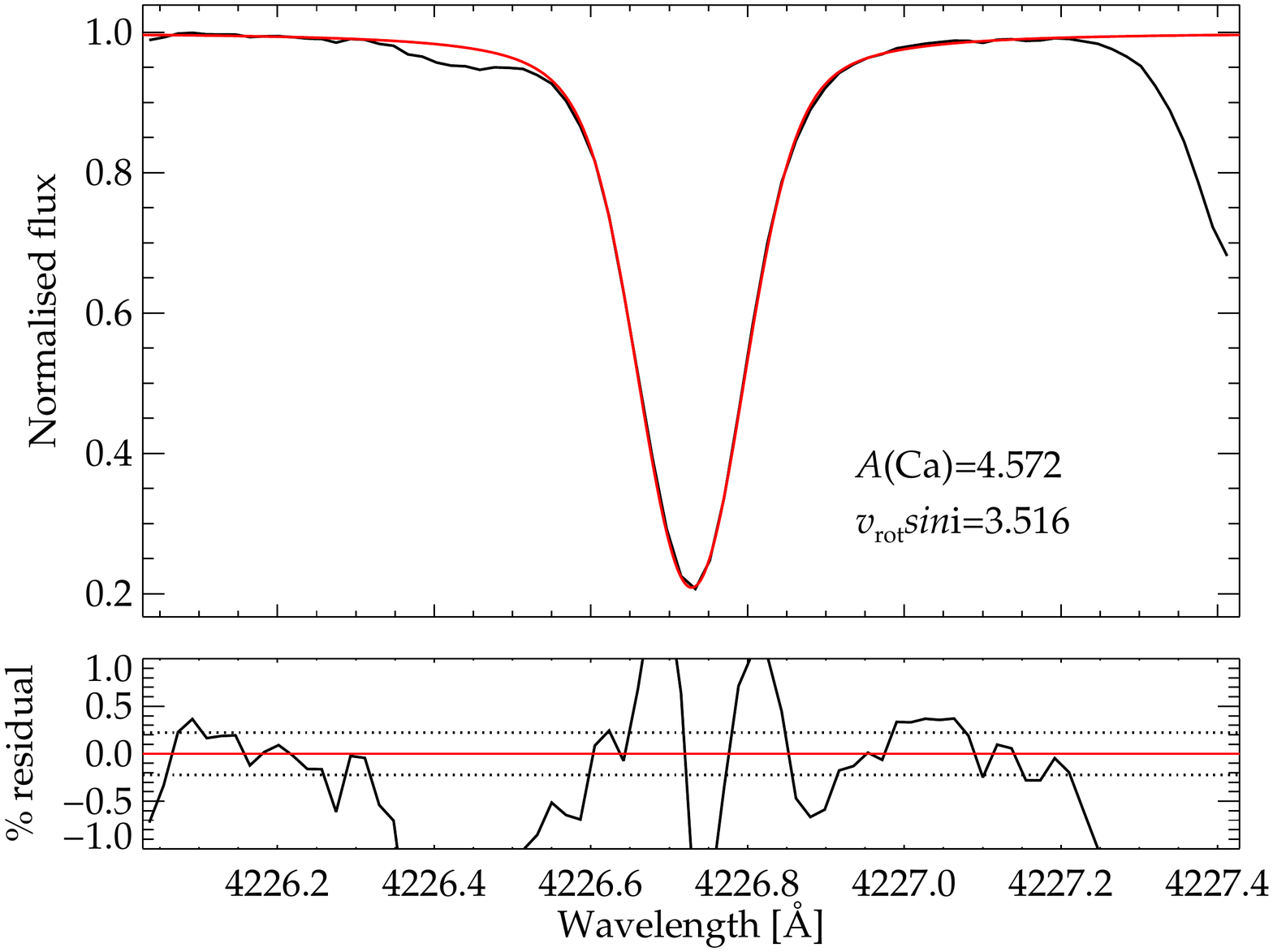}
\end{minipage}
\begin{minipage}[b]{0.33\linewidth}
\centering
\includegraphics[scale=0.23,viewport=5cm 0cm 25cm 21cm]{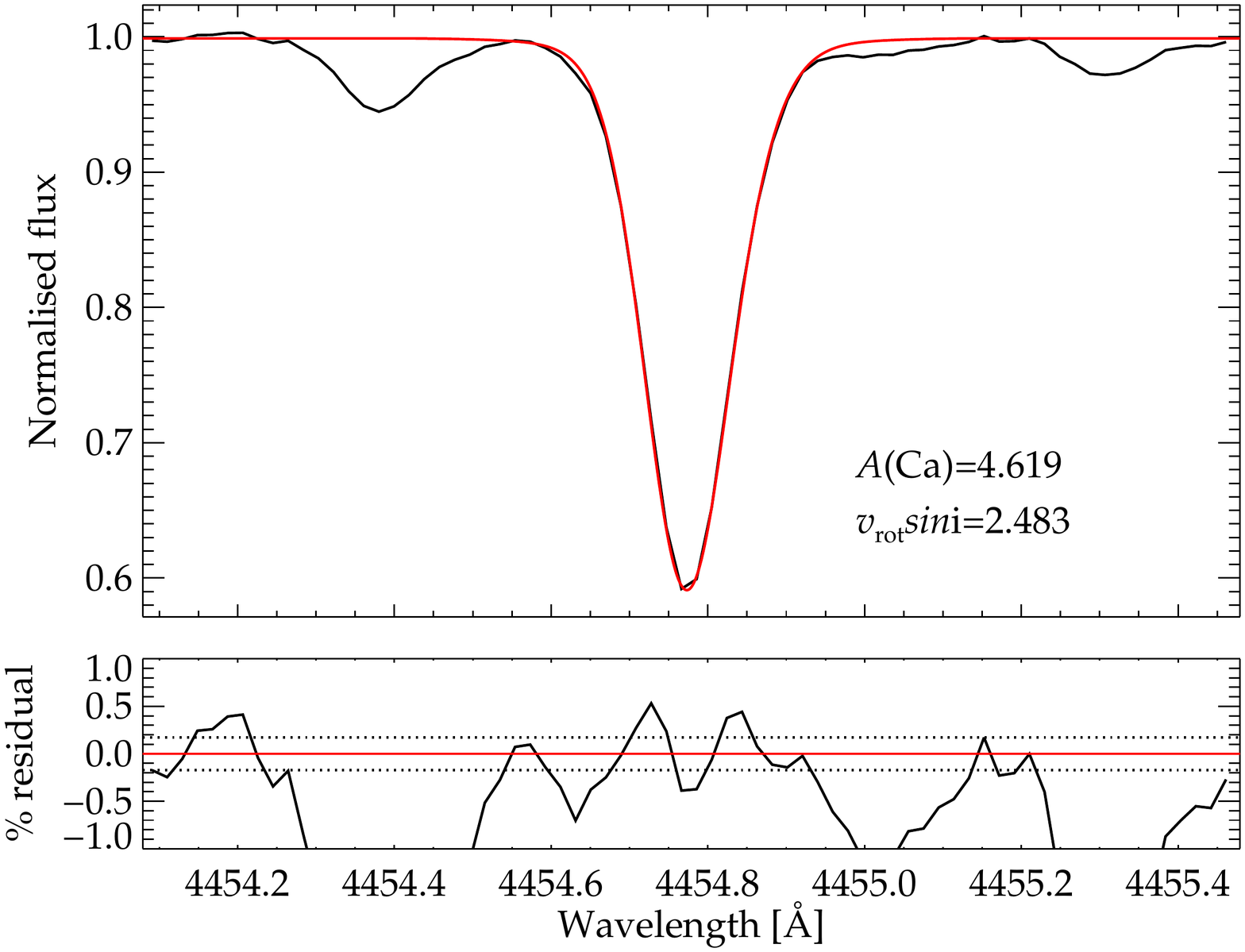}
\end{minipage}
\begin{minipage}[b]{0.33\linewidth}
\centering
\includegraphics[scale=0.23,viewport=5cm 0cm 25cm 21cm]{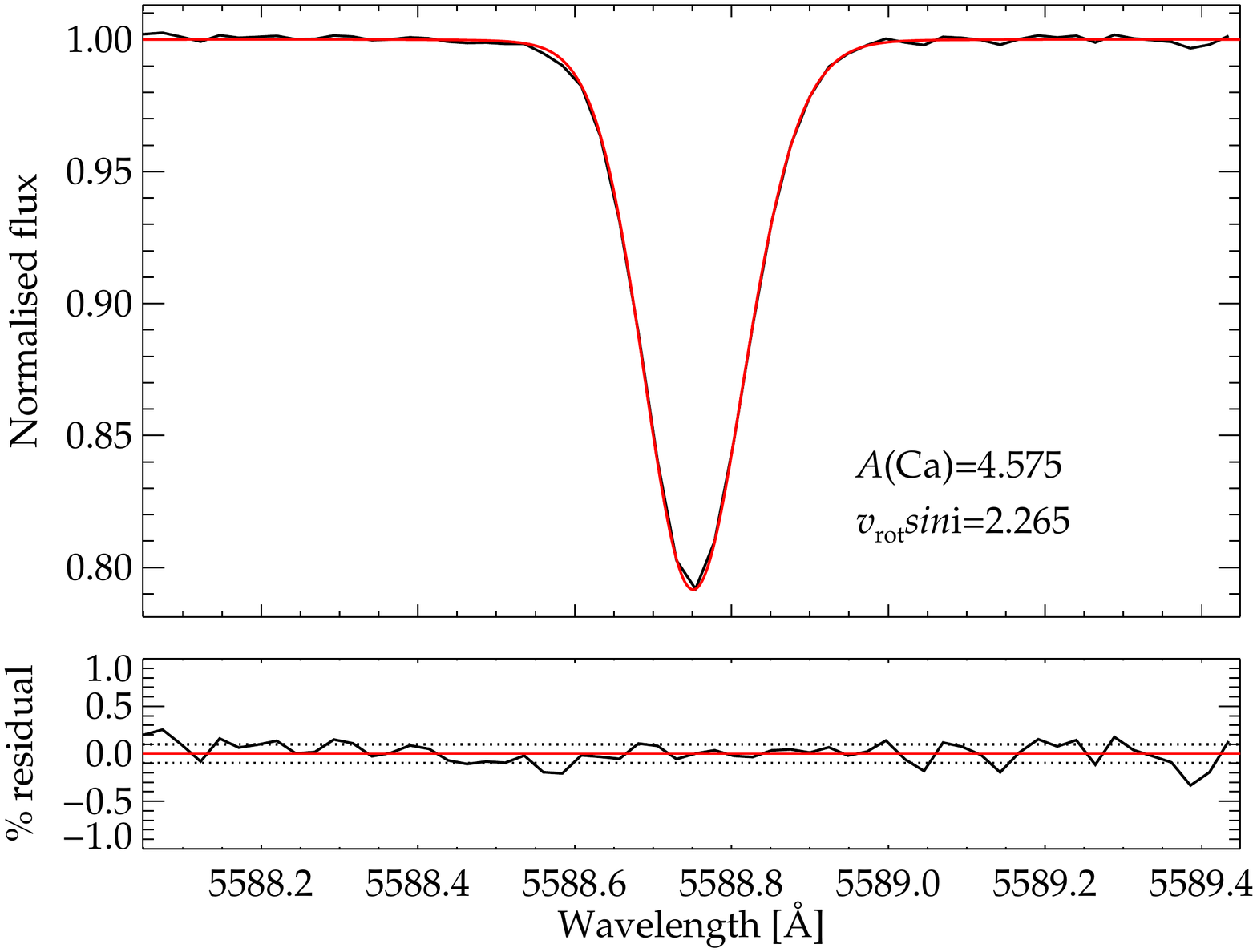}
\end{minipage}
\begin{minipage}[b]{0.33\linewidth}
\centering
\includegraphics[scale=0.23,viewport=5cm 0cm 25cm 21cm]{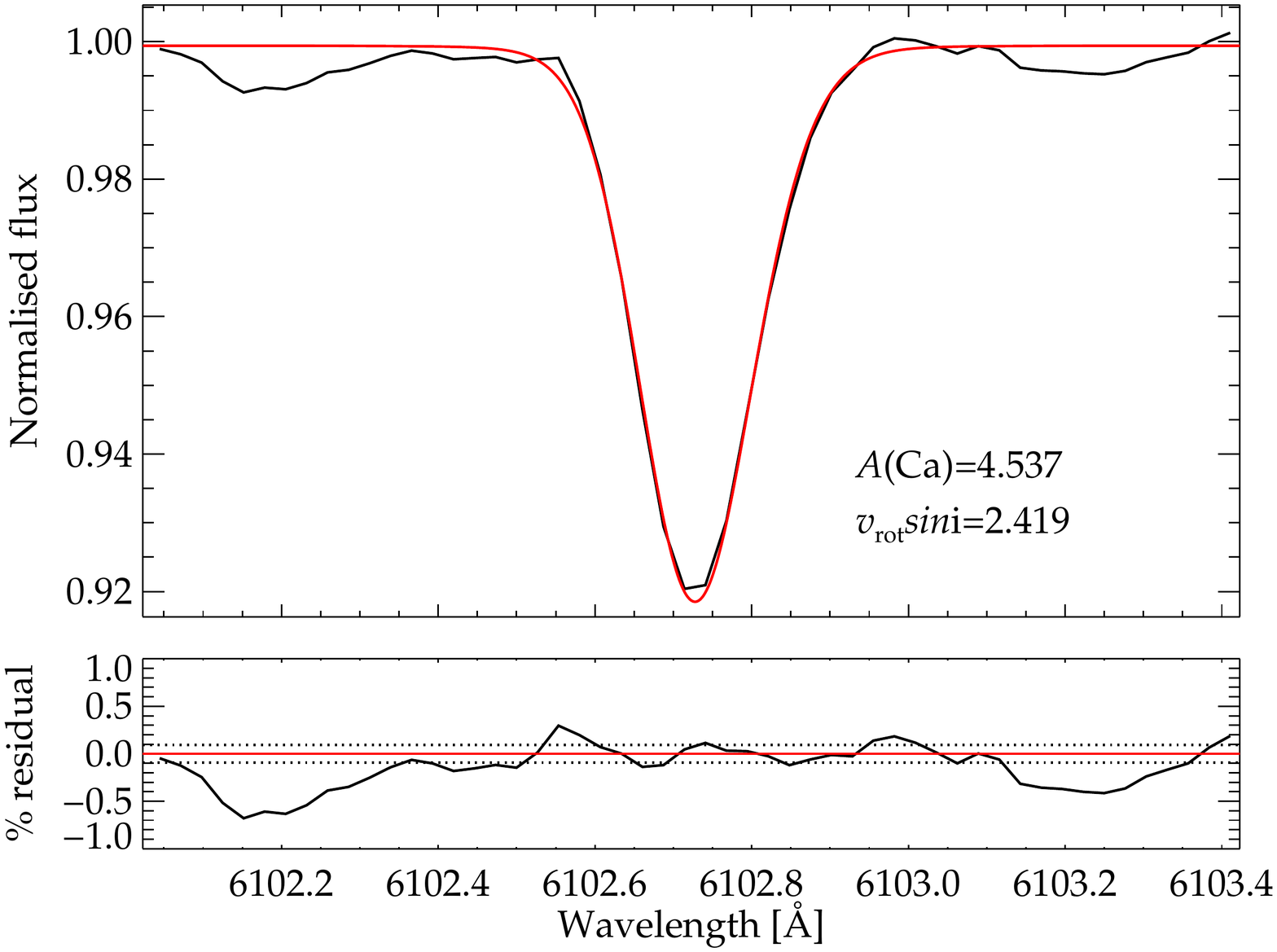}
\end{minipage}
\begin{minipage}[b]{0.33\linewidth}
\centering
\includegraphics[scale=0.23,viewport=5cm 0cm 25cm 21cm]{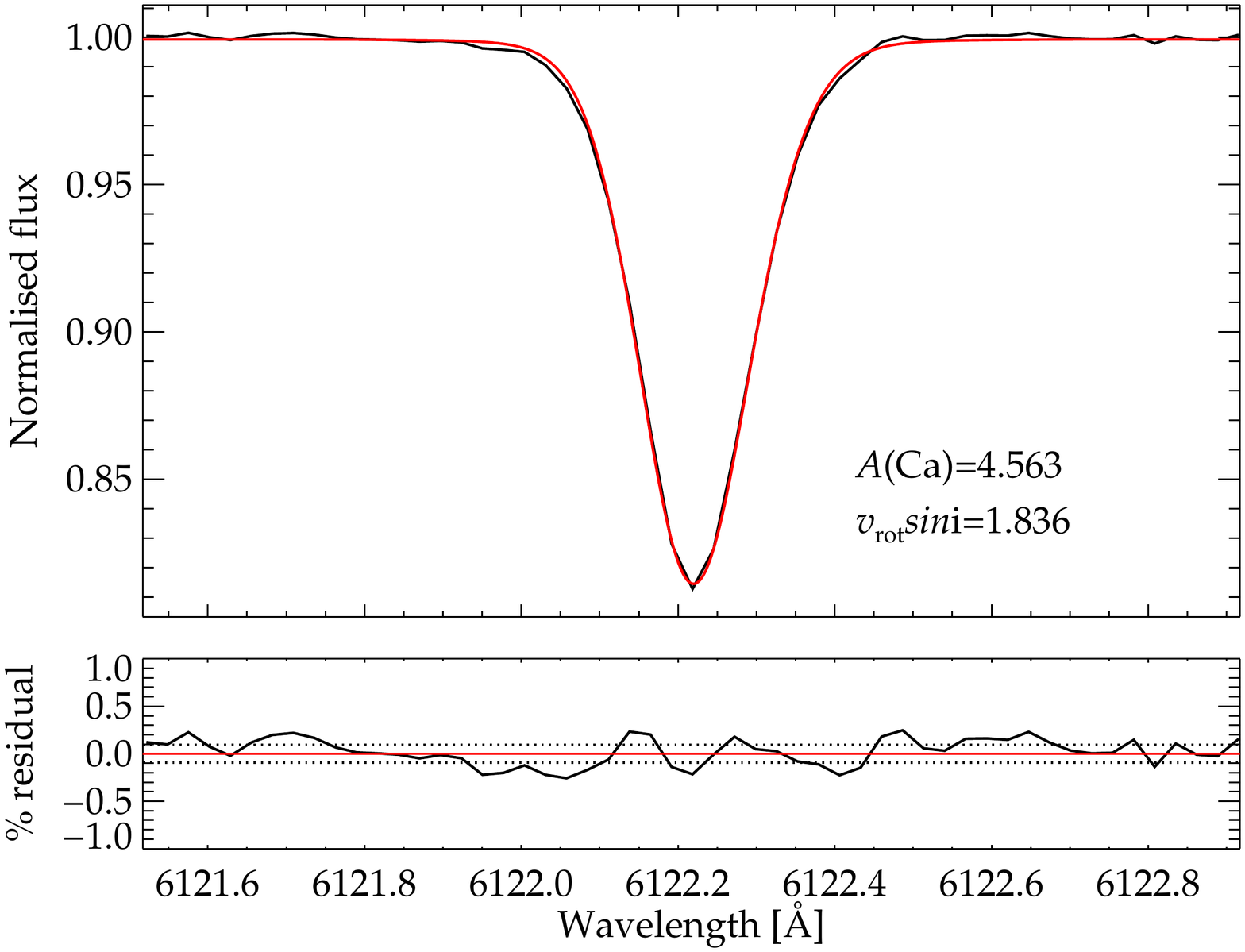}
\end{minipage}
\begin{minipage}[b]{0.33\linewidth}
\centering
\includegraphics[scale=0.23,viewport=5cm 0cm 25cm 21cm]{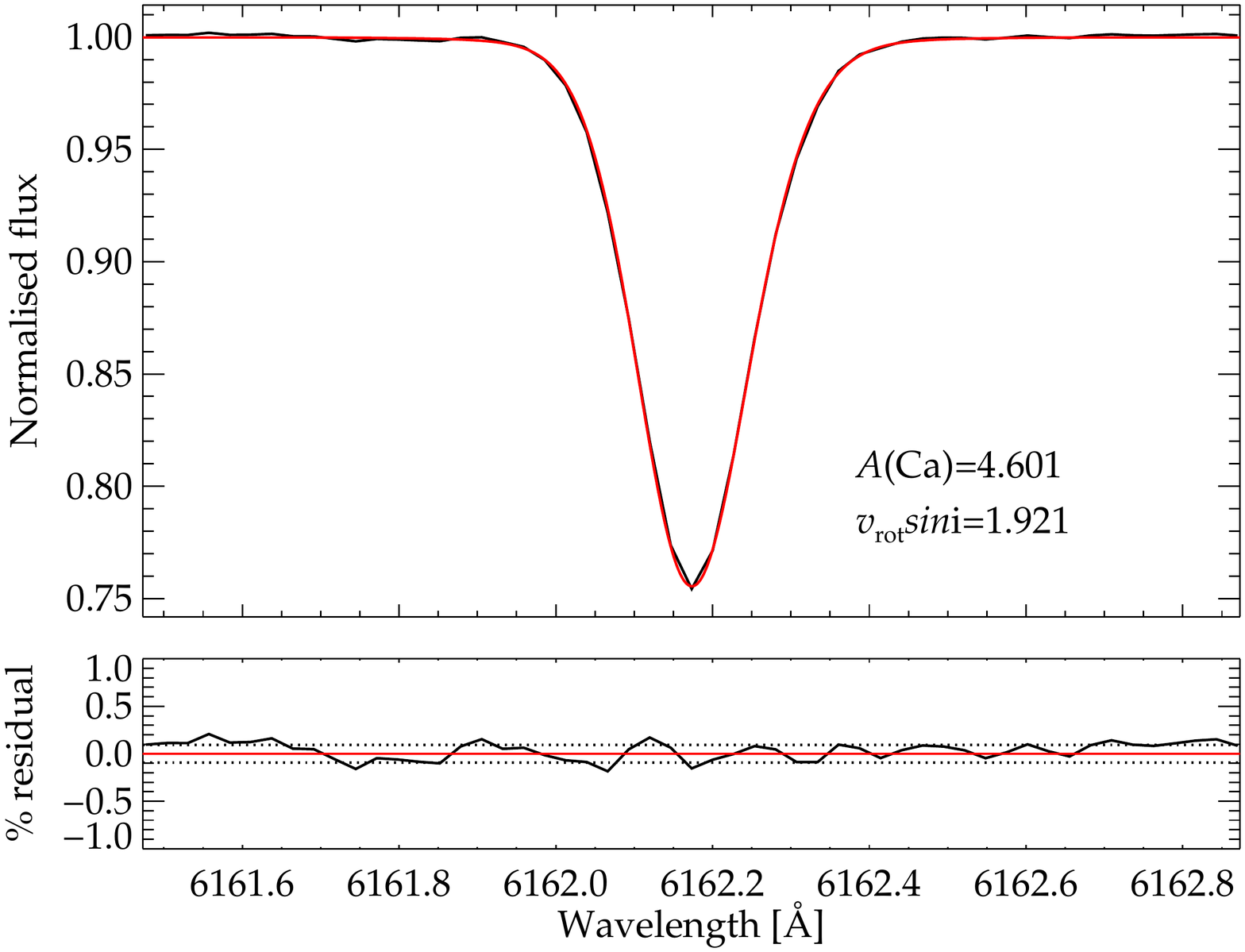}
\end{minipage}
\begin{minipage}[b]{0.33\linewidth}
\centering
\includegraphics[scale=0.23,viewport=5cm 0cm 25cm 21cm]{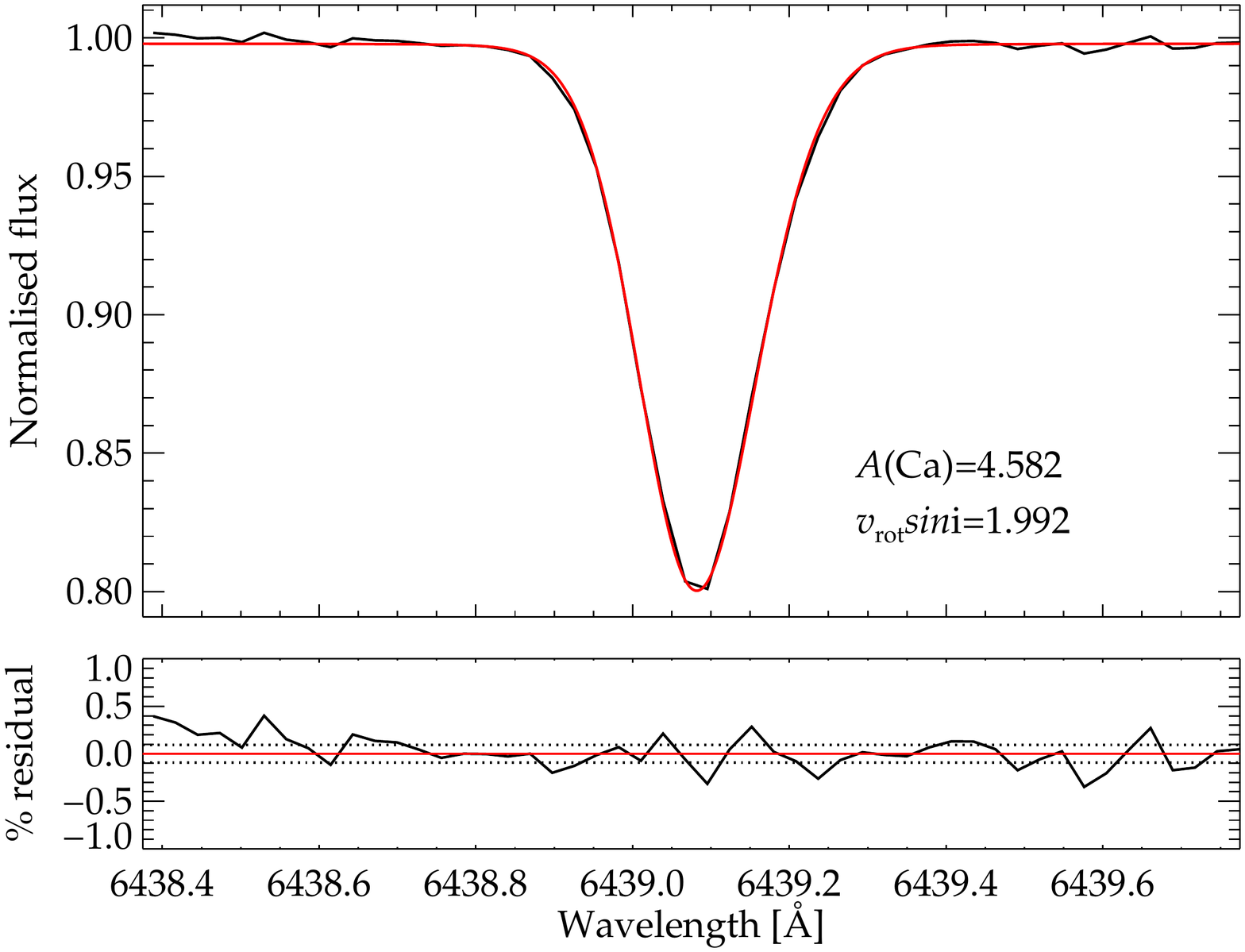}
\end{minipage}
\begin{minipage}[b]{0.33\linewidth}
\centering
\includegraphics[scale=0.23,viewport=5cm 0cm 25cm 21cm]{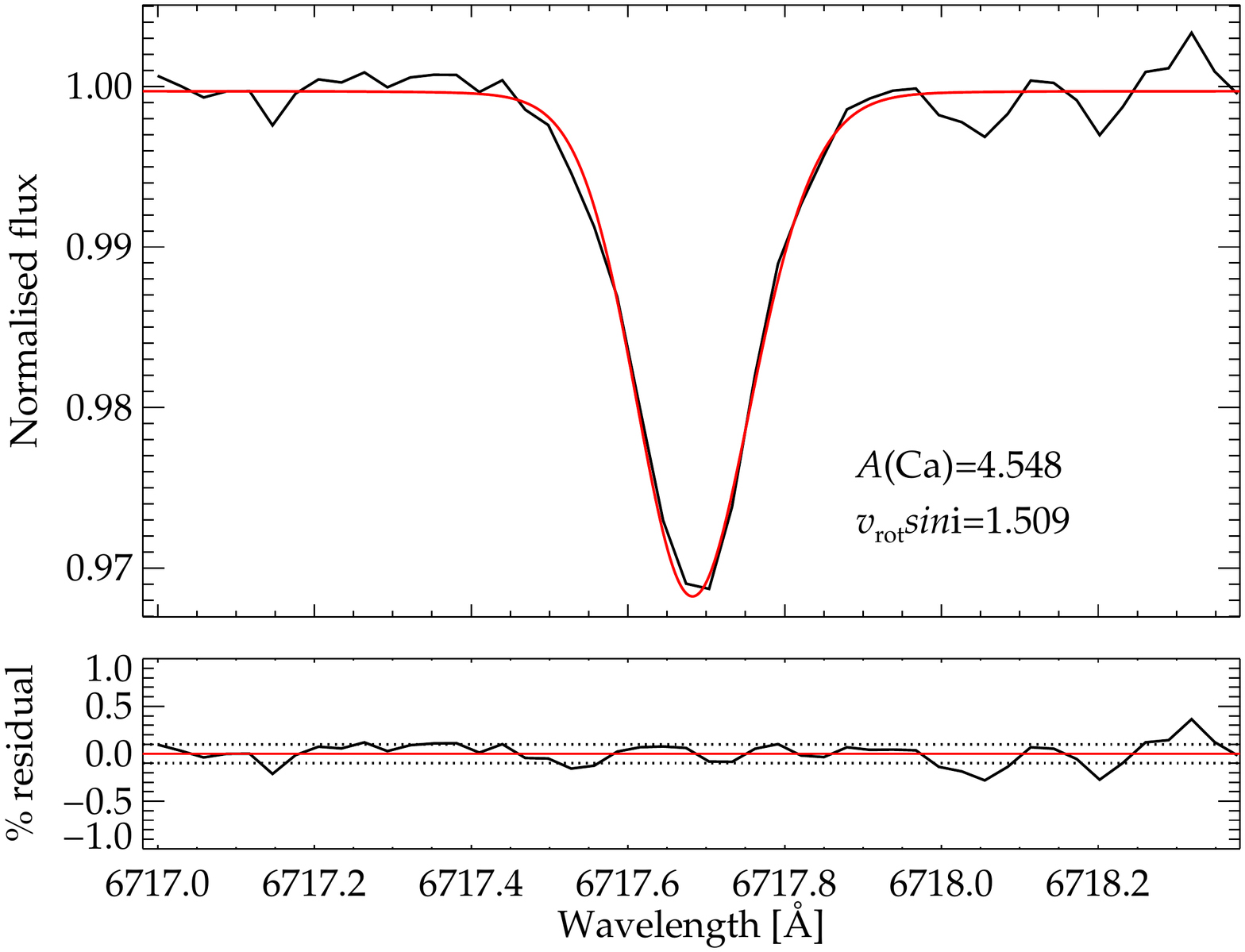}
\end{minipage}
\begin{minipage}[b]{0.33\linewidth}
\centering
\includegraphics[scale=0.23,viewport=5cm 0cm 25cm 21cm]{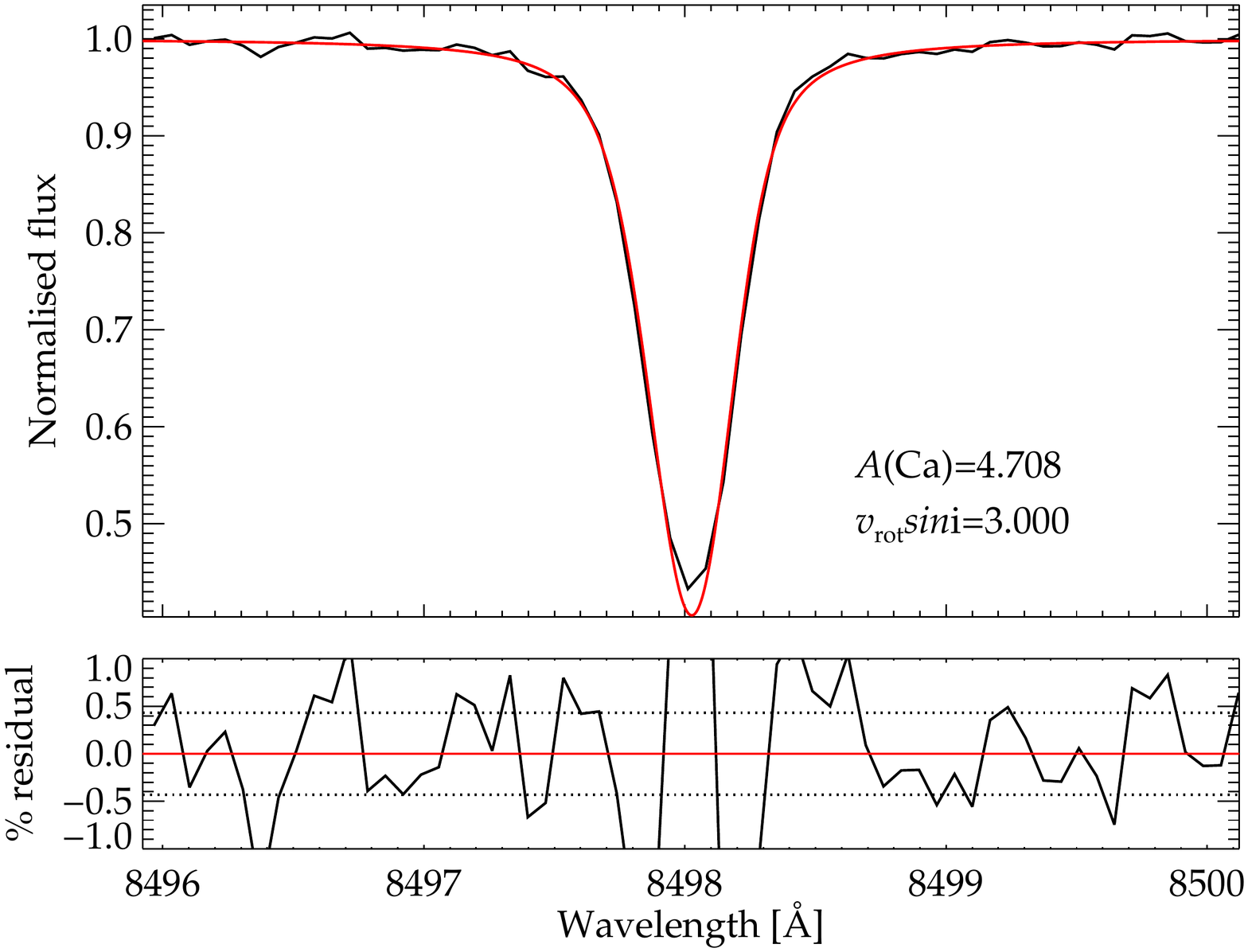}
\end{minipage}
\begin{minipage}[b]{0.33\linewidth}
\centering
\includegraphics[scale=0.23,viewport=5cm 0cm 25cm 21cm]{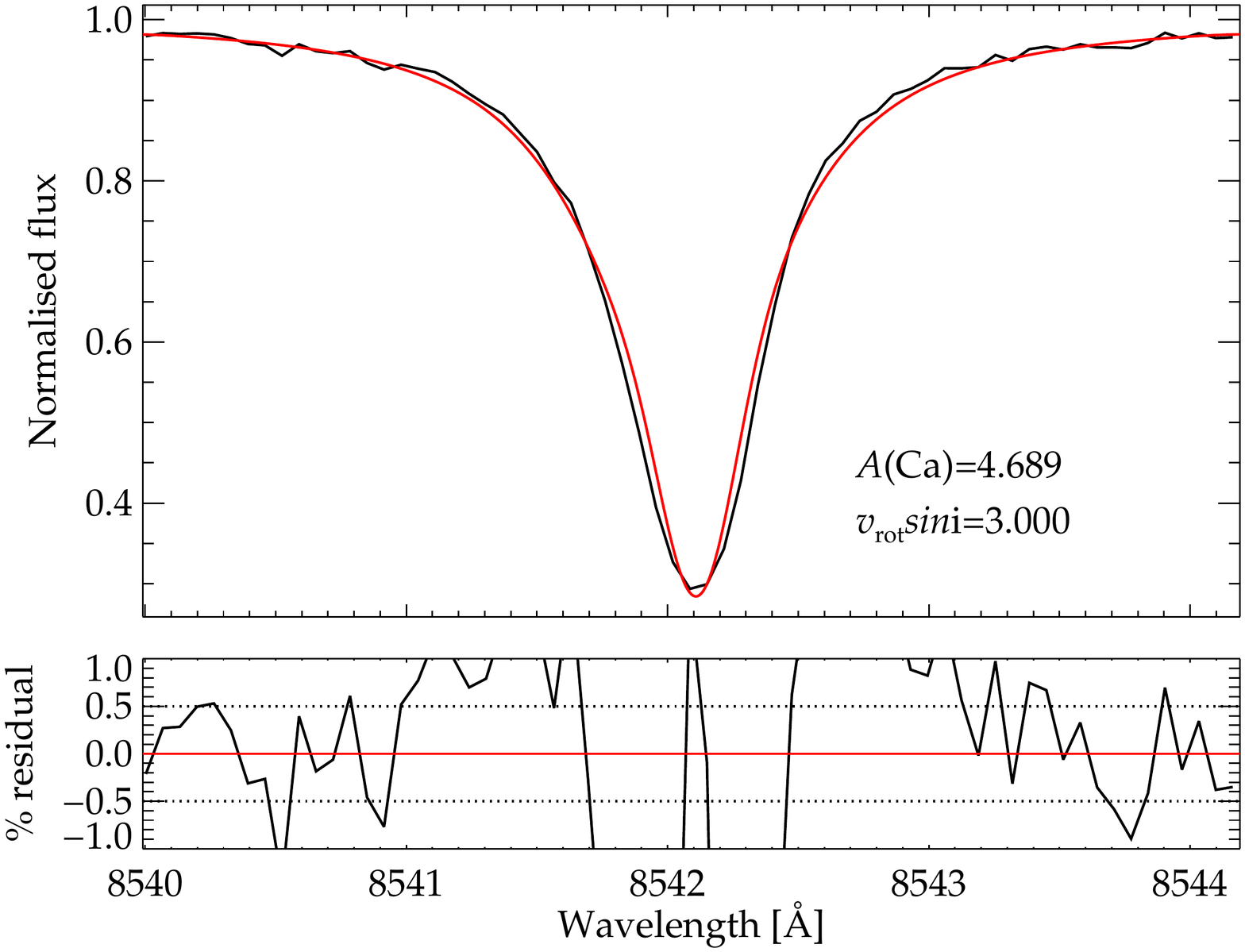}
\end{minipage}
\begin{minipage}[b]{0.33\linewidth}
\centering
\includegraphics[scale=0.23,viewport=5cm 0cm 25cm 21cm]{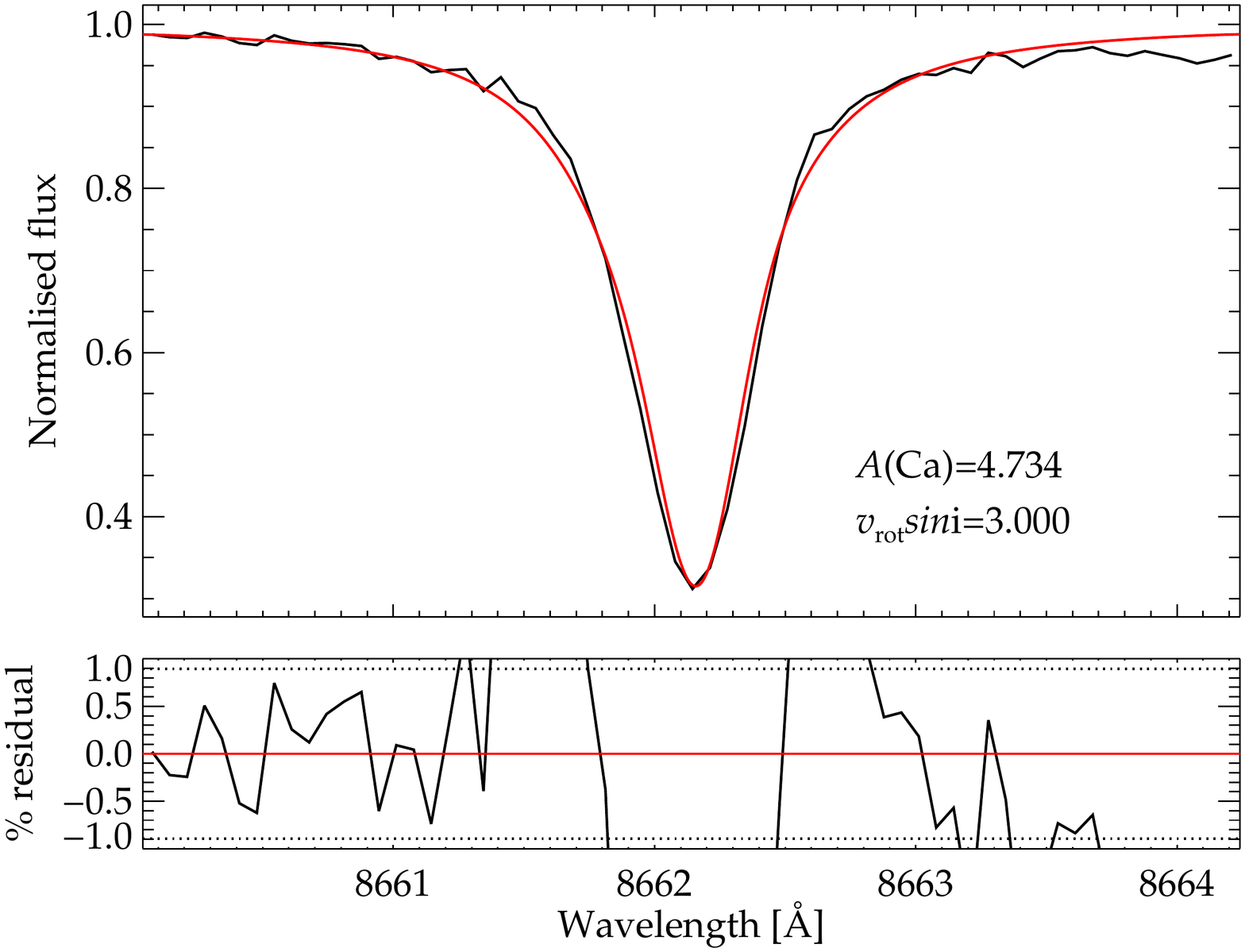}
\end{minipage}
\caption{Same as Fig.\,\ref{fig:hd19445}, but for HD84937.}
\label{fig:hd84937}
\end{figure*}

\begin{figure*}[htb]
\begin{minipage}[b]{0.33\linewidth}
\centering
\includegraphics[scale=0.23,viewport=5cm 0cm 25cm 21cm]{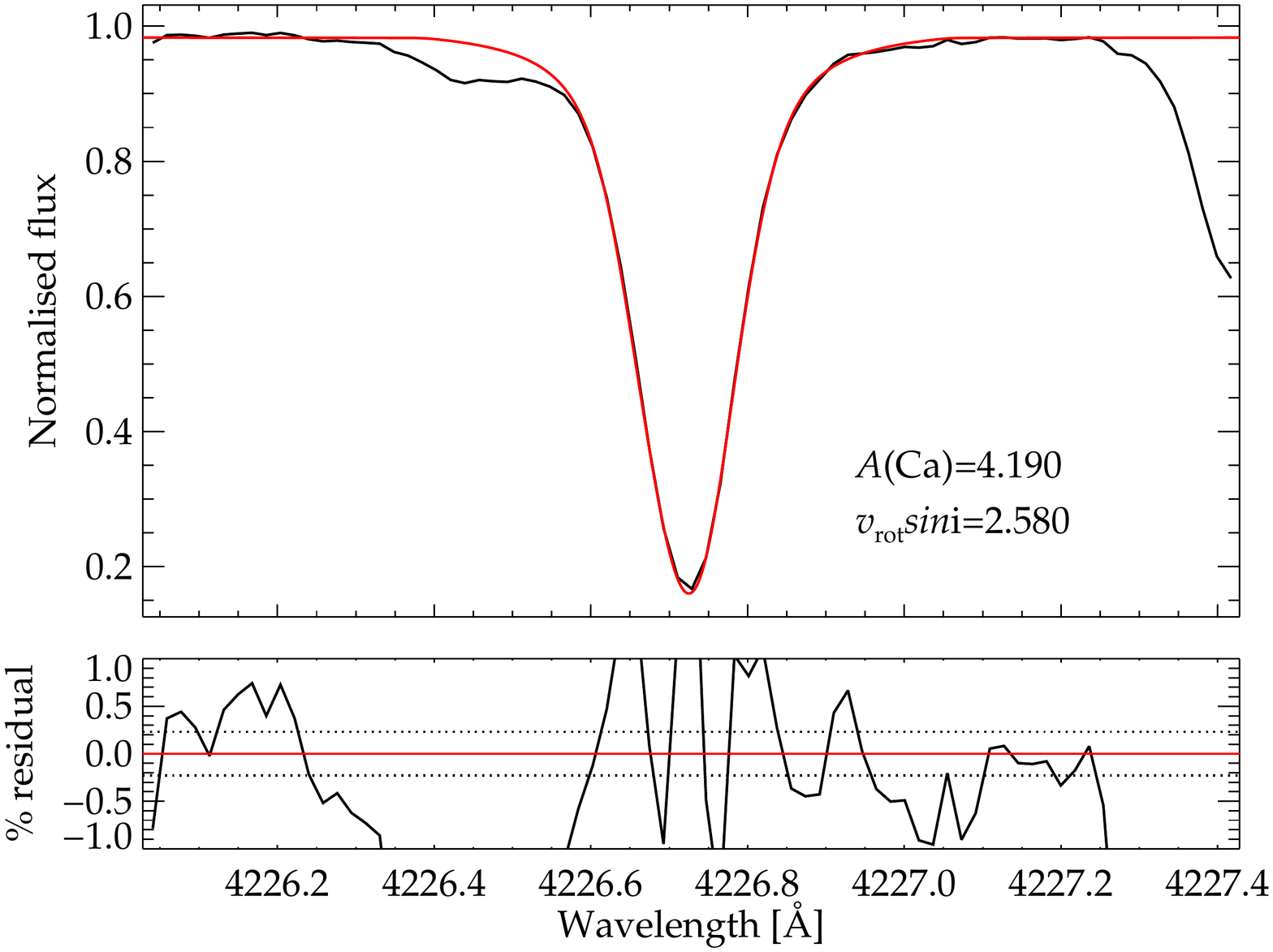}
\end{minipage}
\begin{minipage}[b]{0.33\linewidth}
\centering
\includegraphics[scale=0.23,viewport=5cm 0cm 25cm 21cm]{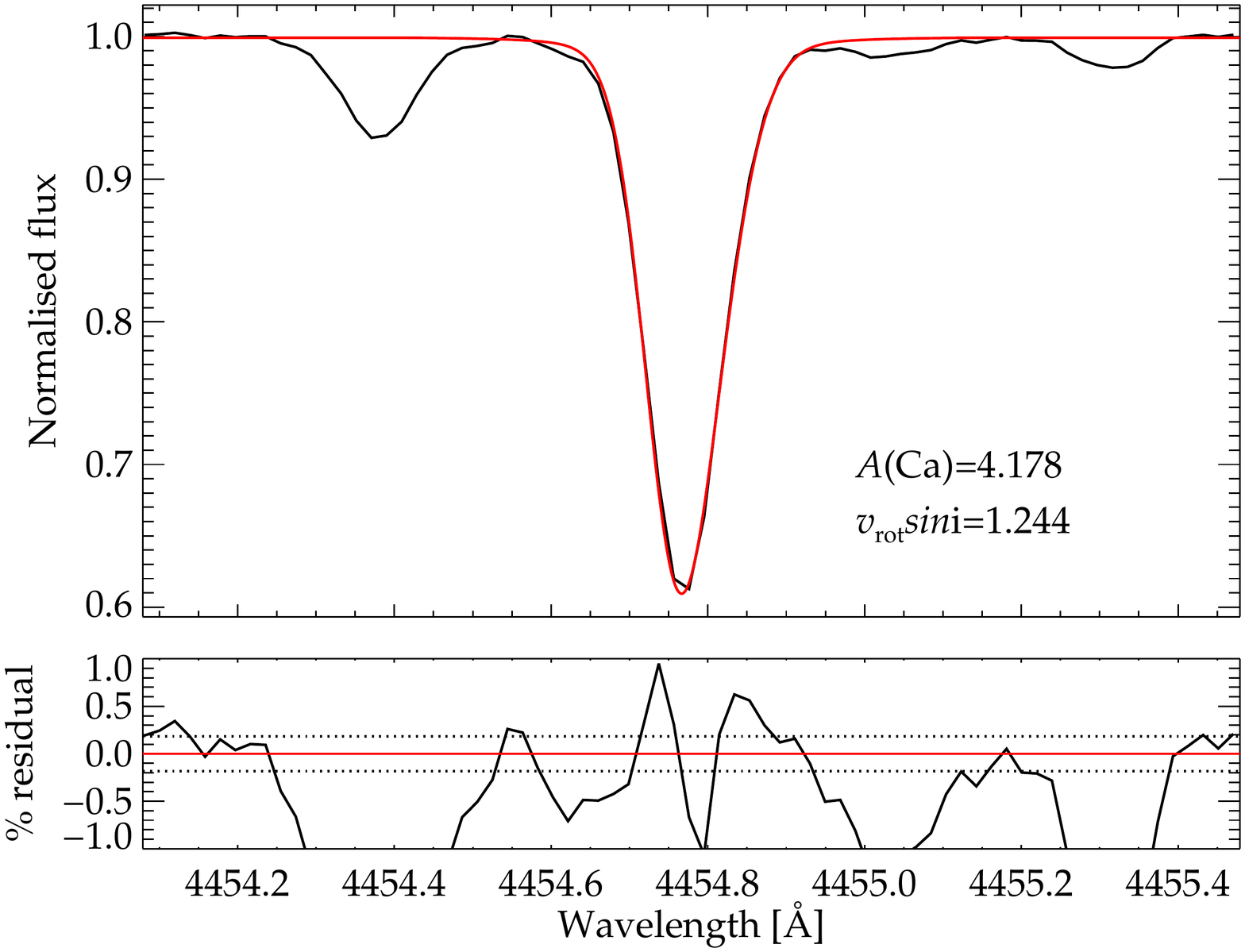}
\end{minipage}
\begin{minipage}[b]{0.33\linewidth}
\centering
\includegraphics[scale=0.23,viewport=5cm 0cm 25cm 21cm]{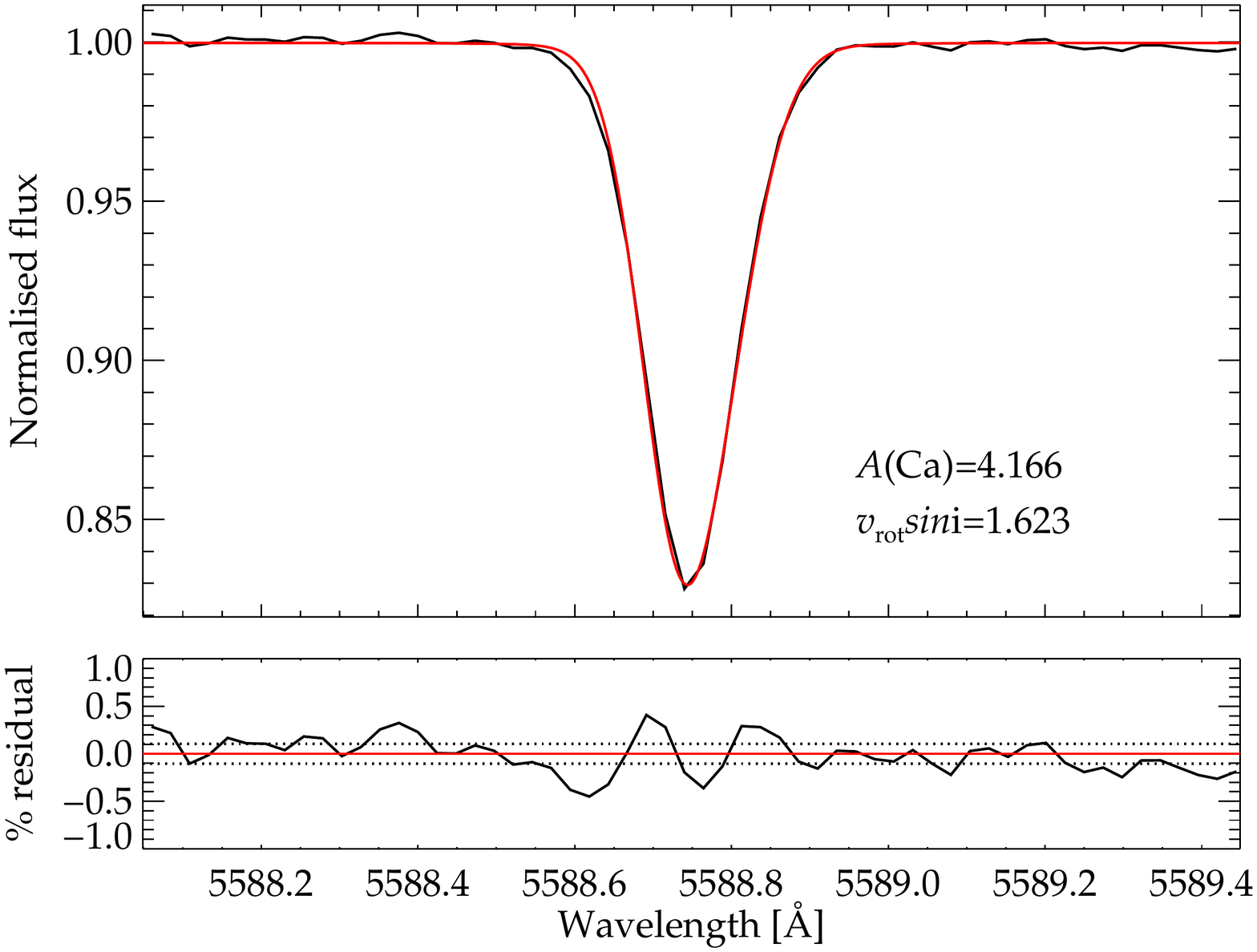}
\end{minipage}
\begin{minipage}[b]{0.33\linewidth}
\centering
\includegraphics[scale=0.23,viewport=5cm 0cm 25cm 21cm]{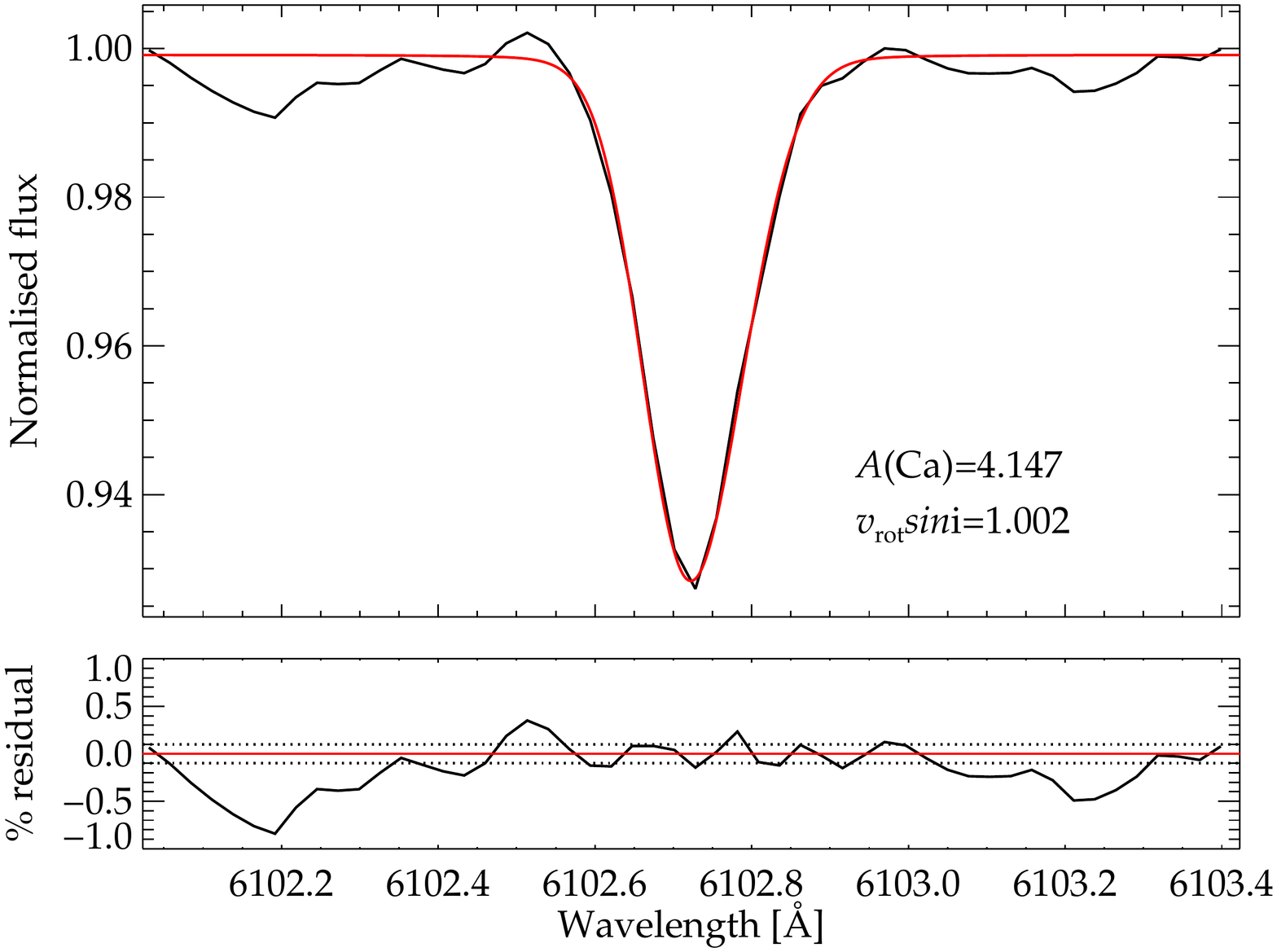}
\end{minipage}
\begin{minipage}[b]{0.33\linewidth}
\centering
\includegraphics[scale=0.23,viewport=5cm 0cm 25cm 21cm]{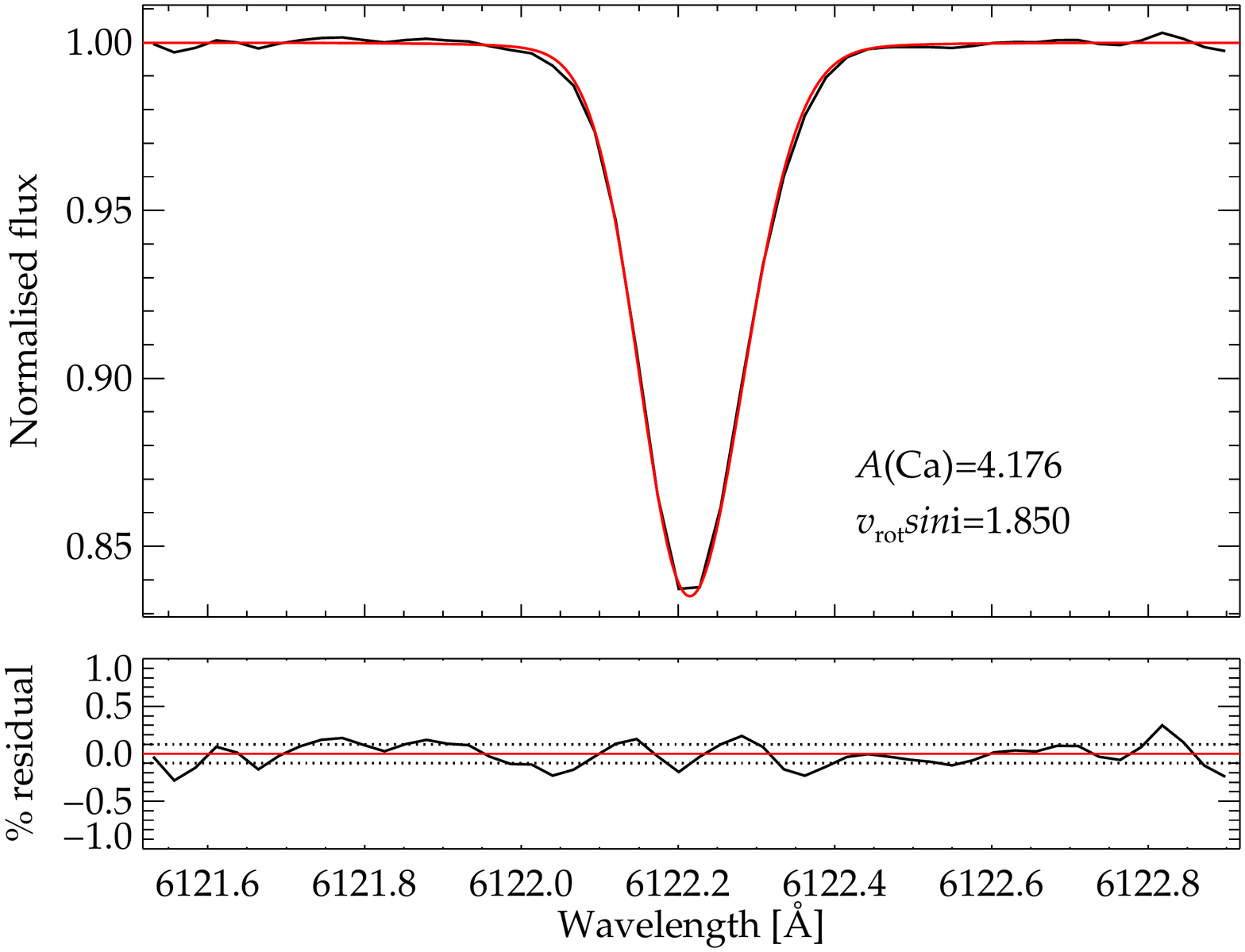}
\end{minipage}
\begin{minipage}[b]{0.33\linewidth}
\centering
\includegraphics[scale=0.23,viewport=5cm 0cm 25cm 21cm]{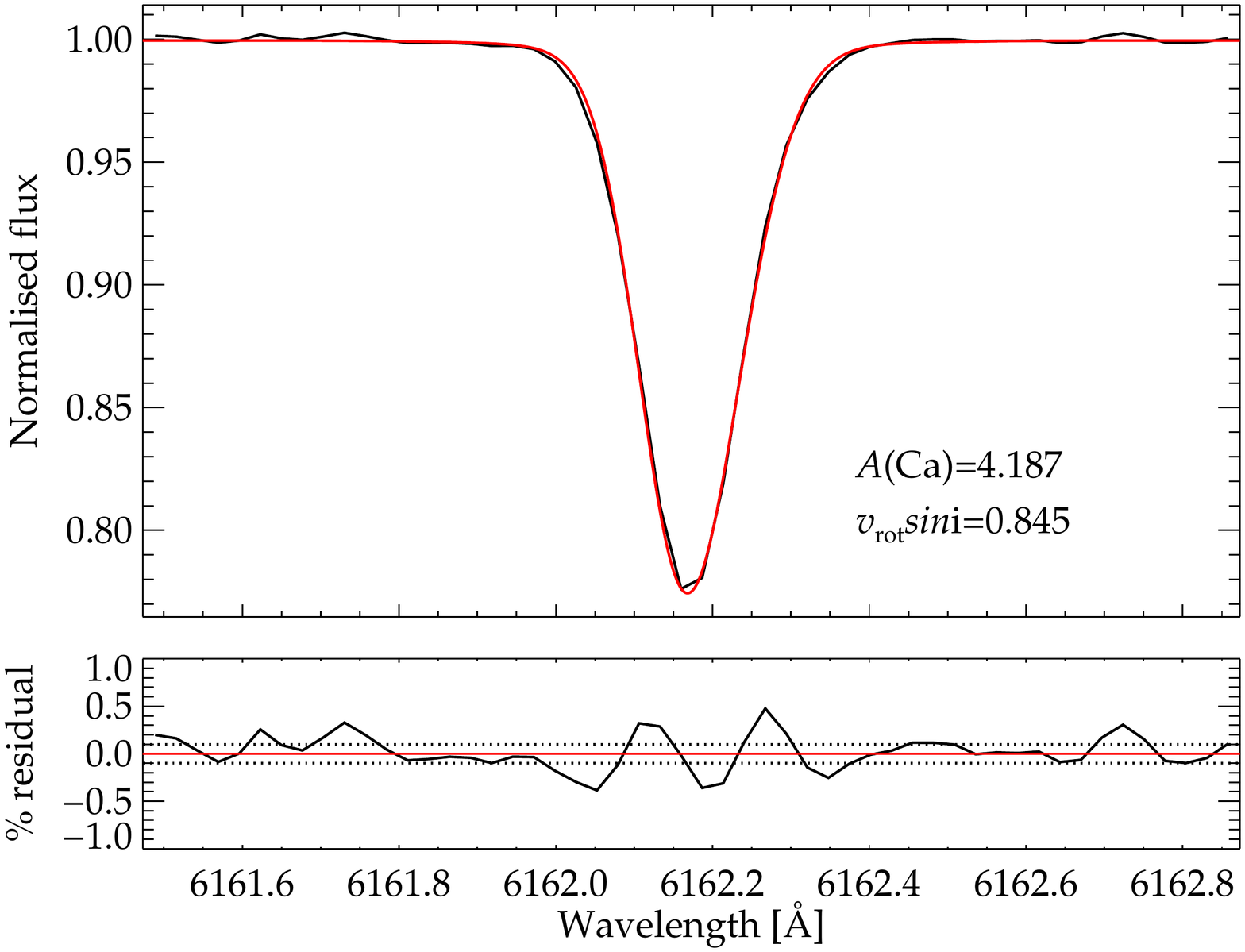}
\end{minipage}
\begin{minipage}[b]{0.33\linewidth}
\centering
\includegraphics[scale=0.23,viewport=5cm 0cm 25cm 21cm]{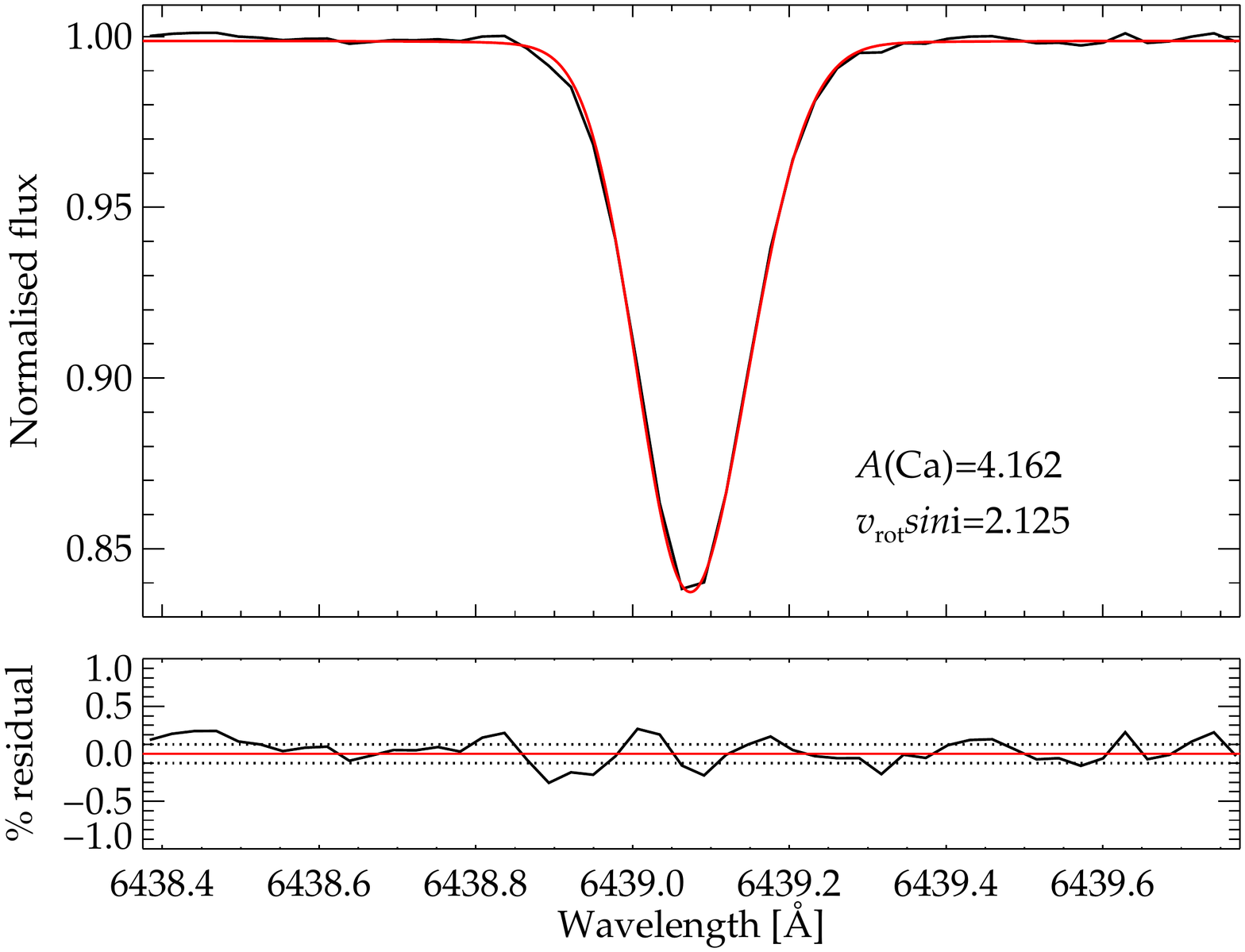}
\end{minipage}
\begin{minipage}[b]{0.33\linewidth}
\centering
\includegraphics[scale=0.23,viewport=5cm 0cm 25cm 21cm]{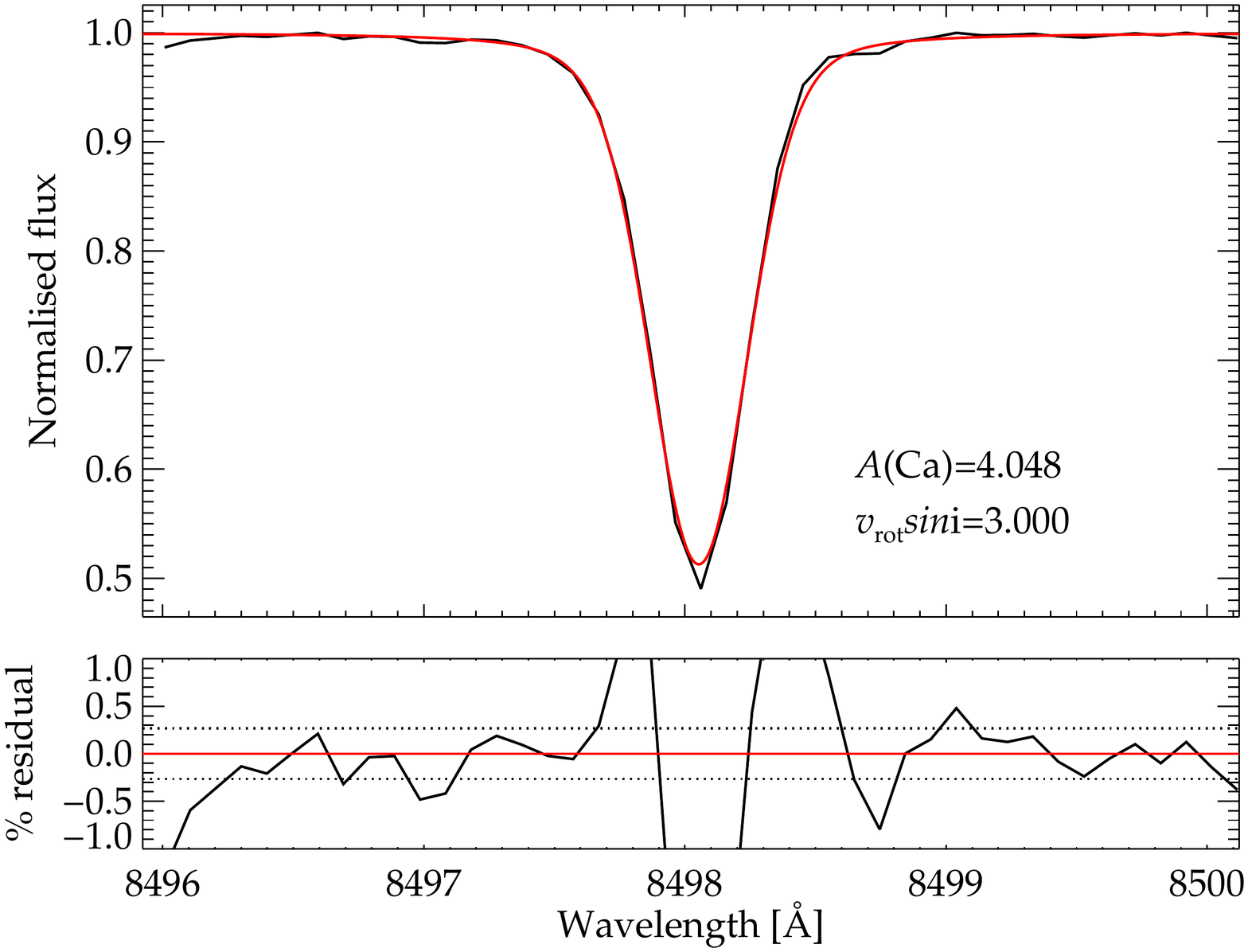}
\end{minipage}
\begin{minipage}[b]{0.33\linewidth}
\centering
\includegraphics[scale=0.23,viewport=5cm 0cm 25cm 21cm]{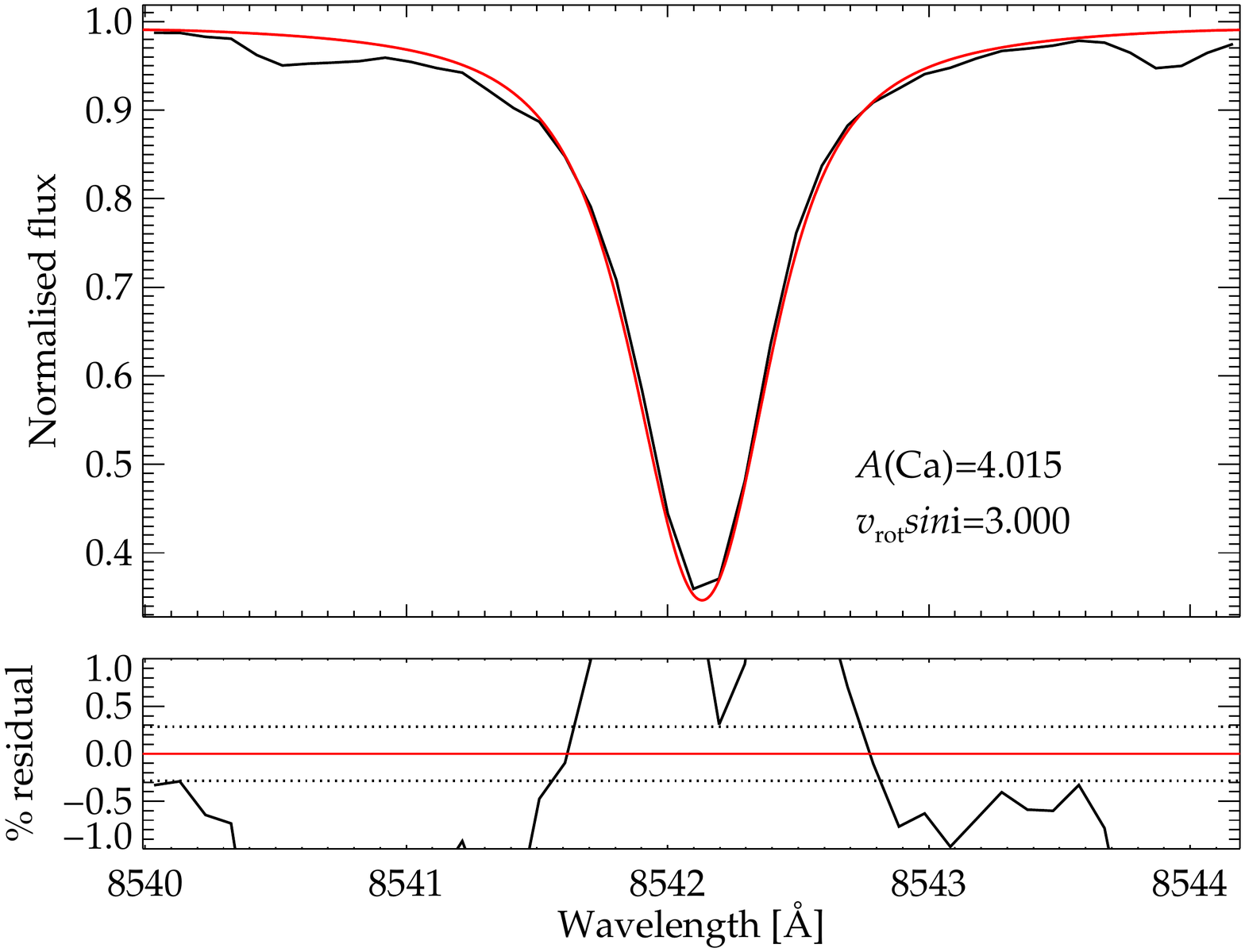}
\end{minipage}
\begin{minipage}[b]{0.33\linewidth}
\centering
\includegraphics[scale=0.23,viewport=5cm 0cm 25cm 21cm]{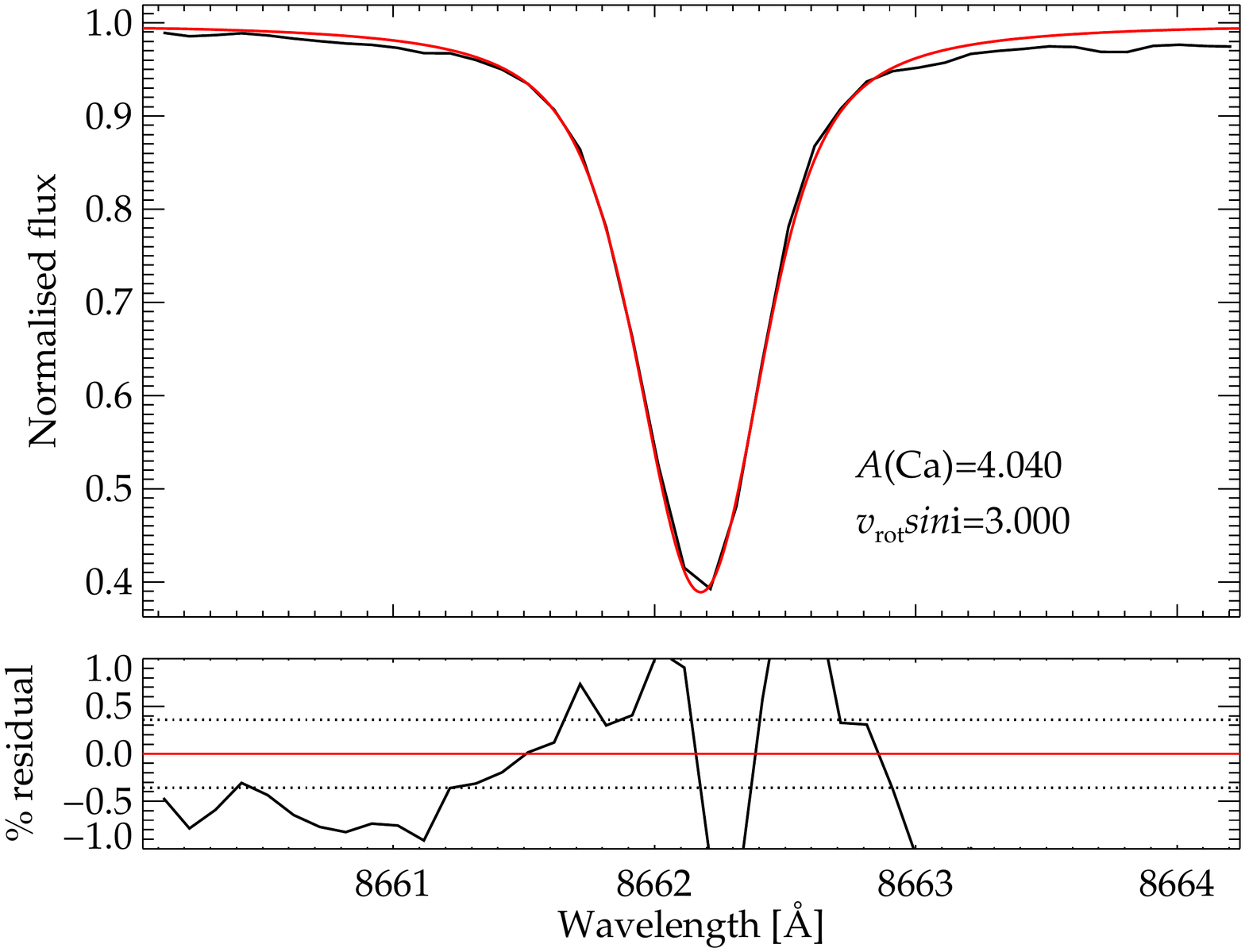}
\end{minipage}
\caption{Same as Fig.\,\ref{fig:hd19445}, but for HD140283.}
\label{fig:hd140283}
\end{figure*}

\begin{figure*}[htb]
\begin{minipage}[b]{0.33\linewidth}
\centering
\includegraphics[scale=0.23,viewport=5cm 0cm 25cm 21cm]{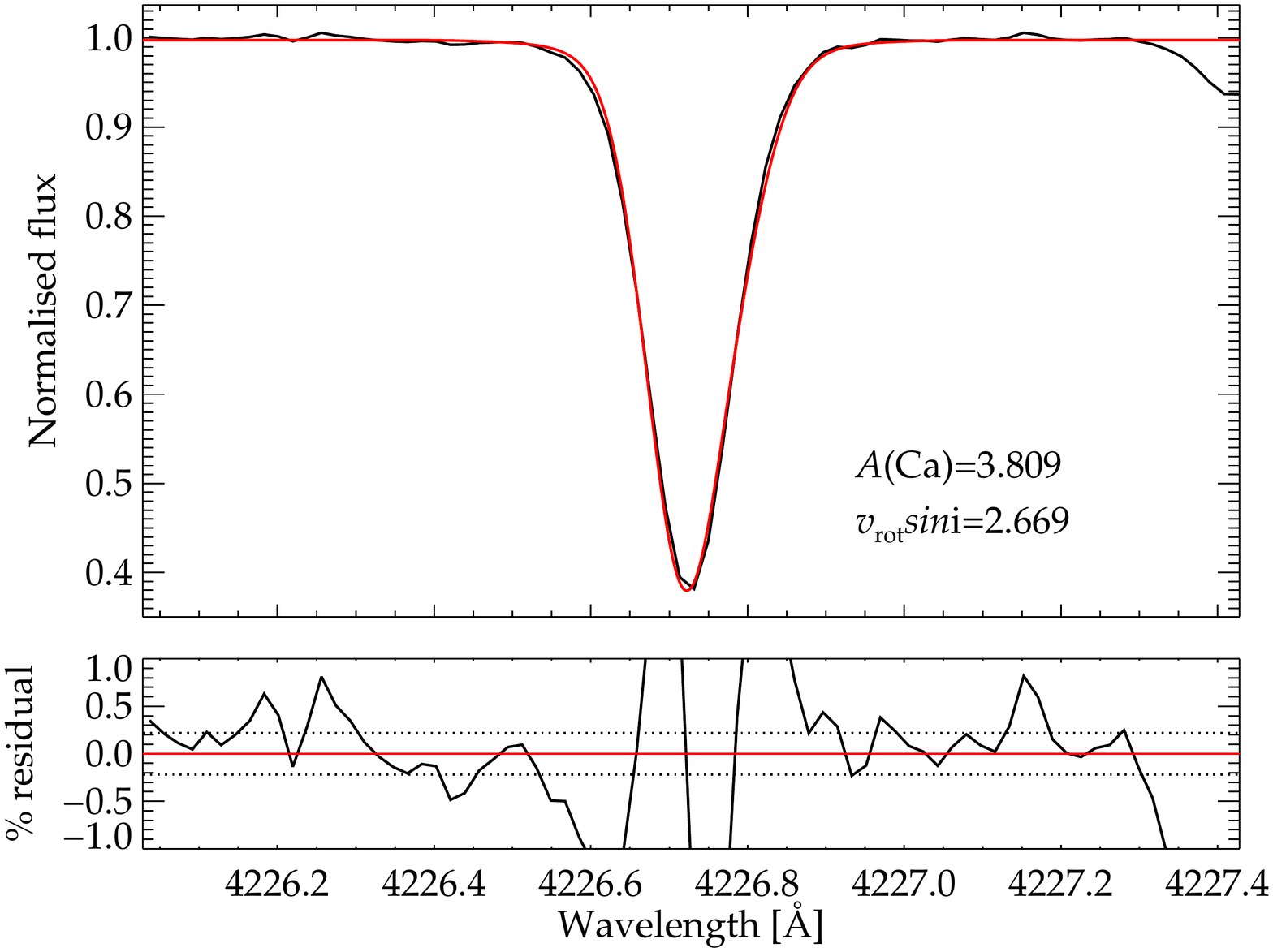}
\end{minipage}
\begin{minipage}[b]{0.33\linewidth}
\centering
\includegraphics[scale=0.23,viewport=5cm 0cm 25cm 21cm]{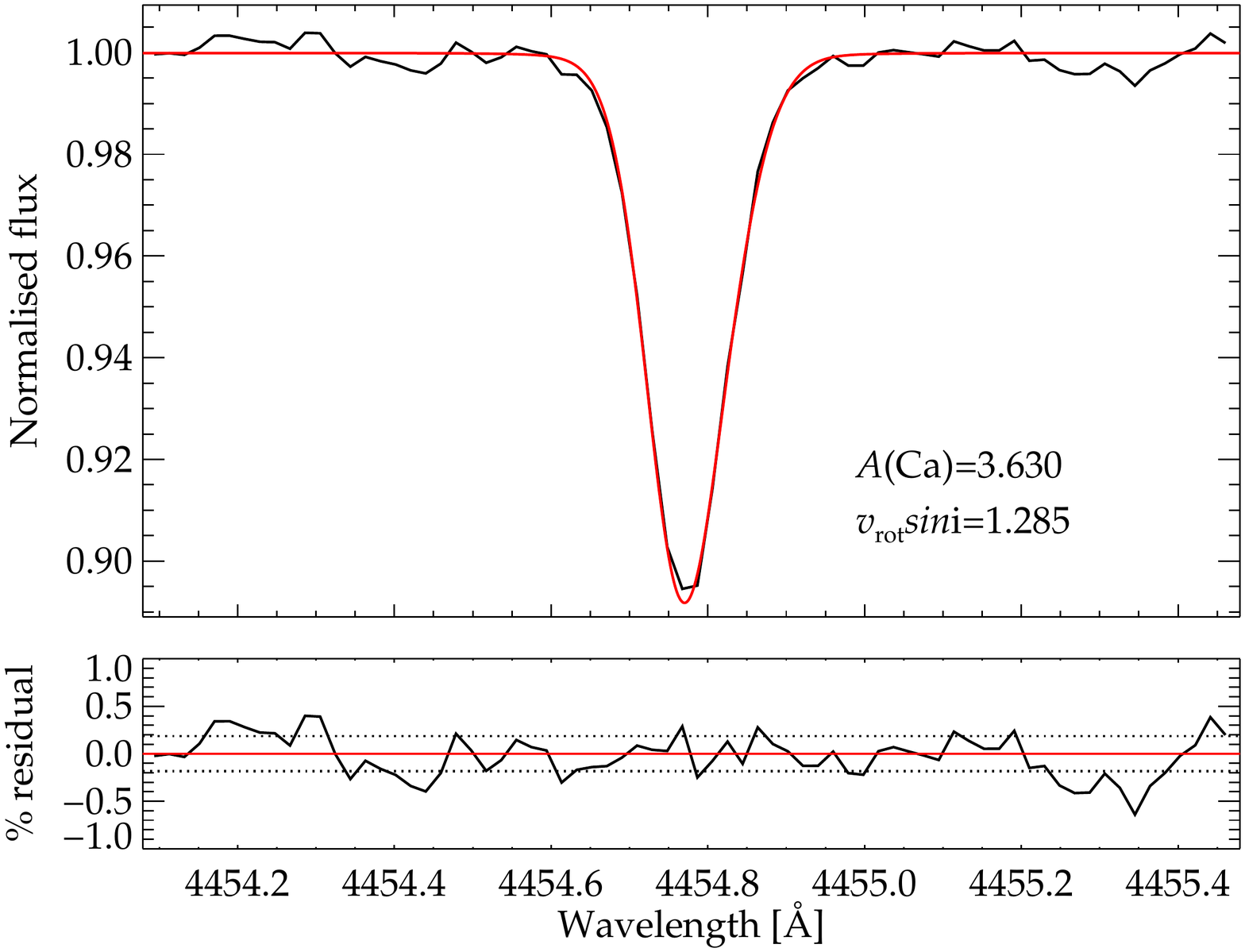}
\end{minipage}
\begin{minipage}[b]{0.33\linewidth}
\centering
\includegraphics[scale=0.23,viewport=5cm 0cm 25cm 21cm]{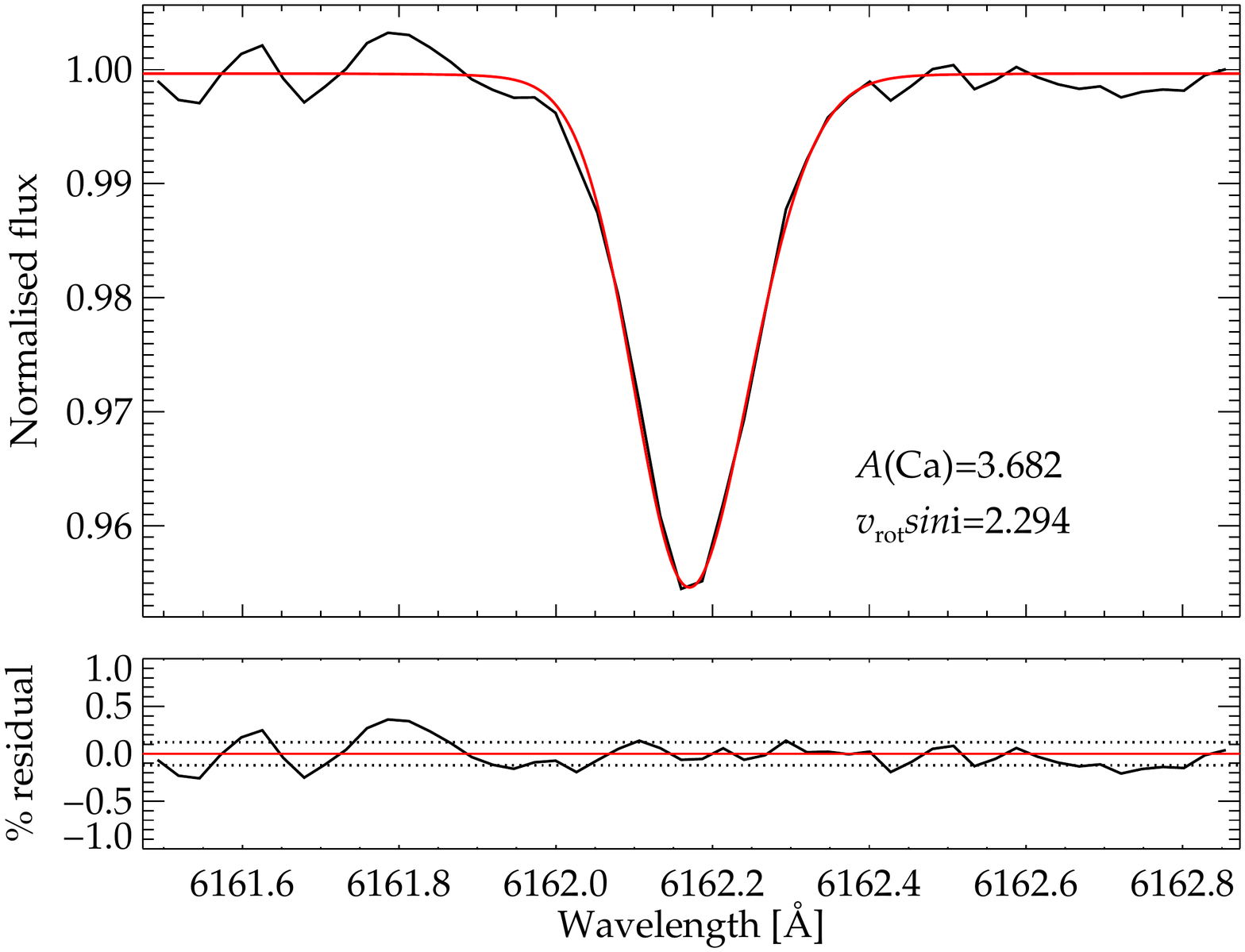}
\end{minipage}
\begin{minipage}[b]{0.33\linewidth}
\centering
\includegraphics[scale=0.23,viewport=5cm 0cm 25cm 21cm]{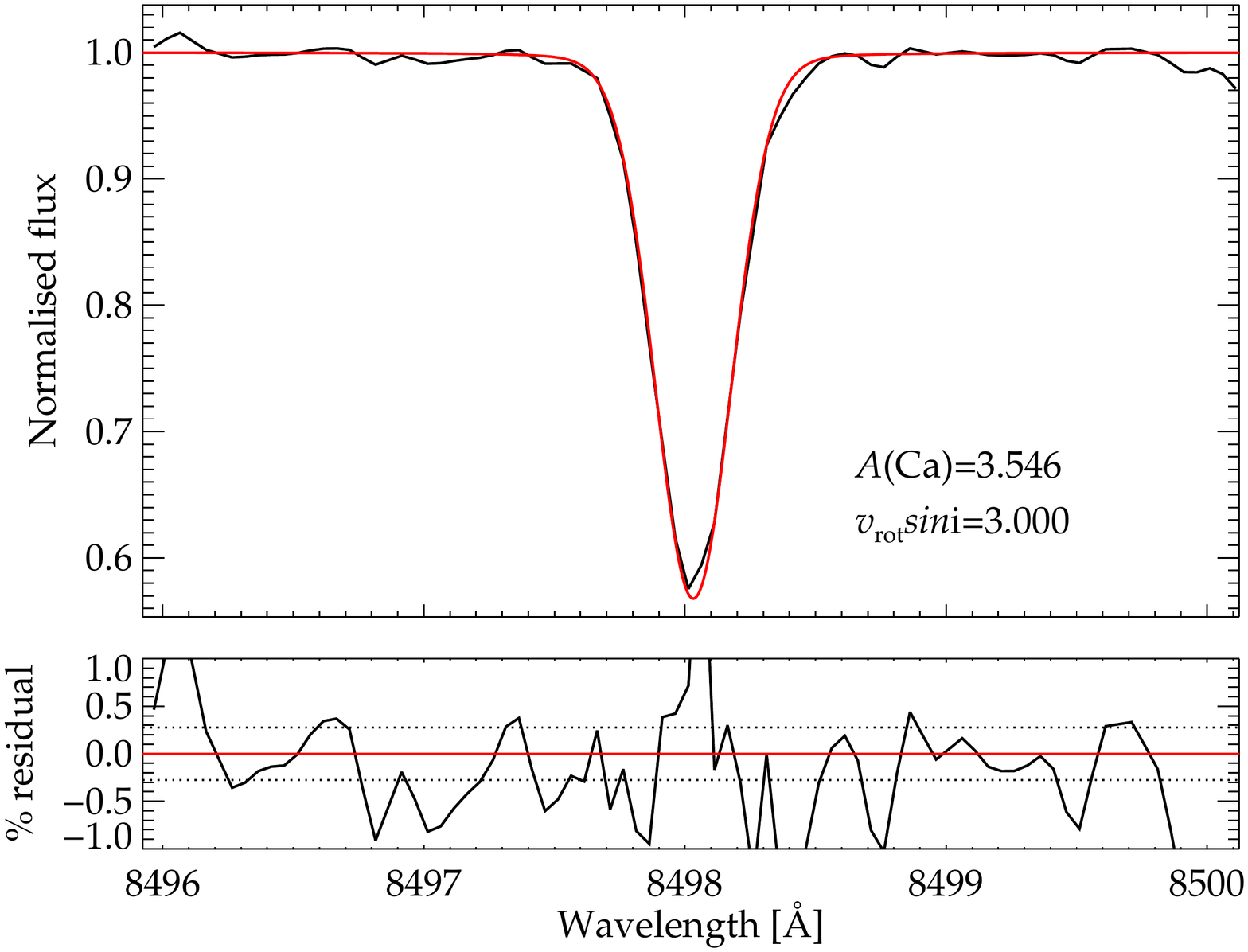}
\end{minipage}
\begin{minipage}[b]{0.33\linewidth}
\centering
\includegraphics[scale=0.23,viewport=5cm 0cm 25cm 21cm]{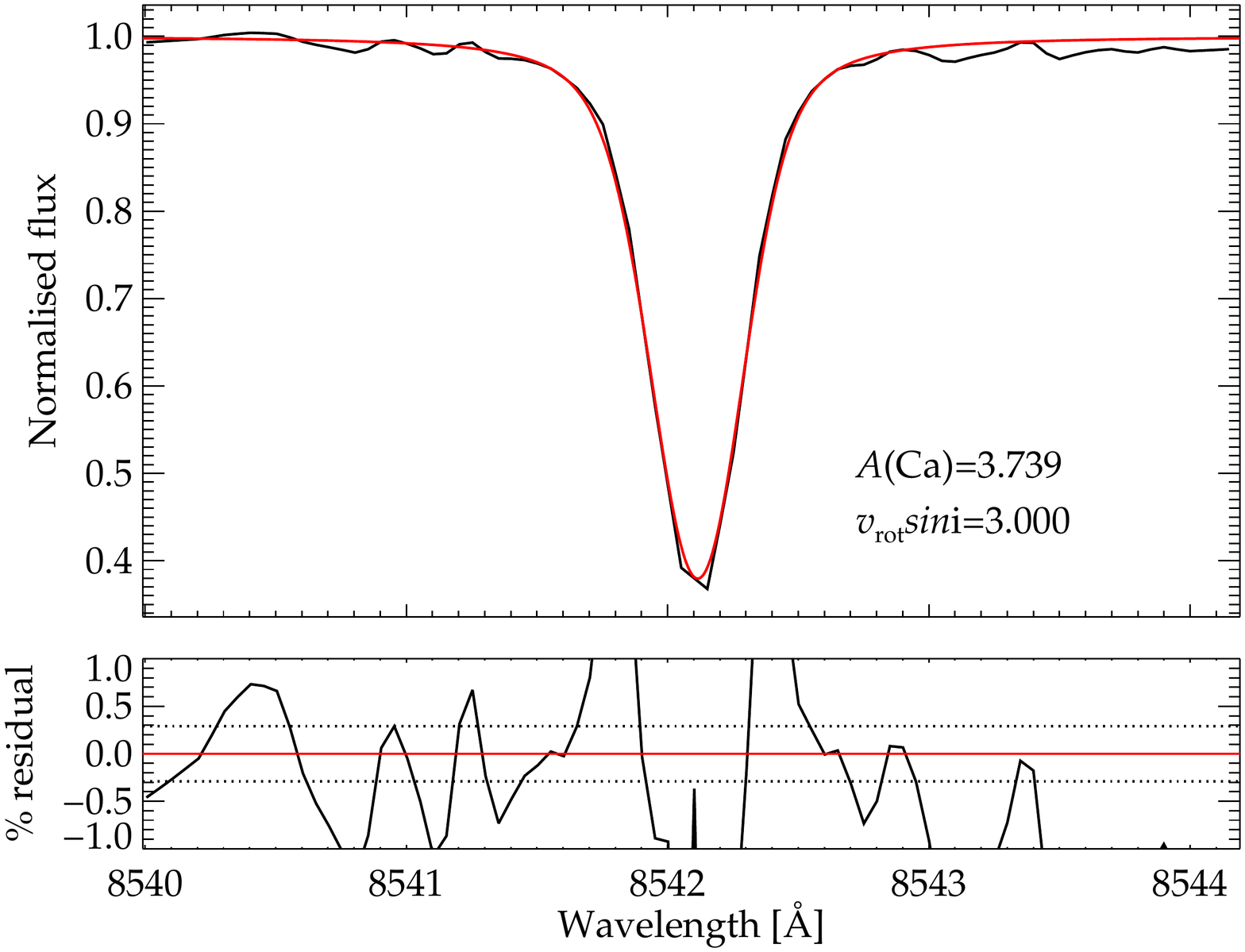}
\end{minipage}
\begin{minipage}[b]{0.33\linewidth}
\centering
\includegraphics[scale=0.23,viewport=5cm 0cm 25cm 21cm]{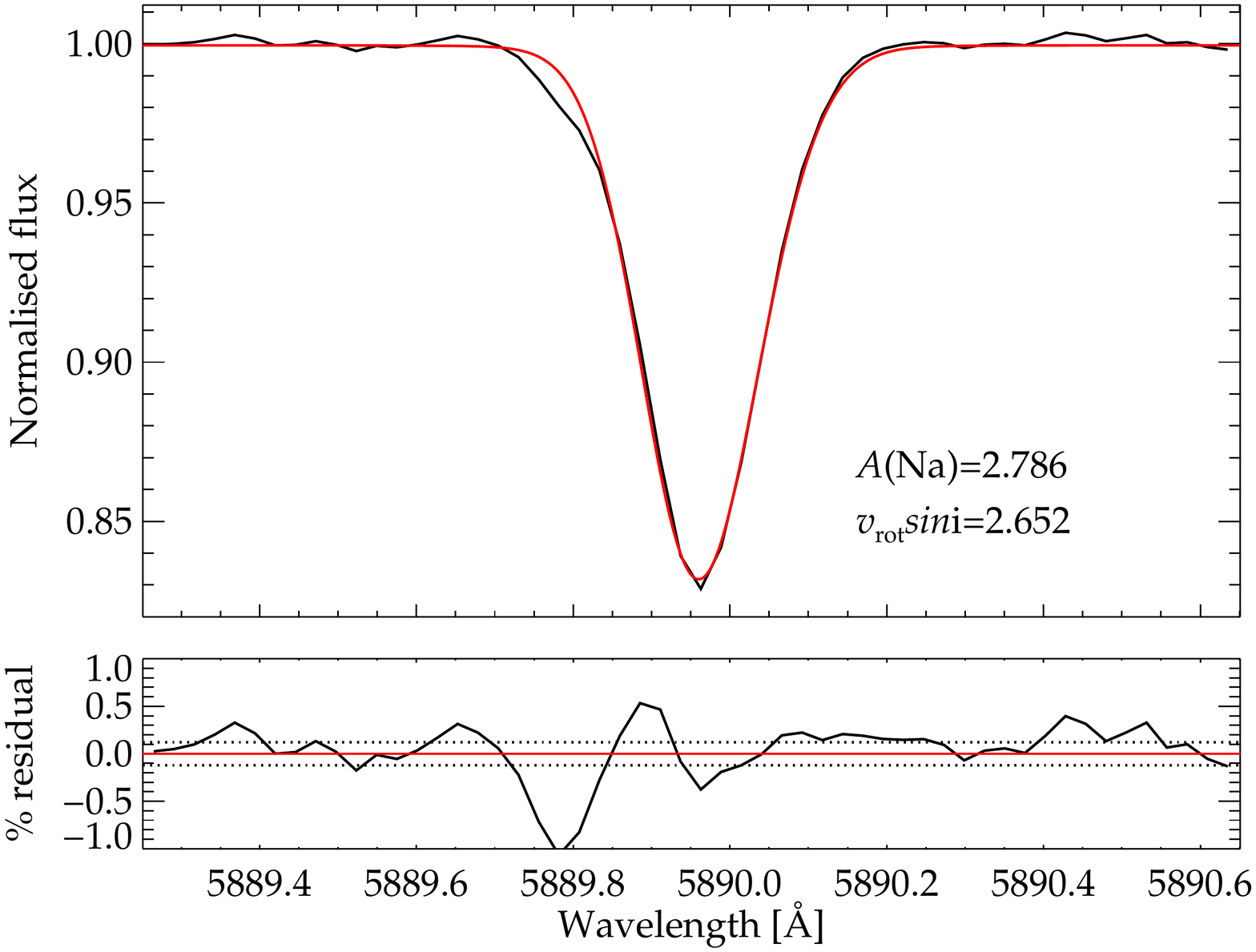}
\end{minipage}
\begin{minipage}[b]{0.33\linewidth}
\centering
\includegraphics[scale=0.23,viewport=5cm 0cm 25cm 21cm]{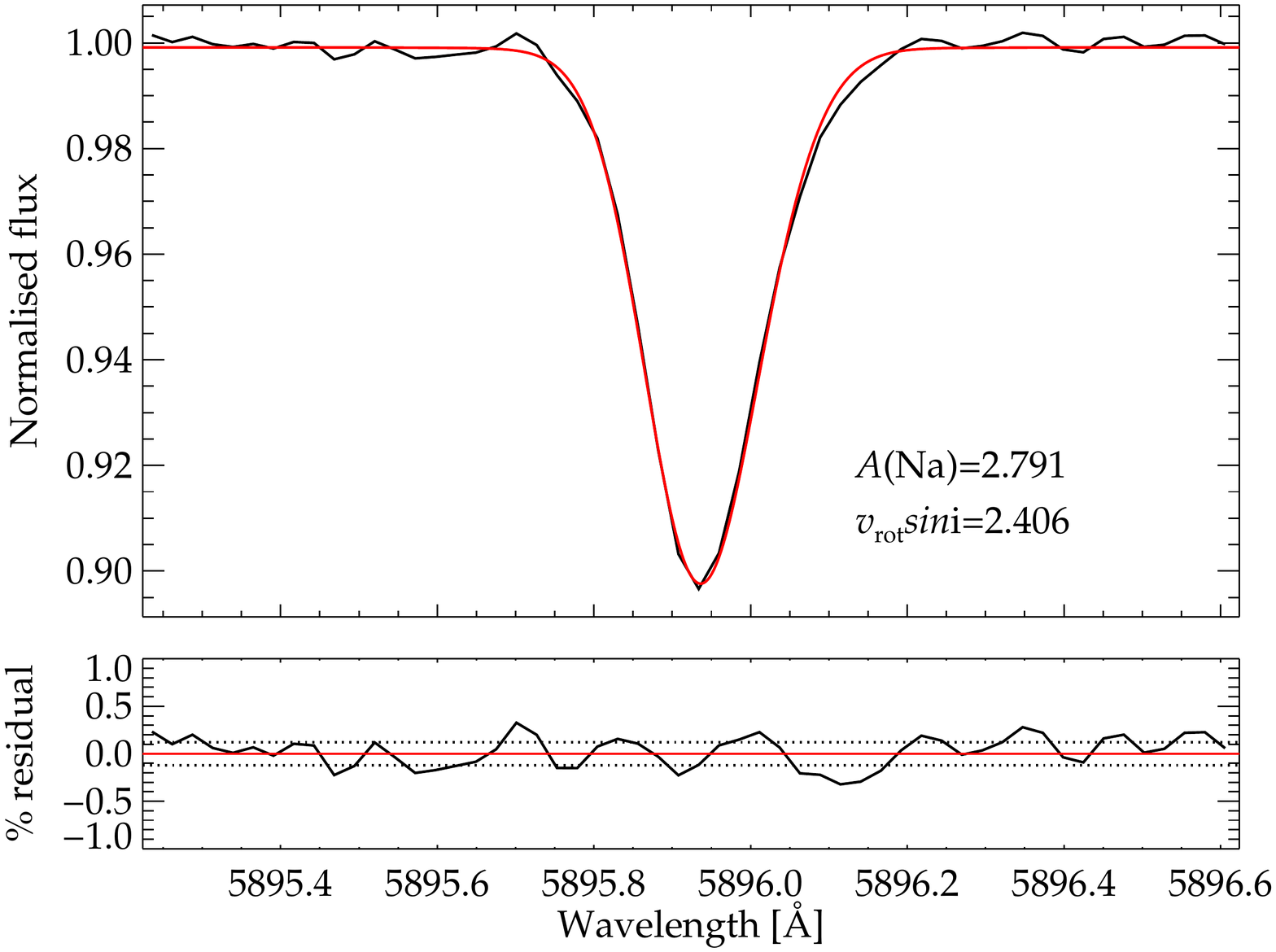}
\end{minipage}
\caption{Same as Fig.\,\ref{fig:hd19445}, but for G64-12.}
\label{fig:g6412}
\end{figure*}
 
\end{appendix}

\end{document}